\documentclass[useAMS,usenatbib]{mn2e}
\usepackage{graphicx}

\title[Dark matter distribution in nearby galaxies]{GHASP :
an H$\alpha$ kinematic survey of spiral and irregular galaxies.
V. Dark matter distribution in 36 nearby spiral galaxies}

\author[M. Spano et al.]   {Spano M.$^{1}$, Marcelin M.$^{1}$, Amram, P.$^{1}$,
Carignan C.$^{2}$, Epinat B.$^{1}$, Hernandez O.$^{2}$
\footnotemark[1]\thanks{Based on observations collected at the Observatoire de Haute Provence}\\
 $^{1}$Laboratoire d'Astrophysique de Marseille, OAMP, 2 Place Le Verrier, 13248 Marseille Cedex 04 France\\
 $^{2}$Laboratoire d'Astrophysique Exp\'erimentale, Universit\'e de Montr\'eal,
 C.P. 6218, Succ. Centre-ville, Montr\'eal, Qc, Canada H3C 3J7}
\begin{document}

\date{Accepted. Received; in original form }

\pagerange{\pageref{firstpage}--\pageref{lastpage}} \pubyear{2006}

\maketitle

\label{firstpage}

\begin{abstract}

The results obtained from a study of the mass distribution of 36
spiral galaxies are presented. The galaxies were observed using
Fabry-Perot interferometry as part of the GHASP survey. The main aim
of obtaining high resolution H$\alpha$ 2D velocity fields is to
define more accurately the rising part of the rotation curves which
should allow to better constrain the parameters of the mass
distribution. The H$\alpha$ velocities were combined with low
resolution HI data from the literature, when available. Combining
the kinematical data with photometric data, mass models were derived
from these rotation curves using two different functional forms for
the halo: an isothermal sphere and an NFW profile. For the galaxies
already modeled by other authors, the results tend to agree. Our
results point at the existence of a constant density core in the
center of the dark matter halos rather than a cuspy core, whatever
the type of the galaxy from Sab to Im. This extends to all types the
result already obtained by other authors studying dwarf and LSB
galaxies but would necessitate a larger sample of galaxies to
conclude more strongly. Whatever model is used (ISO or NFW), small
core radius halos have higher central densities, again for all
morphological types. We confirm different halo scaling laws, such as
the correlations between the core radius and the central density of
the halo with the absolute magnitude of a galaxy: low luminosity
galaxies have small core radius and high central density. We find
that the product of the central density with the core radius of the
dark matter halo is nearly constant, whatever the model and whatever
the absolute magnitude of the galaxy. This suggests that the halo
surface density is independent from the galaxy type.


\end{abstract}


\section{Introduction}

Rotation curves are a fundamental tool for studying the dynamics
and mass distribution in galaxies. The distribution of the total
mass can then be compared with the distribution of visible light,
assuming a certain mass-to-light (M/L) ratio. Observations have
clearly established that dark matter is needed for explaining the
rotation velocities observed in the outer parts of spirals. The
dark matter is most often considered as being distributed in a
spherical dark halo but its density profile, especially at small
radii, is still a matter of debate (Blais-Ouellette et al. 2001,
de Blok \& Bosma 2002, Swaters et al. 2003, Navarro 2004, Graham
et al. 2006, Kusio de Naray 2006, Hayashi et al. 2007).

However, mass models of spiral and dwarf galaxies have well-known
degeneracies (Barnes et al. 2004) that prevent a unique mass
decomposition, the most important being the unknown value of the
stellar mass-to-light ratio (Dutton et al. 2005). Unfortunately,
stellar population models (e.g. de Jong \& Bell 2006) still cannot
predict accurately the (M/L) values of the stellar disks based on
their colors. The main problem comes from the fact that the
normalization of this relation (color vs M/L) depends critically
on the shape of the stellar IMF at the low mass end. It is well
known that the faint stars contribute significantly to the mass
but not to the luminosity and color of the stellar disks. This
means that it will prove difficult to tighten this relation and
lessen the effect of the disk-halo degeneracy.

Cosmological numerical simulations favor cuspy dark halos,
although the value of the inner slope $\gamma$ of the radial
density profile, where $\rho$ $\propto$ r $^{\gamma}$ (see
equation 6 of section 4) depends on the authors, with $\gamma$ =
-1.5 for Moore et al. 1999 as well as Fukushige \& Makino 2001 or
-1.0 for Navarro, Frenk \& White 1997 (hereafter NFW). Recent
simulations (Graham et al. 2006) have extrapolated inner
logarithmic profile slope ranging from -0.2 to -1.5, with a
typical value at 0.1 kpc around -0.7.

Most observers conclude however that $\gamma$ is closer to 0.0
than to -1.0 and that mass models give better results with an
isothermal (or pseudo-isothermal) sphere halo rather than a NFW
profile (e.g. Blais-Ouellette et al. 2001, de Blok \& Bosma 2002,
Swaters et al. 2003, de Blok et al. 2003, de Blok 2005, Kassin et
al. 2006a). Most of these observations are based on the rotation
curves of dwarf and Low Surface Brightness (LSB) galaxies (Kuzio
de Naray et al. 2006). It is still unclear if this is true also
for High Surface Brightness (HSB) systems and for all
morphological types.

One problem in this cusp-core debate is that numerical simulations
mainly predict the halo density profile shape at the time of
formation while galaxies are observed after many Gyrs of
evolution. The problem is that internal dynamics and interaction
between the dark halo and the luminous disk (e.g. adiabatic
contraction: Dutton et al. 2005) and interaction with the
environment (Maccio et al. 2007) may have altered the shape of the
halo density profile.

Moreover, the shape of the gravitational potential in CDM halos
may explain the core-like halo density profiles seen in LSB
systems. Hayashi et al. 2007 suggest that galactic disks may
be forming in elliptical gravitational potential. This could
result in significant non-circular motions in systems such as LSBs
which would mimic constant density cores. Thus, taking into
account the 3D shape of the dark mass distribution could reconcile
the constant density cores observed in LSBs with the predicted
cuspy mass profiles of CDM halos.

Finally, it is unclear if CDM simulations, mainly obtained to
compare with the large scale structure, have sufficient resolution
to reliably probe the kpc scales necessary to compare with the
observationally derived dark matter halo density profiles on the
scale of a few halo core radii. As discussed by Navarro (2004),
unfortunately rotation curves constraints are strongest where
numerical simulations are the least reliable. In fact, rotation
curves are usually compared with extrapolations of the simulation
data that rely heavily on the applicability and accuracy of fitting
formulae such as the NFW profile to regions that may be compromised
by numerical artifact. This is especially true for LSB and dwarf
galaxies (for instance Dutton et al. 2005).

Blais-Ouellette et al. 1999 and Barnes et al. 2004 have shown the
necessity of optical integral field spectroscopy to accurately
determine the rotation curves in the inner parts of spiral and
dwarf galaxies, for which the HI data are affected by beam
smearing (Swaters et al. 2000, van den Bosch et al. 2000). While
the use of two-dimensional data does not necessarily alter the
halo parameters derived from optical long-slit data it however
decreases the uncertainties by roughly a factor of 2 (de Naray et
al. 2006). Blais-Ouellette et al. 2001 and Blais-Ouellette et al.
2004 also pointed out the great sensitivity of the mass
distribution parameters to the inner rotation curve. The ideal
rotation curve will therefore combine high resolution optical
data, for the inner part, with radio data, for the outer part
extending well beyond the luminous disk.

\section{The sample}

\begin{table*}
\caption{Data used for the analysis of our sample.}
\begin{tabular}{cccccccccccccccc}
\noalign{\medskip} \hline
N$^{\circ}$ & N$^{\circ}$ & Type & Distance & M$_{B}$ & D$_{25}$/2 & b/a & i & A$_{G}$(R) & $\Sigma$HI  & HI RC & Phot./Band & h & Rlast \\
 UGC        &  NGC   &    & Mpc & & arcmin & & $^{\circ}$ & & & & & kpc & kpc  \\
 (1)&(2)&(3)&(4)&(5)&(6)&(7)&(8)&(9)&(10)&(11)&(12)&(13)&(14)\\
\hline

2034 & & Im/LSB/dS & 10.1 (Sw02) & -16.41 & 1.3 & 0.79 & 19 & 0.15 & Sw02a & Sw99 & Sw02b/R & 1.26$\pm$0.02 & 5.9 \\
2455 & 1156 & IB(s)m & 7.8 (Sw02) & -18.05 & 1.7 & 0.68 & 51 & 0.60 & Sw02a & Sw99 & Sw02b/R & 0.87$\pm$0.07 & 4.5 \\
2503 & 1169 & SABb & 34.4 (Koo06) & -21.82 & 2.1 & 0.74 & 53 & 0.58 & & vD94 & vD94/R & 5.14$\pm$0.01 & 38  \\
3876 & & SA(s)d & 14.5 (Jam04) & -17.65 & 1.1 & 0.58 & 59 & 0.12 & & & Cou96/R & 1.37$\pm$0.02 & 5.1 \\
4256 & 2532 & SAB(rs)c & 71.7 & -21.65 & 1.1 & 0.83 & 38 & 0.14 & & & Dej94/R & 5.15$\pm$0.01 & 18 \\
4274 & 2537 & SB(s)m & 6.9 (Kar04) & -17.17 & 0.9 & 0.85 & 40 & 0.14 & Sw02a & S02 & Sw02b/R & 0.68$\pm$0.17 & 4.0\\
4325 & 2552 & SA(s)m & 10.1 (Sw02) & -17.76 & 1.7 & 0.66 & 41 & 0.12 & Sw02a & Sw99 & Sw02b/R & 1.73$\pm$0.01 & 7.1 \\
4456 & & SA(rs)c & 74.0 & -20.93 & 0.9 & 0.89 & 27 & 0.10 & & & OHP120/R & 4.71$\pm$0.01 & 29 \\
4499 & & SABdm & 13.0 & -17.36 & 1.3 & 0.74 & 50 & 0.09 & Sw02a & Sw99 & Sw02b/R & 1.40$\pm$0.02 & 8.5 \\
4555 & 2649 & SAB(rs)bc & 58.0 & -20.91 & 0.8 & 0.95 & 35 & 0.09 & & & OHP120/R & 4.25$\pm$0.01 & 17 \\
5175 & 2977 & Sb & 44.1 & -20.62 & 0.9 & 0.44 & 62 & 0.08 & & & OHP120/R & 2.36$\pm$0.01 & 9.6 \\
5272 & & Im/LSB/dS & 7.1 (Kar04) & -15.27 & 1.0 & 0.38 & 59 & 0.06 & Sw02a & Sw99 & Sw02b/R & 0.89$\pm$0.12 & 2.2 \\
5279 & 3026 & Im/LSB & 21.2 & -18.53 & 1.3 & 0.30 & 78 & 0.06 &  &  & OHP120/R & 2.44$\pm$0.01 & 8.8 \\
5721 & 3274  & SABd/dS & 6.5 (Kar04) & -16.38 & 1.1 & 0.48 & 61 & 0.06 & Sw02a & Sw99 & Sw02b/R & 0.43$\pm$0.22 & 7.1 \\
5789 & 3319 & SB(rs)cd/LSB & 14.1 (Saha06) & -19.62 & 3.1 & 0.55 & 65 & 0.04 & Mo98 & Mo98 & Kas06/R & 6.19$\pm$0.01 & 24 \\
5842 & 3346 & SB(rs)cd & 15.2 (Sha01) & -18.67 & 1.4 & 0.87 & 47 & 0.07 & & & Dej94/R & 1.75$\pm$0.02 & 5.8 \\
6537 & 3726 & SAB(r)c & 14.3 & -20.11 & 3.1 & 0.69 & 50 & 0.04 & Ver01 & Ver01 & Kas06/R & 3.07$\pm$0.01 & 26 \\
6778 & 3893 & SAB(rs)c & 15.5 (Shap01) & -20.27 & 0.5 & 0.62 & 49 & 0.06 & Ver01 & Ver01& Kas06/R & 2.06$\pm$0.04 & 17  \\
7045 & 4062 & SA(s)c & 11.4 (Mou06) & -19.00 & 2.0 & 0.43 & 69 & 0.07 & Bro94 & & Kas06/R & 1.89$\pm$0.01 & 7.6 \\
7323 & 4242 & SAB(s)dm & 8.1 (Sw02) & -18.24 & 2.5 & 0.76 & 50 & 0.03 & Sw02a & Sw99 & Sw02b/R & 2.04$\pm$0.01 & 5.9  \\
7524 & 4395 & SAB(s)m/LSB & 4.6 (Kar04) & -17.74 & 6.6 & 0.83 & 46 & 0.05 & Sw02a & Sw99 & Sw02b/R & 3.00$\pm$0.01 & 10 \\
7699 & & SBcd/LSB & 9.3 & -17.51 & 1.9 & 0.27 & 78 & 0.03 & Bro94 & & Jan00/R & 0.67$\pm$0.16 & 6.3  \\
7876 & 4635 & SAB(s)d & 12.8 & -17.68 & 1.0 & 0.70 & 44 & 0.07 & &  & Dej94/R & 1.17$\pm$0.04 & 3.0  \\
7901 & 4651 & SA(rs)c & 20.6 (Sha01) & -20.53 & 2.0 & 0.66 & 53 & 0.07 & War88 & & Dej94/R & 2.69$\pm$0.01 & 15 \\
8490 & 5204 & SA(s)m/dS & 4.7 (Kar04) & -16.88 & 2.5 & 0.60 & 50 & 0.03 & Sw02a & Sw99 & Sw02b/R & 0.67$\pm$0.07 & 10  \\
9179 & 5585 & SAB(s)d & 5.7 (Kar04) & -17.73 & 2.9 & 0.65 & 54 & 0.04 & Cot91 & Cot91 & Cot91/R & 1.07$\pm$0.05 & 8.8 \\
9219 & 5608 & Im/dS & 10.2 (Jam04) & -16.35 & 1.3 & 0.51 & 59 & 0.03 & & & Jan00/R & 0.69$\pm$0.23 & 3.7  \\
9248 & 5622 & Sb & 54.9 & -20.11 & 0.9 & 0.58 & 55 & 0.06 & & & Cou96/R & 2.79$\pm$0.01 & 11  \\
9465 & 5727 & SABdm/LSB & 26.4 (Jam04) & -18.25 & 1.1 & 0.52 & 73 & 0.04 & & & Her96/I & 2.29$\pm$0.01 & 12  \\
9866 & 5949 & SA(r)bc & 7.4 (Jam04) & -17.07 & 1.1 & 0.47 & 57 & 0.06 & & & Cou96/R & 0.54$\pm$0.27 & 1.6  \\
10075 & 6015 & SA(s)cd & 14.7 (Jam04) & -19.78 & 2.7 & 0.40 & 65 & 0.03 & & & Her96/I & 2.49$\pm$0.01 & 12  \\
10310 & & SB(s)m & 12.7 (Jam04) & -17.00 & 1.4 & 0.79 & 34 & 0.03 & Sw02a & Sw99 & VZ00/B & 1.50$\pm$0.03 & 7.4  \\
11557 & & SAB(s)dm & 19.7 (Jam04) & -19.12 & 1.1 & 0.80 & 30 & 0.63 & Sw02a & Sw99 & Sw02b/R & 1.70$\pm$0.02 & 8.6  \\
11707 & & SAdm/LSB & 15.9 (Sw02) & -17.54 & 1.8 & 0.52 & 55 & 0.44 & Sw02a & Sw99 & Sw02b/R & 3.04$\pm$0.01 & 15  \\
11914 & 7217 & SA(r)ab & 15.0 (Koo06) & -20.35 & 1.9 & 0.83 & 28 & 0.24 & VM95 & VM95 & VM95/R & 1.76$\pm$0.20 & 9.1  \\
12060 & & IBm/LSB & 15.7 (Sw02) & -16.60 & 0.8 & 0.65 & 40 & 0.24 & Sw02a & Sw99 & Sw02b/R & 1.58$\pm$0.01 & 12  \\
\hline
\end{tabular}
\\(1) name of the galaxy in the UGC catalog. (2) name in the NGC
catalog when available. (3) morphological type from the RC3
catalog, completed by a mention LSB and dS as explained in section
2 of the text. (4) Distance and source. When no reference is
given, the distance has been computed as explained in section 2 of
the text. Jam04=James et al. 2004 ; Kar04=Karachentsev et al. 2004
; Koo06=Koopmann et al. 2006 ; Mou06=Moustakas \& Kennicutt 2006 ;
Saha06=Saha et al. 2006 ; Sha01=Shapley et al.2001 ; Sw02=Swaters
\& Balcells 2002. (5) Absolute B magnitude computed from the
corrected apparent B magnitude B$_{T0}$ from the RC3 using the
distance of column 4. (6) radius at 25 B mag/$arcsec^{2}$ from the
RC3. (7) axis ratio at 25 B mag/$arcsec^{2}$ from the RC3.(8)
inclination adopted for deriving the rotation curve. (9) Galactic
absorption in the R band, computed as explained in section 2 of
the text. (10) source of total HI distribution. Bro94=Broeils \&
van Woerden 1994 ; Cot91=Cot\'{e} et al. 1991 ; Mo98= Moore 1998 ;
Sw02a=Swaters \& Balcells 2002 ; Ver01=Verheijen and Sancisi 2001
; VM95=Verdes-Montenegro et al. 1995 ; War88=Warmels 1988 (11)
source of HI rotation curve. Cot91=Cot\'{e} et al. 1991 ; Mo98=
Moore 1998 ; S02=Stil 2002 ; Sw99=Swaters 1999 ; vD94=van Driel
1994 ; Ver01=Verheijen \& Sancisi 2001 (12) source of photometry
and corresponding band. Cot91=Cot\'{e} et al. 1991 ;
Cou96=Courteau 1996 ; Dej94=De Jong 1994 ; Her96=H\'{e}raudeau and
Simien 1996 ; Jan00=Jansen et al.2000 ; Kas06=Kassin et al.2006b ;
OHP120=Observed by Spano with 1.20m OHP in March 2006 ;
Sw02b=Swaters et al. 2002 ; vD94=van Driel 1994 ; VZ00=van Zee
2000 (13) Disk scale length, in kpc, derived from the photometric
profile of column 12, with distance of column 4. (14) Radius of
the last point, in kpc, of the rotation curve.

\end{table*}

The sample of 36 galaxies studied in this paper is composed of
nearby galaxies observed for the GHASP survey. These 36 galaxies
are simply the GHASP galaxies already reduced (113 out of a total
of $\sim$ 200 observed galaxies) for which there is photometry
available in the literature and for which we could draw an
acceptable rotation curve at H$\alpha$ wavelength. Half of them
has been already published and another half has been reduced
recently and only available on the GHASP website. For 19 of them,
the velocity fields and rotation curves have been published in
Garrido et al. 2002, 2003, 2004 and 2005; 17 other galaxies
(observed in 2004) have been reduced recently and the data will be
published later (Epinat et al. 2008). However, their monochromatic
images, velocity fields and rotation curves are already available
on the GHASP web site :
http://www.oamp.fr/interferometrie/GHASP/ghasp.html.

Eight other galaxies from the same observing run (2004) are also
presented on the GHASP website but they are not analyzed here,
because no photometry is available (UGC 4770, 5319, 5373 and IC
2542) or because of problems with the rotation curve (UGC 4393,
6277, 8863 and 9358), as explained in the next paragraph.

The GHASP survey provides high resolution velocity fields of
nearby galaxies at the H$\alpha$ wavelength. For the analysis done
in this paper, we selected the galaxies for which photometric data
were available in the literature. Then we excluded all the
galaxies with strong differences between the red and the blue side
of the rotation curve or displaying high velocity dispersions
hence difficult to fit properly with mass models. These galaxies
are: UGC 4278 (seen edge-on), UGC 4305 (strong velocity dispersion
and counter rotation in the center), UGC 4393 (anomalous rotation
curve with counter rotation in the center), UGC 5414 (strong
dispersion and asymmetry of the velocities between receding and
approaching side), UGC 6277 (rotation curve abnormally flat in the
central part), UGC 7971 (strong dispersion and asymmetry of the
velocities between receding and approaching side), UGC 8863 (not
enough H$\alpha$ emission to get a correct rotation curve) and UGC
9358 (very strong asymmetry : plateau immediately reached for the
approaching side and solid body rotation for the receding side).

In Table 1, the morphological type is completed by an additional
mention: LSB for low surface brightness galaxies and dS for dwarf
spirals. We have adopted the same criterion as Pizzella et al.
2005, considering that LSB have a disk with a central face-on
surface brightness $\mu_{B}$ $\geq$ 22.6 mag/arcsec$^{2}$. The
criteria for dwarf spirals are those defined by Schombert et al.
1995. We thus find that 9 galaxies of our sample are LSB and 5 are
dwarf spirals, UGC 2034 and 5272 being both. All of them are late
type galaxies: 2 Scd, 2 Sdm, 1 Sm and 4 Im for the 9 LSB, 1 Sd, 1
Sm and 3 Im for the 5 dS.

Since there are only 9 LSB and 5 dS in our sample, it is
not possible however to make any significant study for these
classes. Our whole sample is in fact not big enough to reach
strong conclusions and most results presented here are trends to
be confirmed.

When HI data were available (which is the case for more than half
of the selected sample) we derived an hybrid rotation curve, by
combining our optical data with the HI data from the literature.
Table 1 gives the origin of the HI data (when available) and of
the photometry data used to derive our mass models. The
morphological types are also given in that table, together with
the main parameters adopted for the galaxies analyzed in this
paper (distance, radius at 25 B mag/arcsec$^{2}$, axis ratio and
Galactic absorption).

The distances for each galaxy are also given together
with the corresponding references in the literature. Some of these
distances are based on local indicators such as the brightest
supergiants, cepheids, red giant branch position or group
membership but most of them are based on the Hubble law, assuming
$H_{o}$=75 km s$^{-1}$ Mpc$^{-1}$ and taking into account the
Virgo Infall. When no precise value of distance was found in the
literature, we simply divided by $H_{o}$=75 km s$^{-1}$ Mpc$^{-1}$,
using the velocity correction for infall of the Local Group towards Virgo
(vvir) given in the HyperLeda data base.

The galactic absorption in the R band has been determined from
that in the B band given in the HyperLeda data base, assuming
A(R)/A(B)=0.62 (from the values of A(R)/A(V) and A(B)/A(V) given
in Table 6 of Schlegel et al. 1998, evaluated using the
$R_{V}$=3.1 extinction laws of Cardelli et al. 1989 and O'Donnell
1994). The value A(R)/A(B)=0.62 has been confirmed by Choloniewski
\& Valentijn 2003 using a new method for the determination of the
extinction in our Galaxy, based on the observation of surface
brightnesses of external galaxies in the B and R bands. Only 4
galaxies of our sample are affected by a relatively strong
Galactic absorption in the R band (between 0.4 and 0.6 (see column
8 of Table1) which should not affect our conclusions.


\section{The data set}
\subsection{The hybrid rotation curve}

The hybrid rotation curve was obtained with the following method.

In the central part of the galaxy, we preferred the H$\alpha$
curve because of its higher spatial resolution. The curve was
extended in the outer parts with the HI data when available.

The rotation curves given on figures 6 to 14, in the annex, show
the hybrid rotation curve used for the fit of the models, with
full dots for the H$\alpha$ velocities and open circles for the
HI velocities. The HI data in the inner parts are displayed as
crosses, for comparison with our H$\alpha$ data, but they are not
used when fitting the models.

The H$\alpha$ rotation curves of the GHASP survey are derived from
the 2D velocity fields obtained from the scan of the H$\alpha$
line with a Fabry-Perot. Each velocity field is composed of a lot
of individual radial velocity points (hereafter mentioned as
velocity pixels) since a velocity is computed for each pixel as
soon as the signal to noise ratio is sufficient. The error bars of
our rotation curves were computed up to now, for every point of
the curve, from the velocity dispersion found among the velocity
pixels contributing to that very point. Each galaxy being
considered as a rotating disk, the velocity pixels contributing to
a given point of the rotation curve are found to be lying within
an elliptical ring in the sky. However, we do not take into
account the points too close from the minor axis (otherwise the
dispersion increases rapidly because the non circular motions are then
amplified when converting radial velocities into rotational
velocities). Also, when computing the final rotation curve, both
approaching and receding sides are averaged.

The velocity points of the rotation curve resulting from the
average of a large number of velocity pixels are much more
accurate and reliable than those obtained from a limited number of
pixels. That is why we weight the original dispersion by the
number of velocity pixels contributing to a given point of the
rotation curve. To do that, we simply divide the original error
bar by the square root of this number.

Whenever a rotation velocity point is computed from a too small
number of velocity pixels we simply attribute to this point the
average error bar found for the other points of the rotation
curve.

The inclination adopted for deriving the rotation curves from the
radial velocities (given in Table 1) are generally close from the
values found in the literature and deduced from the photometric
axis ratio (also given in Table 1). However, in some cases we
found a significantly smaller dispersion of the rotation
velocities when adopting a different inclination. Note also that
the inclination has been changed for two GHASP galaxies previously
published in Garrido et al. 2005: for UGC 11557 it is now
30$^{\circ}$ instead of 37$^{\circ}$ and for UGC 11914 it is now
28$^{\circ}$ instead of 35$^{\circ}$. The highest rotation
velocity dispersions are found for the smallest inclinations since
the original radial velocities are then divided by the sine of the
inclination, leading to large error bars on the rotation curves of
almost face-on galaxies. This effect remains marginal for
inclinations between 40$^{\circ}$ and 30$^{\circ}$ (five galaxies
of our sample are concerned) but becomes noticeable for
inclinations smaller than 30$^{\circ}$ because it more than
doubles any velocity fluctuation. Only three galaxies of our
sample have such small inclinations:  UGC 11914 at 28$^{\circ}$,
UGC 4456 at 27$^{\circ}$ and, the closest to face-on, UGC 2034 at
19$^{\circ}$. Note that it is one of the three irregular galaxies
excluded from the fits in Fig.2 to Fig.5 because of chaotic
behavior and large error bars of the rotation curve.

Most of the galaxies of our sample are rather nearby. However, 6
of them have distances larger than 30 Mpc. They are of Sb, Sbc or
Sc type and large enough on the sky so that there is no problem of
sampling or spatial resolution for deriving their rotation curve.
Indeed these 6 galaxies exhibit well behaved rotation curves with
rather small error bars.

We defined a "quality parameter" for each rotation curve,
going from 1 (for best quality curves) to 4 (for very low quality
curves). This parameter was estimated from characteristics such as
asymmetry, velocity dispersion, bar, etc. Surprisingly, no
significant difference was found on our plots between low and high
quality points (indeed the dispersion was about the same, whatever
the quality parameter) and we finally decided not to use this
parameter for the plots. Nevertheless, it is worth being mentioned
that 12 galaxies of our sample have good quality RC (class 1): UGC
3876, 4499, 5175, 5721, 6778, 7323, 7524, 8490, 9866, 11557,
11707, 11914; 11 galaxies have acceptable quality RC (class 2):
UGC 4256, 4456, 4555, 5279, 7876, 9179, 9219, 9248, 9465, 10075,
12060; 9 galaxies have low quality RC (class 3): UGC 2034, 2455,
2503, 4325, 5272, 5842, 7045, 7901, 10310; 4 galaxies have very
low quality RC (class 4): UGC 4274, 5789, 6537, 7699.

\subsection{The luminosity profile}

We used the photometric data in the R band from the literature in
order to derive the distribution of stellar mass. Additionally 4
galaxies were observed in March 2006, with the 1.20m telescope at
OHP, in order to get their luminosity profile in the R band (see
Table 1 for details). For UGC 9465 and UGC 10075 only the I band
was available in the literature and we used the color indexes to
derive the R band. For this purpose we used the effective color
indexes V-I and I-R given by Prugniel \& H\'{e}raudeau 1998 as a
function of the morphological type of galaxies. Same thing for
UGC10310 for which only the B band was found in the literature, we
derived the R band now using B-V and V-R from Prugniel \&
H\'{e}raudeau 1998. An exponential was then fitted to the outer
parts of the photometric profile and subtracted in order to check
for the presence of a central bulge. Whenever a bulge was seen on
the residual luminosity profile we decomposed the profile into two
components. However, some of these residuals were found to be a
mere nucleus and not kept into account. Also, we assumed that
galaxies later than Scd type had no bulge and discarded any
central component for these galaxies. As a result, only 4 galaxies
of our sample were decomposed into bulge and disk components (see
Table 2). Best fit models for extracting the disk and bulge
components were carried out by minimizing the $\chi$$^{2}$ using
the Minuit package, with the Simplex and Milgrad routines from
Fletcher 1970.

The bulge component has been extracted from the observed
luminosity profile in the central part of the galaxy, after
extrapolating the exponential profile of the underlying disk by an
exponential law :

\begin{equation}
\mu_{disk}(r)= \mu_{o}+1.0857(\frac{r}{h})
\end{equation}

where $\mu_{o}$ is the central surface brightness of the disk and
$h$ the disk scale length.

The remaining luminosity profile has been fitted by a de
Vaucouleurs law :

\begin{equation}
\mu_{bulge}(r)=
\mu_{e}+8.3268[(\frac{r}{r_{e}})^{1/4}+(\frac{r}{r_{t}})^{4}]
\end{equation}

where $\mu_{e}$ is the central surface brightness of the bulge and
$r_{e}$ the bulge scale factor. $r_{t}$ is the truncation radius
(limit beyond which the bulge is no more detected, taken here as
the radius where the luminosity of the bulge reaches the magnitude
30/arcsec$^{2}$).

This bulge profile was then subtracted from the total luminosity
profile observed in order to get the true disk component alone. It
is this last component that was used for the mass modelling and
not the pure exponential component given above (which was merely
an intermediate step to get the bulge profile as best as
possible). Some discrepancies with other authors must be mentioned
however. For instance we found no bulge component for UGC 7045,
whereas Baggett et al. 1998 found one. Conversely we found a bulge
for UGC 7901 and 11914, while Kassin et al. 2006a considered the
bulge as negligible for both of them (which is especially
surprising for UGC 11914 since it is an Sab galaxy).

\subsection{The distribution of neutral gas}

The distribution of neutral hydrogen (found in the literature) has
been multiplied by 1.33 in order to take into account helium. The
thickness of the disk was calculated for each galaxy, depending on
its type, with the formula given by Bottinelli et al. 1983.

Then, the halo contribution is deduced from the contribution of
the different components to the final rotation curve we observe :

\begin{equation}
V_{rot}(r)=(V_{gaz}^{2}+V_{disk}^{2}+V_{bulge}^{2}+V_{halo}^{2})^{0.5}
\end{equation}

For 14 galaxies of our sample there is no HI distribution in the
literature (see Table 1). However, since the influence of the
neutral gas is most often negligible in the central part of the
rotation curves of galaxies, it should not change our conclusions
about the shape of the dark matter halo in the center.

In order to check that, we have run our models for the 6 galaxies
of our sample having a strong HI contribution and set the HI
component at zero. The comparison of the results, with and without
HI, shows that the change in both $R_{o}$ and $\rho_{o}$ ranges
from 0\% to 33\%, with an average around 15\%. The change in
$\chi^{2}$ is slightly smaller, with an average around 10\%.

\section{Density profiles and fitting procedure}

The method used in this paper to model the mass distribution is
the same as described in Blais-Ouellette et al. 2001.

Since we have no a priori knowledge of the dark matter halo shape,
we assumed the simplest shape of a spherical and symmetric
distribution of matter (isothermal sphere):

\begin{equation}
\rho(r)=\frac{\rho_{o}}{{{(1+(\frac{r}{R_{o}})^{2})}}^{3/2}}
\end{equation}

where $R_{o}$ is the core radius, and  $\rho_{o}$ the central
density of dark matter. This kind of profile is flat in the
center. However, N-body simulations, in the framework of the
$\Lambda$CDM theory, favor cuspy halo profiles, peaked in the
center (e.g. Navarro, Frenk \& White, 1997, Moore et al. 1999,
Fukushige \& Makino 2001)

For instance, the NFW profile is :
\begin{equation}
\rho_{NFW}(r)=\frac{\rho_{o}}{(\frac{r}{R_{o}})(1+\frac{r}{R_{o}})^{2}}
\end{equation}
where $R_{o}$ is the core radius and $\rho_{o}$ is the central
density of dark matter. The density $\rho(r)$ is thus proportional
to $r^{-1}$ when r is small.

We used for our fits the density profiles defined by Zhao 1996 :
\begin{equation}
\rho(r)=\frac{\rho_{o}}{(c+(\frac{r}{R_{o}})^{-\gamma})(1+(\frac{r}{R_{o}})^{\alpha})^{\frac{\beta+\gamma}{\alpha}}}
\end{equation}
where $\rho_{o}$ is the central density and $R_{o}$ the core
radius. The density profiles are then defined by the set of
parameters (c,$\alpha$,$\beta$,$\gamma$). Our fitting procedure
was led with the two following sets : (0,2,3,0) corresponding to
the isothermal sphere (equation 4) with a constant central density
and (0,1,3,-1) corresponding to the NFW profile (equation 5), with
a peaked central density.

For each of the 36 galaxies studied here, we have applied the best
fit model technique for these two profiles (isothermal sphere,
hereafter ISO, and NFW). It consists in minimizing the $\chi^{2}$
in the 3 dimensions of the parameter space defined by
((M/L)$_{disk}$, $\rho_{o}$, $R_{o}$). Whenever a bulge component
is taken into account, a 4th dimension is added with
(M/L)$_{bulge}$.

Some constraints were added to these parameters in order to avoid
non physical values. We thus limited M/L to a minimum value of 0.1
for both disk and bulge, $R_{o}$ to a minimum value of 1kpc and
$\rho_{o}$ to a minimum value of 10$^{-3}$ solar masses pc$^{-3}$.
These minima values were arbitrarily chosen (although
suggested by the current values encountered for these
parameters).

Table 2 summarizes the results of the best fits obtained with our
mass models (for both NFW and ISO profiles of the dark matter
halo) in terms of luminosity ratios of the visible components,
central density and core radius of the dark matter halo together
with the corresponding $\chi^{2}$.

For the calculation of our mass luminosity ratios, we adopted 4.28
for the absolute magnitude of the sun in the R band (Allen
Astrophysical Quantities, 4th edition, 2000).

\begin{table*}
\caption{Results of the best fits of rotation curves with
isothermal sphere and NFW models of halo profiles}
\begin{tabular}{cccccccccccccccc}
\noalign{\medskip} \hline

N$^{\circ}$ & N$^{\circ}$ & M/L & M/L & $R_{o}$ & $\rho_{o}$ & $\chi^{2}$ & M/L & M/L & $R_{o}$ & $\rho_{o}$ & $\chi^{2}$ & Best & Class & \\
 UGC &  NGC   & Disk & Bulge & kpc & 10$^{-3}$M$_{\odot}$pc$^{-3}$ & & Disk & Bulge & kpc & 10$^{-3}$M$_{\odot}$pc$^{-3}$ & & model & & \\
 &    & ISO & ISO & ISO & ISO & ISO & NFW & NFW & NFW & NFW & NFW & & & \\
 (1)&(2)&(3)&(4)&(5)&(6)&(7)&(8)&(9)&(10)&(11)&(12)&(13)&(14)& \\
\hline

2034 & & 1.9 & & 184 & 1 & 0.97 & 0.83 & & 10 & 1 & 0.90 & Both & 4 & \\
2455 & 1156 & 0.10 & & 30 & 5 & 1.24 & 0.10 & & 5.8 & 1 & 2.33 & ISO & 3 & \\
2503 & 1169 & 6.8 & 1.5 & 188 & 1 & 1.27 & 6.0 & 0.73 & 57 & 1 & 1.91 & Both & 2 & \\
3876 & & 7.9 & & 100 & 23 & 0.60 & 0.19 & & 33 & 5 & 0.47 & Both & 4 & \\
4256 & & 0.27 & & 3.4 & 49 & 0.60 & 0.10 & & 6.8 & 19 & 0.74 & Both & 1 & \\
4274 & & 0.42 & & 1.0 & 253 & 2.34 & 0.43 & & 2.3 & 60 & 1.73 & NFW & 0 & \\
4325 & 2552 & 2.6 & & 1.9 & 121 & 2.50 & 9.00 & & 2.7 & 1 & 2.90 & Both & 4 & \\
4456 & & 1.0 & & 15 & 1 & 1.54 & 0.20 & & 14 & 2 & 1.48 & Both & 1 & \\
4499 & & 1.7 & & 2.2 & 43 & 0.90 & 0.15 & & 4.3 & 19 & 0.76 & Both & 4 & \\
4555 & 2649 & 8.1 & & 100 & 5 & 5.23 & 0.21 & & 20 & 14 & 3.07 & NFW & 4 & \\
5175 & 2977 & 5.3 & & 2.0 & 200 & 2.16 & 6.4 & & 3.8 & 54 & 4.52 & ISO & 2 & \\
5272 & & 2.6 & & 25 & 14 & 1.10 & 0.21 & & 25 & 1 & 1.44 & Both & 4 & \\
5279 & 3026 & 0.10 & & 3.2 & 95 & 1.23 & 0.85 & & 35 & 3 & 2.78 & ISO & 0 & \\
5721 & 3274 & 0.10 & & 1.0 & 408 & 1.58 & 0.10 & & 2.6 & 84 & 3.57 & ISO & 0 & \\
5789 & 3319 & 0.10 & & 7.4 & 11 & 3.20 & 0.10 & & 35 & 1 & 5.04 & ISO & 0 & \\
5842 & 3346 & 0.10 & & 1.7 & 251 & 1.58 & 0.23 & & 5.0 & 43 & 2.52 & ISO & 0 &  \\
6537 & 3726 & 0.21 & & 4.7 & 8.0 & 7.04 & 0.10 & & 16 & 12 & 15.57 & ISO & 1 & \\
6778 & 3893 & 0.10 & 2.2 & 2.8 & 282 & 2.04 & 0.25 & 0.47 & 5.8 & 95 & 2.59 & ISO & 1 & \\
7045 & 4062 & 5.4 & & 12 & 10 & 2.77 & 0.74 & & 5.2 & 75 & 1.90 & NFW & 4 & \\
7323 & 4242 & 1.5 & & 7.7 & 12 & 1.99 & 0.10 & & 20 & 3 & 2.58 & ISO & 4 & \\
7524 & 4395 & 0.77 & & 3.4 & 24 & 1.10 & 0.40 & & 19 & 2 & 1.70 & Both & 1 & \\
7699 & & 0.10 & & 2.7 & 29 & 0.69 & 0.10 & & 15 & 2 & 1.82 & ISO & 0 & \\
7876 & 4635 & 0.45 & & 1.6 & 239 & 0.87 & 0.10 & & 9.3 & 18 & 2.41 & ISO & 0 & \\
7901 & 4651 & 0.84 & 2.5 & 5.8 & 48 & 4.47 & 0.10 & 1.2 & 4.0 & 179 & 4.01 & Both & 1 \\
8490 & 5204 & 1.1 & & 1.9 & 95 & 0.62 & 0.16 & & 4.5 & 26 & 2.09 & ISO & 1 & \\
9179 & 5585 & 0.31 & & 2.8 & 69 & 3.41 & 0.10 & & 16 & 5 & 9.51 & ISO & 1 & \\
9219 & 5608 & 0.10 & & 1.8 & 40 & 2.88 & 0.10 & & 5.0 & 6 & 6.86 & ISO & 0 & \\
9248 & 5622 & 4.1 & & 1.7 & 75 & 5.13 & 4.9 & & 1.0 & 10 & 5.18 & Both & 2 & \\
9465 & 5727 & 3.6 & & 13 & 10 & 1.32 & 0.10 & & 61 & 1 & 1.63 & ISO & 4 & \\
9866 & 5949 & 3.5 & & 1.0 & 426 & 1.01 & 4.1 & & 12 & 10 & 1.41 & ISO & 2 & \\
10075 & 6015 & 5.6 & & 2.8 & 182 & 2.96 & 11 & & 11 & 16 & 2.82 & Both & 1 & \\
10310 & & 2.1 & & 4.8 & 10 & 1.71 & 0.74 & & 14 & 3 & 1.72 & Both & 4 & \\
11557 & & 0.22 & & 5.0 & 18 & 1.53 & 0.10 & & 37 & 1 & 2.16 & ISO & 1 & \\
11707 & & 0.10 & & 2.4 & 84 & 2.59 & 0.25 & & 6.1 & 19 & 3.81 & ISO & 0 & \\
11914 & 7217 & 4.2 & 6.2 & 13 & 63 & 2.12 & 2.7 & 5.9 & 16 & 32 & 2.78 & Both & 3 & \\
12060 & & 0.76 & & 4.6 & 26 & 2.74 & 0.10 & & 14 & 5 & 3.15 & Both & 0 & \\
\hline
\end{tabular}
\\(1) name of the galaxy in the UGC catalog. (2) name in the NGC
catalog when available. (3) M/L of the disk in the R band for the
isothermal sphere model. (4) M/L of the bulge in the R band for
the isothermal sphere model. (5) Core radius of the dark matter
halo for the isothermal sphere model. (6) Central density of the
dark matter halo for the isothermal sphere model. (7) $\chi^{2}$
for the isothermal sphere model. (8),(9),(10),(11),(12), same as
columns (3) to (7) but now with NFW profile for the dark matter
halo model. (13) Best model: ISO, NFW or Both. (14) Classification
of the RC according to the following code : 0 when Halo is the
main component whatever the model, 1 when Halo is the main
component except in the very center (and for ISO only), 2 when
Disk is the main component at all radii whatever the model, 3 when
Disk (or Disk + Bulge) is dominating in the center whatever the
model, 4 when the main component is changing with the model (ISO
or NFW).

\end{table*}

Figures 6 to 14, given in the annex, show the plots of the best
fits obtained  using the two different density profiles for the
dark matter halo (isothermal sphere and NFW) for the 36 rotation
curves of our sample. The observed rotation curves are coded as
explained in section 3.1. The dotted blue (light) line is for the
HI component. The dashed-dotted blue line is for the disk
component. The dashed-dotted pink line is for the bulge component.
The dashed green line is for the halo component. The continuous
red line is for the total (quadratic sum of all components).

\section{Results of the fits}

\subsection{Best fit model}

For 10 galaxies, the halo is clearly the dominating component at
all radii, whatever the model (NFW or ISO). They are UGC 4274,
5279, 5721, 5789, 5842, 7699, 7876, 9219, 11707 and 12060. All of
them are late type spirals (from Scd to Im).

For 10 galaxies, the halo is dominating except in the very center
(where the disk - or bulge - is the main component) and for ISO
model only. They are UGC 4256, 4499, 6537, 6778, 7524, 7901, 8490,
9179, 10075 and 11557 (types ranging from Sc to Sm).

For 4 galaxies the disk is dominating at all radii whatever the
model. They are UGC 2503, 5175, 9248 and 9866. All of them being
rather early type galaxies (Sb or Sbc).

For 2 galaxies the disk (or disk + bulge) is dominating in the
center whatever the model. They are UGC 2455 and UGC 11914. The
result is not so surprising for this last one (being of Sab type)
and more surprising for UGC 2455 (of Im type) but it must be taken
with care because both models (NFW and ISO) led to abnormally low
(non physical) values of M/L for this galaxy (blocked at 0.1
adopted here as the lowest acceptable value).

For the 10 remaining galaxies, the main component changes with the
model but the disk is almost always dominating with the ISO model
whereas the halo is dominating with the NFW model (the only
exception being UGC 4325, for which the main component is the halo
with the ISO model and the disk with the NFW model). These
galaxies are mainly UGC 2034, 3876, 4555, and 7045, ranging from
Sbc type to Im. Things are not so clear for UGC 4456, 5272, 7323,
9465 and 10310 (one Sc and four magellanic galaxies) because the
disk is then overtaken by the halo in the outer parts with the ISO
model.

Following the above discussion, we defined a
classification code given in the last column of Table 2. To
summarize, we find 10 galaxies in class 0, 10 in class 1, 4 in
class 2, 2 in class 3 and 10 in class 4.

This quick review of the results of the best fits  confirms the
well known fact that later type galaxies are more dark matter
dominated.

Some galaxies of our sample have already been observed by other
authors and their rotation curves have been analyzed in terms of
mass models :

Four galaxies have been already observed by de Blok \& Bosma 2002
all of them being LSB type galaxies : UGC 4325, 5272, 5721, 7524.
They also observed UGC 10310 but could not derive mass models
(except a minimum disk one) because they had no surface
photometry. Our results are in good agreement for UGC 4325 and
5272, since they find good fits with their pseudo-isothermal model
and bad ones with the NFW model. We also agree for UGC 7524, where
they find that both models give acceptable fits. As for UGC 5721,
we also find that the isothermal sphere provides a good fit but
the fit we obtain with the NFW model is not satisfying (we find a
$\chi^{2}$ which is more than twice that obtained with the
isothermal sphere) whereas they find it almost as good.

UGC 7323 (a dwarf galaxy of Sdm type) has been already observed by
van den Bosch \& Swaters 2001 who concluded that "no meaningful
fit can be obtained for this galaxy". Our fits do not look so bad
however, even with the NFW model for which we had to push down to
the minimum acceptable value of M/L for the disk (0.1). Our best
fit (smaller $\chi^{2}$) was obtained with the isothermal sphere
model, with a quite normal set of parameters.

Six dwarf or LSB galaxies have been already observed and analyzed
in terms of mass models by Swaters et al. 2003 : UGC 4325, 4499,
5721, 8490, 11557 and 11707. They find better results for four of
them (UGC 4325, 4499, 8490 and 11557) with a pseudo-isothermal
model rather than with NFW (in agreement with our results except
for UGC4499 that we discuss below) but they have a reverse for UGC
5721 (which is in contradiction with de Blok \& Bosma 2002 and
with our results) as well as for UGC 11707. About UGC 5721, we
must confess that our ISO model gives a better solution to the
price of minimal accepted values for both M/L (0.1) and $R_{o}$ (1
kpc) thus casting a doubt on the validity of the result hence to
the comparison that can be done with the NFW model for this
galaxy. As for UGC 11707, we have H$\alpha$ points on the rotation
curve hovering significantly above the HI curve for radii going
from 3 to 6 kpc (see Fig.14), exactly where the pseudo-isothermal
sphere model from Swaters et al. is too low to fit correctly the
observed curve (see their Fig.3). The difference between our
H$\alpha$ curves may be explained by the smaller value of
inclination we adopted for that galaxy (55$^{\circ}$ instead of
68$^{\circ}$) but also because with Fabry-Perot observations we
take into account data from a large part of the disk instead of
getting points only along a slit. We agree with Swaters et al. who
find that NFW overpredicts the inner slope of the rotation curve
for UGC 4325 (also in agreement with de Blok \& Bosma 2002, Fig.
8) and for UGC 8490 (as can be seen on their figure 3) but not for
UGC 4499 where the slope of our H$\alpha$ rotation curve can be
fitted correctly with our models (even slightly better with the
NFW model although it is not very significant). Indeed, the inner
slope of their rotation curve is not much different from the slope
of the HI curve (plotted as crosses on our Fig. 8) whereas our
H$\alpha$ curve is clearly above. This discrepancy is maybe due to
the fact that the authors use slit spectroscopy to get the
H$\alpha$ rotation curve in the central part of the galaxy (we
adopted here the same inclination of 50$^{\circ}$), although they
note that it generally provides a steeper slope than the HI curve,
this last one being biased by beam smearing. As for the three
other galaxies (UGC 5721, 11557 and 11707) we do find that NFW
overpredicts the inner slope, more or less strongly but
systematically.

Six galaxies of our sample have been studied by Kassin et al.
2006b (UGC 5789, 6537, 6778, 7045, 7901 and 11914, of respective
types Scd, Sc, Sc, Sc, Sc and Sab) with another approach since
they decompose the rotation curves found in the literature into
baryonic and dark matter components (stellar mass profiles are
created by applying color-M/L relations to near-infrared and
optical photometry). As a consequence, no straightforward
comparison could be done with our results. However it is
interesting to note that Kassin et al. find that for two of them
(UGC 5789 = NGC 3319 and UGC 7045 = NGC 4062) the baryons
underpredict the rotation curve, although they conclude that the
case is not so clear for NGC4062 when taking into account all the
uncertainties. Indeed, for UGC 5789 our models (NFW and ISO) both
point at an M/L ratio of the disk at the minimum accepted value,
0.1, suggesting that it is dominated by the dark matter halo
component. As for UGC 7045, the disk alone (with M/L = 5.4) is
almost sufficient to explain the rotation curve with the ISO model
(which is then very close to a maximum disk model) whereas it is
the contrary with the NFW model where the halo alone can explain
the rotation curve (M/L is then found to be close to 1 for the
disk). For the four other galaxies (UGC 6537, 6778, 7901 and
11914) we find that the fits obtained with the ISO model are not
far from the fits provided by a maximum disk, more especially for
UGC 6537 and 11914 where the match is perfect (we applied maximum
disk models to the 36 galaxies studied in this paper but the
resulting graphs are not presented here). Our analysis of the
rotation curves in terms of mass models for these six galaxies is
thus in good agreement with the conclusions reached by Kassin et
al.

In summary, in the cases where some of our galaxies have already
been modeled by other authors, the results tend to agree pretty
well. In the few cases which differ, the reason for the
discrepancy has most of the time been identified.

\section{Discussion}

In order to allow a direct comparison, the $\chi^{2}$ given in
Table 2 have been computed using the same variation range for the
parameters and the same step of calculus for both the ISO and the
NFW mass models.

The comparison of the values of the $\chi^{2}$ found for the best
fits of both models clearly shows that the isothermal sphere
profile for the dark matter halo gives better results than the NFW
profile when fitting the observed rotation curves. In Fig.1 we
plotted the minimum value of the $\chi^{2}$ found with the ISO
model as a function of that obtained with the NFW model for each
galaxy. These last $\chi^{2}$ are systematically larger. The
column "best model" in Table 2 indicates which one, from ISO or
NFW, gives the best results as can be judged from the fits
displayed on Figures 6 to 14, with the mention "Both" when they
give comparable results. It appears that there is a clear
difference between ISO and NFW models as soon as there is a 20\%
difference in the corresponding chi2. It also confirms what the
$\chi^{2}$ values suggest, namely that the ISO model provides
better fits than the NFW in most of the cases.

\begin{figure}
\includegraphics[width=9cm,angle=0]{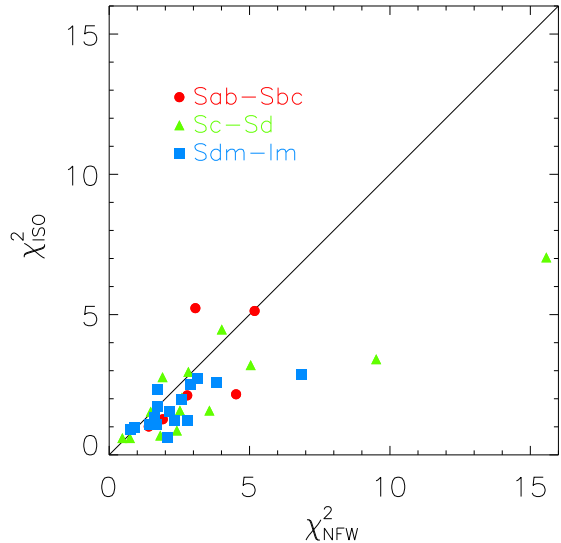} \caption{Minimum
$\chi^{2}$ obtained with an isothermal sphere profile versus a NFW
profile. For figures 1 to 5 we used three different symbols for
the main classes of morphological types of spiral galaxies (early,
late and magellanic). }
\end{figure}

Indeed, the NFW model, because of its steep rise in the center, is
often unable to fit correctly the observed rotation curve. Typical
examples can be seen with UGC 5279, 5842, 7699, 7876, 9219, 9465
or 11557. For these galaxies (all of late type, ranging from Scd
to Im) the central rising part of the rotation curves is
completely missed by the NFW model, although the M/L ratio of the
disk is most often pushed to its lowest acceptable value of 0.1
(except for UGC 5279 and UGC 5842 where it is found to be
respectively 0.56 and 0.16).

In particular, in the case where the $\chi^{2}$ value is lower
for the NFW model, the M/L value found is smaller than for the ISO
model, leading to a poor match between the RC and the disk contribution
in the central region of the galaxy. More generally, because of
the small M/L found for the NFW models, this solution rarely
allows the first 1-2 kpc of the RC to be fitted by the disk
component contribution, even when the shape of the RC should
make it possible. Indeed, the ISO model solutions, which
are often close to maximum disk solutions, lead to better
matches between the stellar mass and the total mass distributions
in the inner region of the galaxies. This is a strong argument
in favor of the ISO solutions.

The NFW profile cannot account for
rotation curves with a linear rise in velocity with radius, as
shown by Hayashi \& Navarro 2006 in their Fig.1. The ISO model, on
the contrary, provides good fits in the central part for such
slowly rising rotation curves and, more generally, systematically
better fits than the NFW model for most of the galaxies of our
sample. The position of the points on Fig. 1 shows no special
trend with morphological type. More generally, it can be seen that
the NFW best fits tend to give systematically unphysical low
values for the disk M/L ratio (0.1). This is even a stronger
argument than the smaller $\chi^{2}$ to favor the ISO solutions.
Also, whenever the $\chi^{2}$ is smaller for the NFW model, it
appears that the M/L is systematically smaller (or equal in one
case) than with the ISO model, leading to abnormally low values of
M/L.

Looking at Tables 1 and 2, one can see that nearly half (16/36) of
the galaxies do not have HI velocities such that the RC is only
defined in the inner parts. Since, as mentioned earlier, the
kinematics in the inner parts can be affected by non-circular
motions (such as the presence of bars), it is interesting to see
if this alters significantly the results when trying to see which
model (ISO or NFW) fits best the kinematics. In fact, for the
whole sample (36) we get that 50\% are best represented by an ISO
model, 8\% by an NFW model and 42\% are equally well represented
by both models while for the sub-sample having only an optical RC
for the inner parts, those numbers are 50\%, 12\% and 38\%. Thus,
the main conclusions hold even when only optical data are
available for the inner parts.

In Figure 2, we plotted the central density of the dark matter
density profile versus the core radius of the dark halo for both
models (top: isothermal sphere profile; bottom: NFW profile). The
central density and the core radius of the halo are clearly
correlated : the higher the central density, the smaller the core
radius, whatever the model and independant of morphological type.
This is in good agreement with the results found by other authors
(Burket, 1995; Kravtsov et al. 1998 and Blais-Ouellette, 2000) who
claim that one free parameter is enough for describing the
distribution of dark matter. More recently, Barnes et al. 2004 and
Kormendy \& Freeman 2004 also found the same type of correlation.
However, the slope of our correlation is close to that found by
Kormendy \& Freeman whereas the slope found by Barnes et al. is
clearly steeper. Indeed, as detailed hereafter, we find a slope of
-0.93 (for isothermal sphere fits) which is very close to that
found by Kormendy \& Freeman (-1.04 for isothermal sphere fits and
-1.20 for pseudo-isothermal fits) while Barnes et al. find a
steeper slope, around $\sim~$-2 (estimated from their Figure 4)
for pseudo-isothermal sphere fits.

All these values are in agrement when considering the
accuracy for the slopes. The error bars on our data show that the
permitted range for the slope deduced from our Fig.2 (top) goes
from 0.5 to 2.0. One can estimate this range from 1.5 to 2.5 for
Barnes et al. and from 0.5 to 1.5 for Kormendy \& Freeman.

When taking into account only the points for which the ISO model
is clearly the best (see column 13 of Table 2) we also find a
steeper slope (-1.38 instead of -0.93) intermediate between that
found by Kormendy \& Freeman and that found by Barnes et al. These
points are surrounded by a circle on Figures 2 to 4. In figure 2
bottom (NFW model) these points also lead to a steeper slope than
the whole set (-1.48 instead of -1.00). In figure 4 too, they
suggest a slightly steeper slope (albeit the least-squares fit
remains very flat and almost horizontal) than the whole set. In
Figure 3 on the contrary, the points for which ISO is the best
suggest a lower slope than the whole set for both graphs (top and
bottom). For all these figures (2,3 and 4) these points tend to
have a smaller dispersion than the whole set.

\begin{figure}
\includegraphics[width=9cm,angle=0]{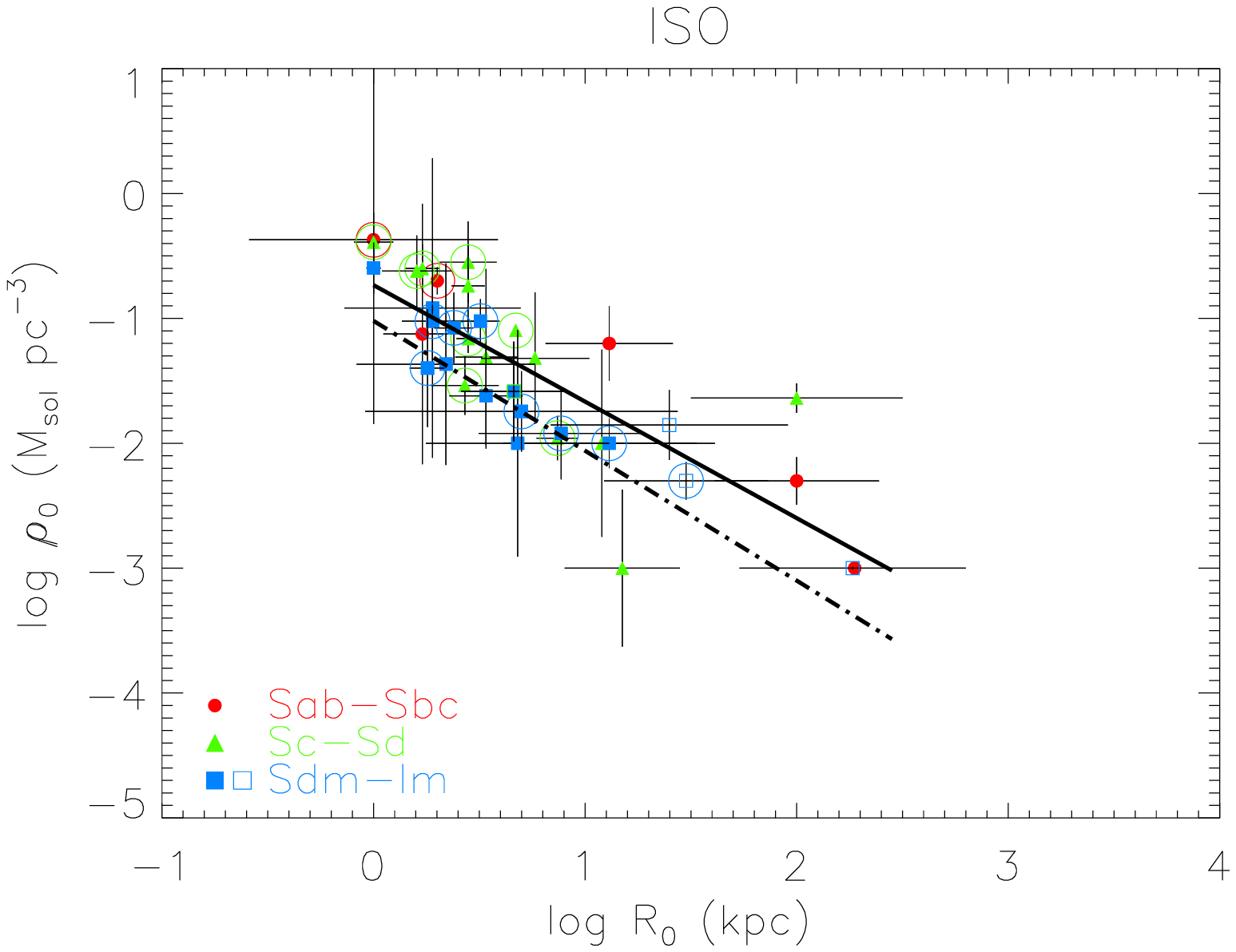} \includegraphics [width=9cm,angle=0]{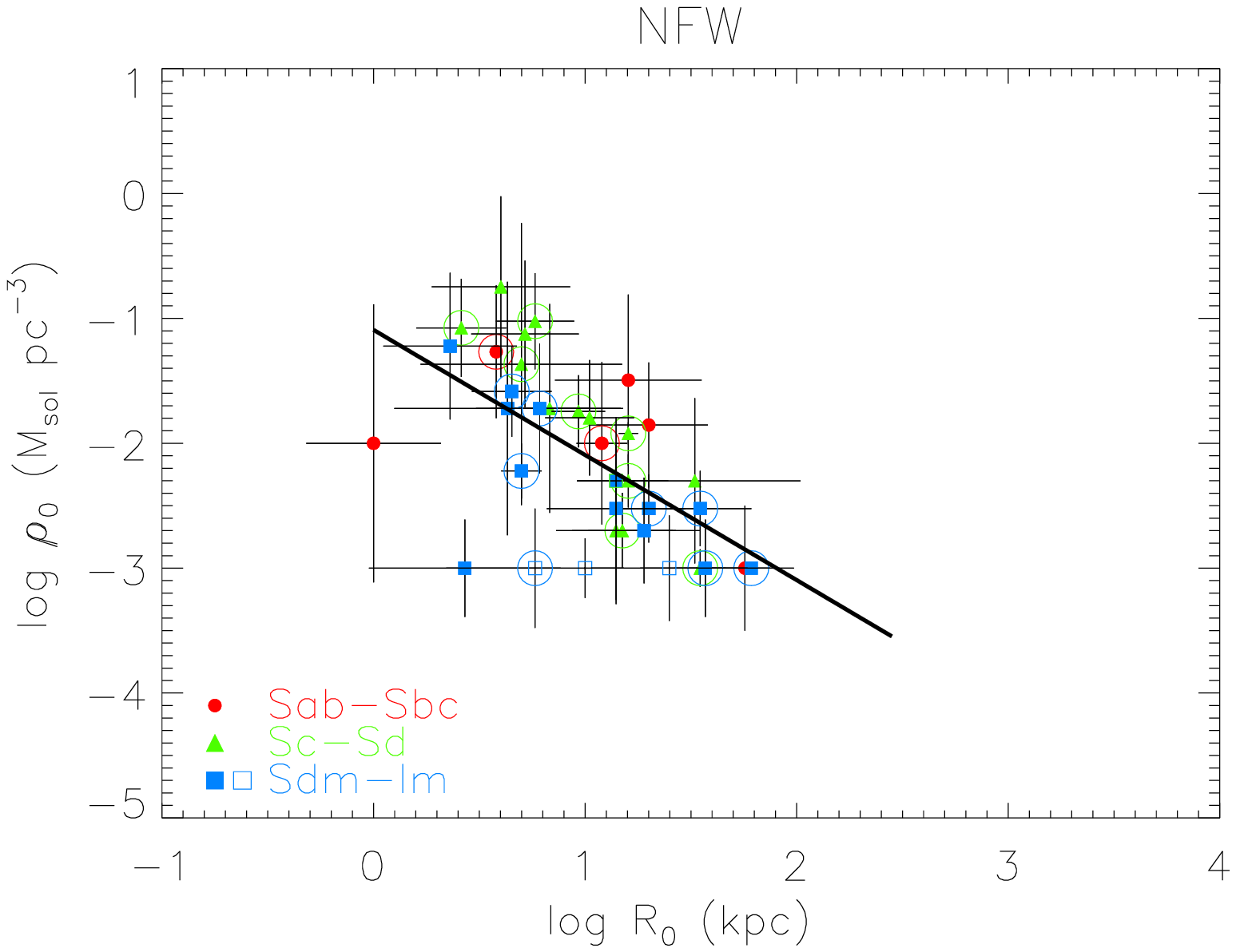}
\caption{Top : Central density (log) of dark matter versus the
core radius (log) of dark matter for the isothermal sphere profile
(ISO). Bottom : Same for the NFW profile. The solid lines are
least-squares fits to our data (excluding the 3 open squares
discussed in section 6). The dashed line (top) is the
least-squares fit extracted from figure 3 of Kormendy \& Freeman
2004. Circles surround the points for which ISO is clearly the
best model}
\end{figure}

Based on published mass models for 37 galaxies, from Sc to Im
type, Kormendy \& Freeman found the following least-squares fit
for isothermal halos:

log$\rho_{o}$ = -1.04 log$R_{o}$ -1.02 (rms=0.17dex)

Our data of Fig. 2 give the following least-squares fits excluding
the 3 points plotted as open squares corresponding to Im galaxies
discussed below about Figure 3):

log$\rho_{o}$ =  -0.93 log$R_{o}$ -0.74 (R$^{2}$=0.61) for ISO

log$\rho_{o}$ =  -1.00 log$R_{o}$ -1.03 (R$^{2}$=0.42) for NFW

We conclude that, on average, log$\rho_{o}$ $\sim$  -log$R_{o}$
-0.8, which implies that the product $\rho_{o}$ $R_{o}$ is almost
constant.

Figure 3 shows the correlation of the central density (top) and of
the core radius (bottom) as a function of absolute magnitude. From
those two plots it is clear that faint galaxies (dwarfs) have
higher central density and smaller core radius. In other words,
dwarf galaxies have a more highly concentrated dark matter content
than massive spirals. This result had already been seen quite
convincingly in the study of the Sculptor group dwarfs by C\^ot\'e
et al. (2000) in their figures 8 and 9 (see also the figures 2 and
3 of Kormendy \& Freeman 2004).

\begin{figure}
\includegraphics[width=9cm,angle=0]{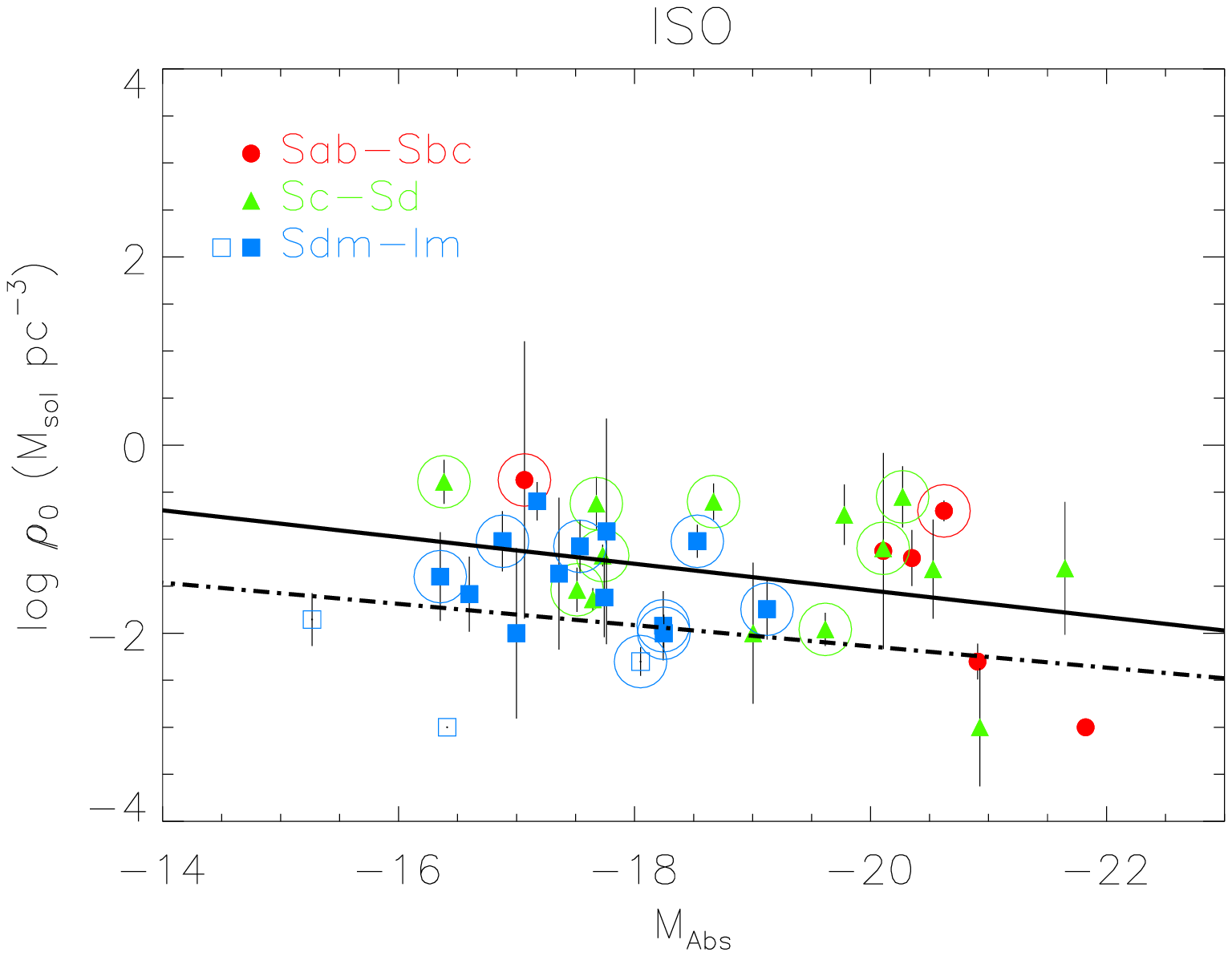} \includegraphics [width=9cm,angle=0]{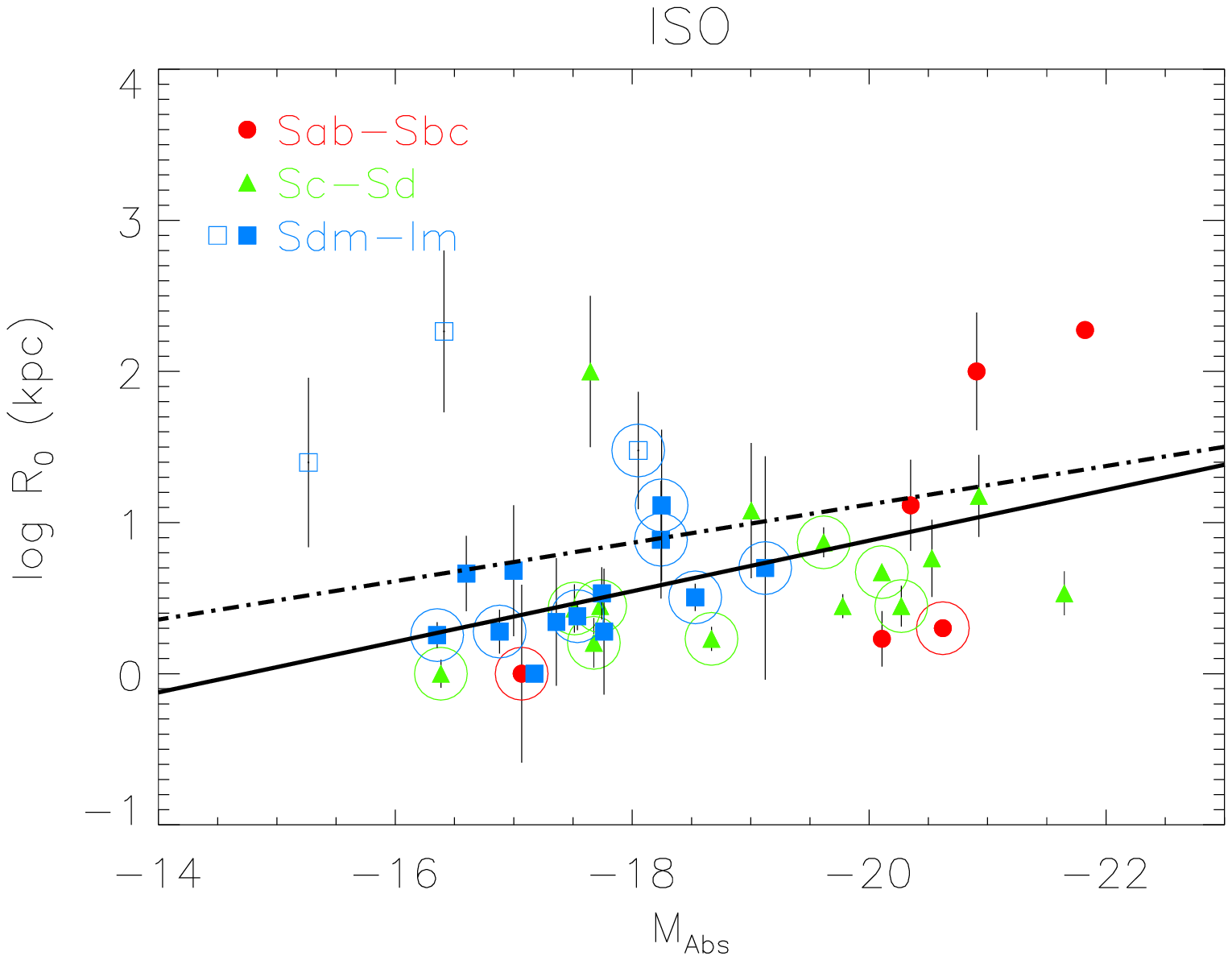}
\caption{Top : Central density (log) of dark halo versus absolute
magnitude. Bottom : Core radius (log) of dark halo versus absolute
magnitude. Both are for the ISO models. The solid lines are
least-squares fits for our data (excluding the 3 open squares
discussed in section 6). The dashed lines are the least-squares
fits extracted from figure 3 of Kormendy \& Freeman 2004. Circles
surround the points for which ISO is clearly the best model}
\end{figure}

For three out of the six Im galaxies of our sample (namely
UGC2034, 2455 and 5272) the optical part of the rotation curve is
very chaotic, with large error bars, and the resulting best fit
models are not very reliable. The correlation is much clearer when
these three galaxies are removed from the sample. Anyway we
plotted them on the graphs as open squares but do not take them
into account when fitting the data points. The same thing has been
done for the two plots of Figure 2, where one can see that the
corresponding points have a normal behavior on the top diagram
(ISO model) whereas all of them appear in the lower part of the
bottom diagram (NFW model) since they reach the minimum accepted
value of $\rho_{o}$.

Kormendy \& Freeman found the following least-squares fit for
isothermal halos:

log$\rho_{o}$ = 0.113 $M_{B}$ + 0.12 (rms=0.28dex)

log$R_{o}$ =  -0.127 $M_{B}$ - 1.42 (rms=0.16dex)

Our plots in Figure 3 give the following least-squares fits (also
for the isothermal models):

log$\rho_{o}$ =  0.142 $M_{B}$ + 1.29 (R$^{2}$=0.12)

log$R_{o}$ =  -0.167 $M_{B}$ - 2.47 (R$^{2}$=0.23)

The agreement is fairly good, as can be seen on Figure 3, and
strengthens the validity of these scaling laws.

Salucci et al. 2007 also get such a relation between the
central density and the core radius of the halo (see Figure 1 of
their paper) although for a different fitting model.

The obvious next step is to look at the product of $R_{o}$ and
$\rho_{o}$ as a function of absolute magnitude. This is done in
Figure 4. Our mass models, applied to galaxies ranging from Sab to
Im, suggest a nearly constant halo surface density around $\sim$
150 M$_{\odot}$pc$^{-2}$, slightly higher than the value found by
Kormendy \& Freeman which is just below $\sim$ 100
M$_{\odot}$pc$^{-2}$ (according to their Fig. 5). Clearly, the
halo surface density is nearly independent of galaxy luminosity. A
small part of the offset ($\sim$0.06 for the log) observed on
Figure 4 between our least-squares fit and that of Kormendy \&
Freeman  may be explained by the adopted value of $H_{o}$=75 km
s$^{-1}$ Mpc$^{-1}$ whereas they use $H_{o}$=70 km s$^{-1}$
Mpc$^{-1}$, anyway the offset is smaller than the error-bars.

\begin{figure}
\includegraphics[width=9cm,angle=0]{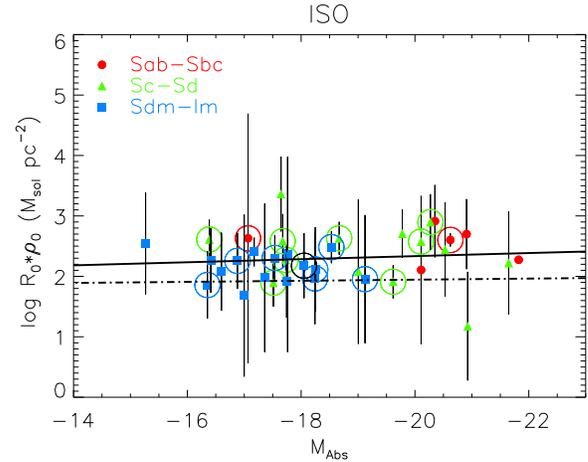}
\caption{Central surface density (log) of dark matter (ISO model)
as a function of absolute magnitude (given in Table 1, determined
from B$_T$(0) found in the RC3 and our adopted distance). The
solid line is a least-squares fit for our data (excluding the 3
open squares discussed in section 6). The dashed line is the
least-squares fit extracted from figure 5 of Kormendy \& Freeman
2004. Circles surround the points for which ISO is clearly the
best model}
\end{figure}

Another interesting correlation is the one found by Donato et al.
2004 between the core radius of the dark matter halo and the
exponential disk scale length h. They find (from a sample of 25
galaxies of different morphological types found in the literature)
the following relation :

log$R_{o}$ = (1.05$\pm$0.11) log h + (0.33$\pm$0.04)

corr=0.90 rms=0.16dex

Our data are in good agreement with that result since our plot of
Figure 5 (top) shows the following relation (least-squares fit
excluding the three Im galaxies already mentioned above and
plotted as open squares) :

log$R_{o}$ = 0.996 log h + 0.40 (R$^{2}$=0.28)

The different types (LSB, dS, HSB) of galaxies are
distinguished in Fig.5 (top) by circles of different sizes
surrounding them (NB For those who are both LSB and dS we prefered
marking only the last classification). Unfortunately 2 of the 5 dS
of our sample are among the group of 3 galaxies already excluded
from the fits in Fig.2 to 5 (where they are plotted as open
squares). As a matter a fact, only 3 "bona fide" dS remain.
Luckily, all of them are on the lower left side of the diagram, as
expected, but of course it is not very significant and prevents to
draw any significant conclusion.

\begin{figure}
\includegraphics[width=9cm,angle=0]{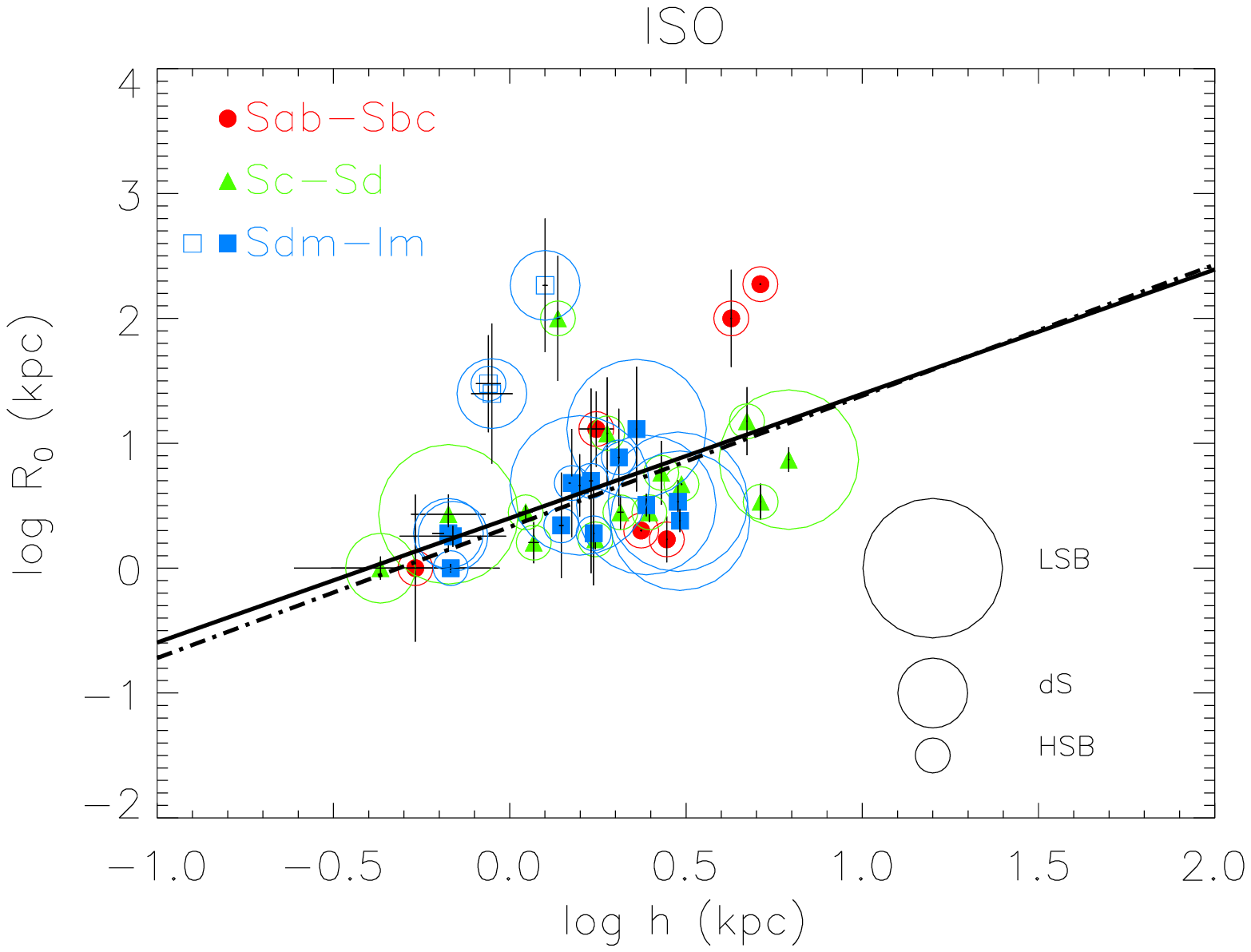} \includegraphics [width=9cm,angle=0]{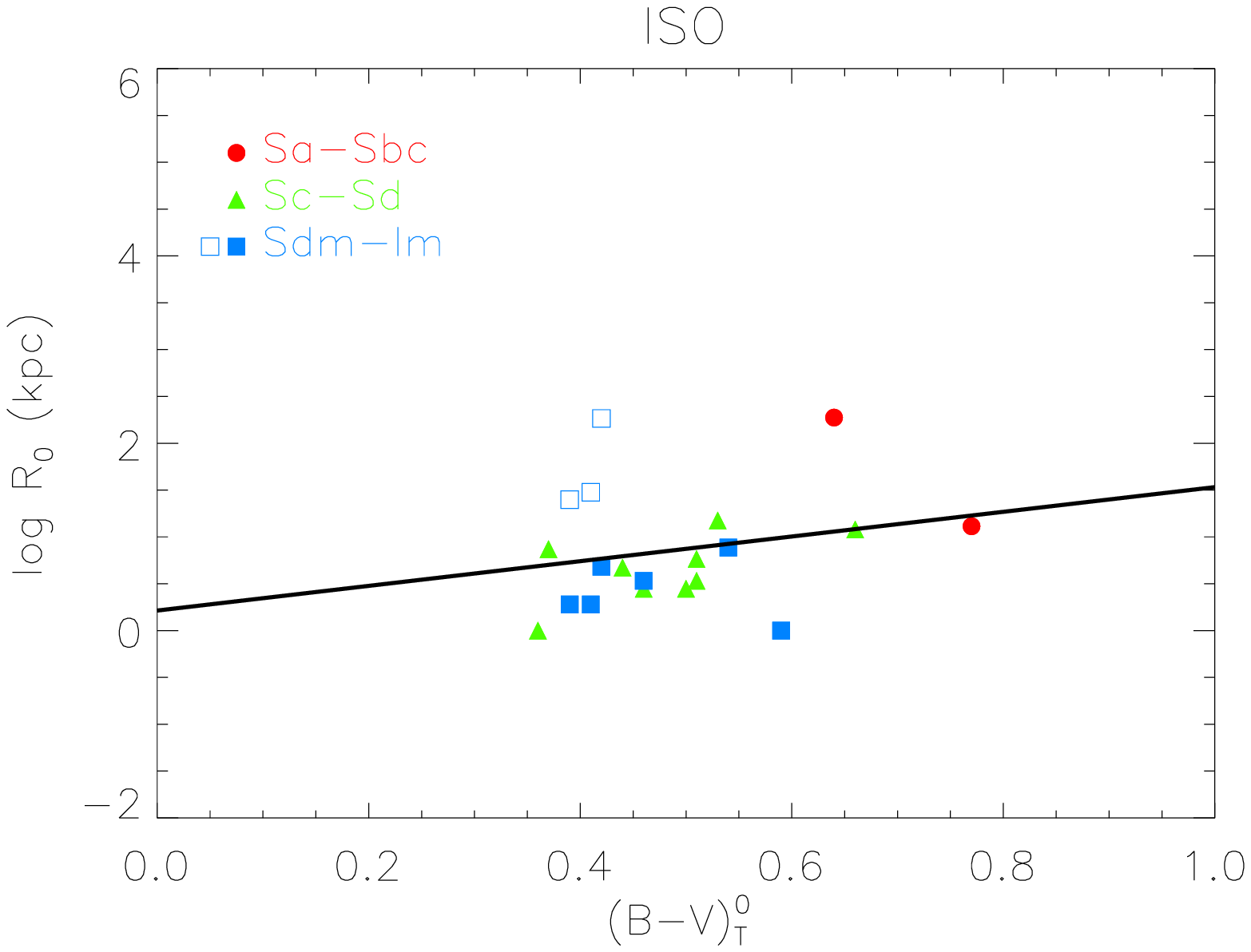}
\caption{Core radius (log) of the dark matter halo (ISO models) as
a function of optical disk scale length (top) and (B-V) color
(bottom). This last parameter, extracted from the RC3 catalog,
could be found for 20 of our galaxies only. The solid lines are
least-squares fits obtained with our data (excluding the 3 open
squares discussed in section 6). The dashed line on the top
diagram is the least-squares fit from figure 1 of Donato et al.
2004. On that diagram we have surrounded HSB, dS and LSB galaxies
by circles of growing sizes.}
\end{figure}

Finally, as seen in C\^ot\'e at al (2000), there is a trend in
Figure 5 (bottom) between the core radius and the (B-V) color.
The different types of galaxies are not distinguished in
that figure because there are too few points to conclude
efficiently about the different subtypes. Furthermore, it is just
a trend that is shown, needing to be confirmed with a larger
sample.


\section{Summary and conclusion}

Our study, based on 36 galaxies of different morphological types,
confirms the result already claimed by other authors about the
shape of the dark matter halo in the center of spiral galaxies,
namely that its density profile is probably closer to an
isothermal sphere profile than to an NFW profile: DM halos are
rather flat than cuspy. Our mass models (ISO and NFW) are compared
when fitting the rotation curve as a whole, however a comparison
of the chi2 values limited to the central regions would be still
clearer since the inner slope of the RC found with the NFW model
is systematically steeper and above the data points, compared with
the ISO model.

Thus far, most of the results found in the literature concerned
dwarf or Low Surface Brightness galaxies. 9 galaxies of our sample
are LSB (UGC 2304, 5272, 5279, 5789, 7524, 7699, 9465, 11707 and
12060, with types going from Scd to Im) and, for all of them, we
systematically find better results with the ISO model (smaller
$\chi^{2}$) than with NFW, except UGC 2304 for which both are
equivalent. Interestingly, our study suggests that this holds for
most spirals since the 36 galaxies of this study have
morphological types ranging from Sab to Im. For almost all of
them, the best fit is obtained with the ISO model (the reverse is
found only for 4 galaxies out of 36). Also, no significant
difference can be seen when comparing the quality of the fits
obtained with the NFW and the ISO model as a function of the
morphological type.

However, a difference can be seen in the way the rotation curves
are decomposed into several components, with the halo being the
main component for late types and the disk (or disk + bulge) the
main component for early types. This result, although
needing to be confirmed with a larger sample, merely reflects the
well known fact that later type galaxies are more dark matter
dominated.

Finally, we confirm different halo scaling laws seen previously by
other authors such as C\^ot\'e et al (2000), Kormendy \& Freeman
2004 and Barnes et al. 2004. Among those, it appears clearly that
low luminosity galaxies have small core radius and high central
density, the product of the two parameters being nearly constant
with absolute magnitude. This means that the galaxy halo surface
density is independent of galaxy type or luminosity.
Trends are also seen for the core radius as a function of
luminosity and color but should be confirmed with a larger
sample.


\section*{acknowledgements} The authors wish to thank the anonymous
referee who helped improve the paper. They thank the Groupement de
Recherche Galaxies (now Programme National Galaxies) for its support
for observing time and the Observatoire de Haute-Provence team for
its technical assistance during the observations. They also thank O.
Garrido, J.L. Gach, O. Boissin and P. Balard for their help during
the observing runs and J. Boulesteix for constantly improving the
ADHOCw software used for the data acquisition and reduction. The
authors thank Chantal Balkowski, Laurent Chemin, Henri Plana and
Olivier Daigle for fruitful discussions. This research has made use
of the NASA/IPAC Extragalactic Database (NED) which is operated by
the Jet Propulsion Laboratory, California Institute of Technology,
under contract with the National Aeronautics and Space
Administration. The authors have also made an extensive use of the
LEDA database (http://leda.univ-lyon1.fr). CC and OH acknowledge
support from the Natural Sciences and Engineering Research Council
of Canada and le Fonds qu\'eb\'ecois de la recherche sur la nature
et les technologies.

\section*{annex}

\begin{figure*}
\begin{center}
\includegraphics[width=5.5cm,angle=-90]{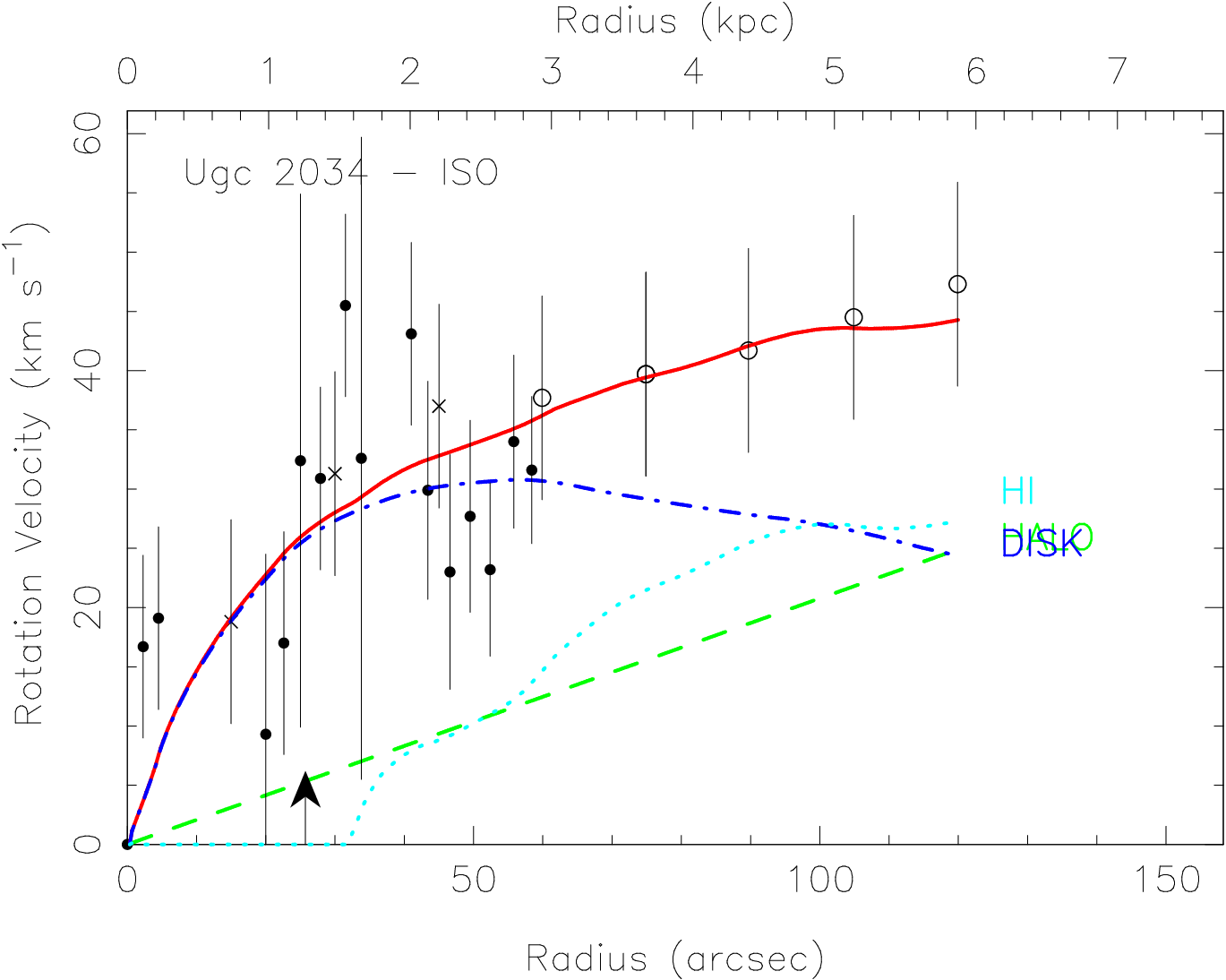}  \includegraphics[width=5.5cm,angle=-90]{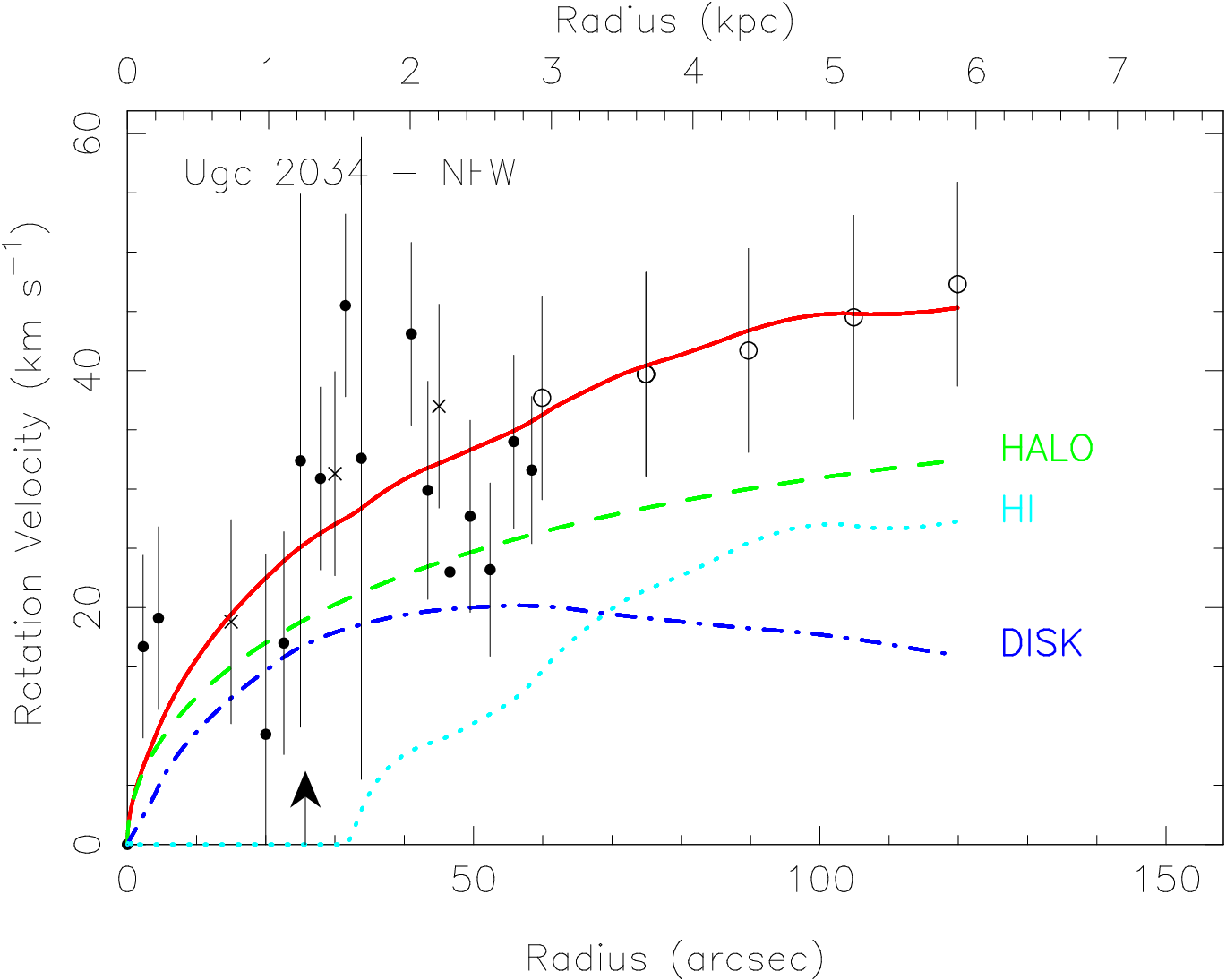} \\
\includegraphics[width=5.5cm,angle=-90]{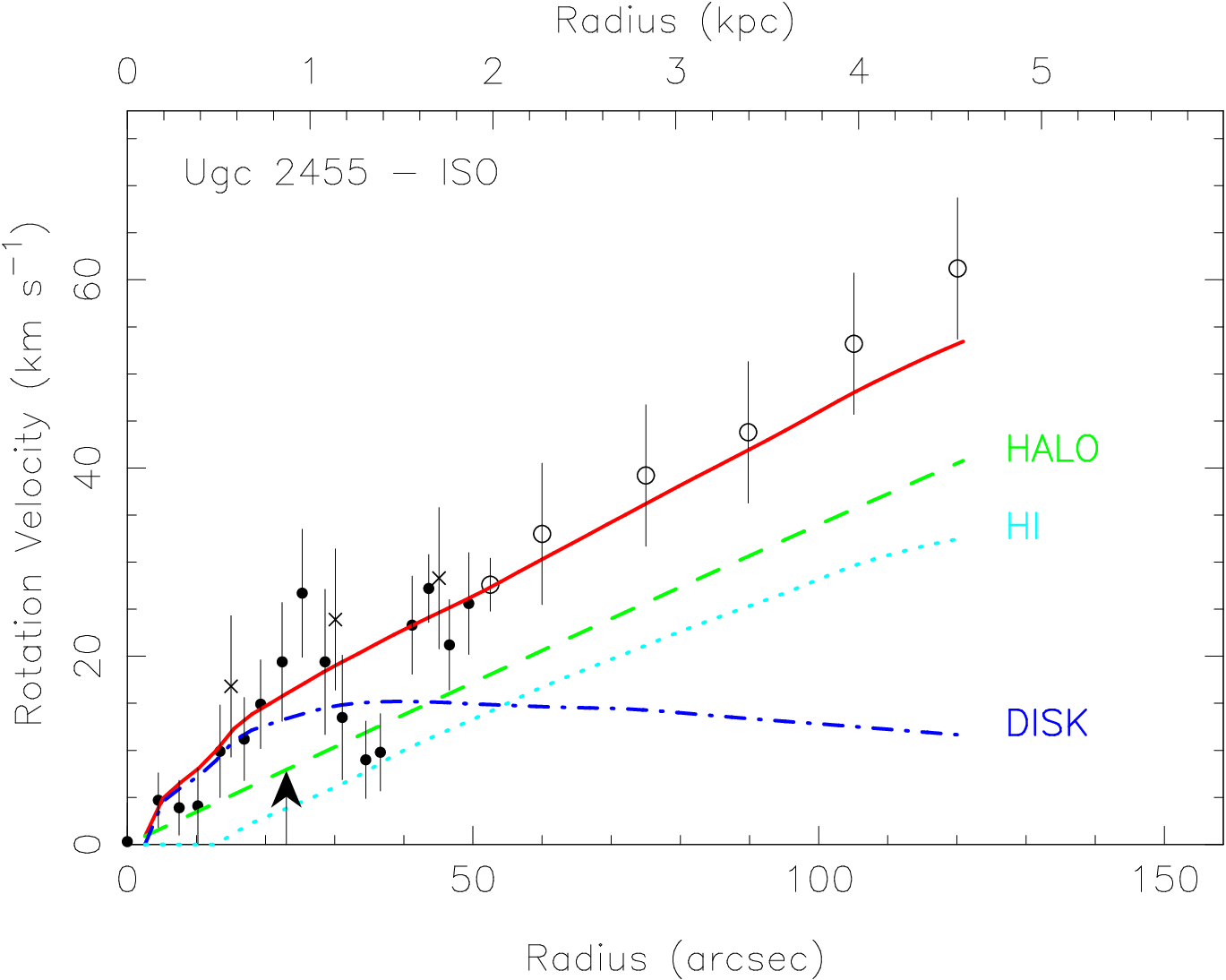}  \includegraphics[width=5.5cm,angle=-90]{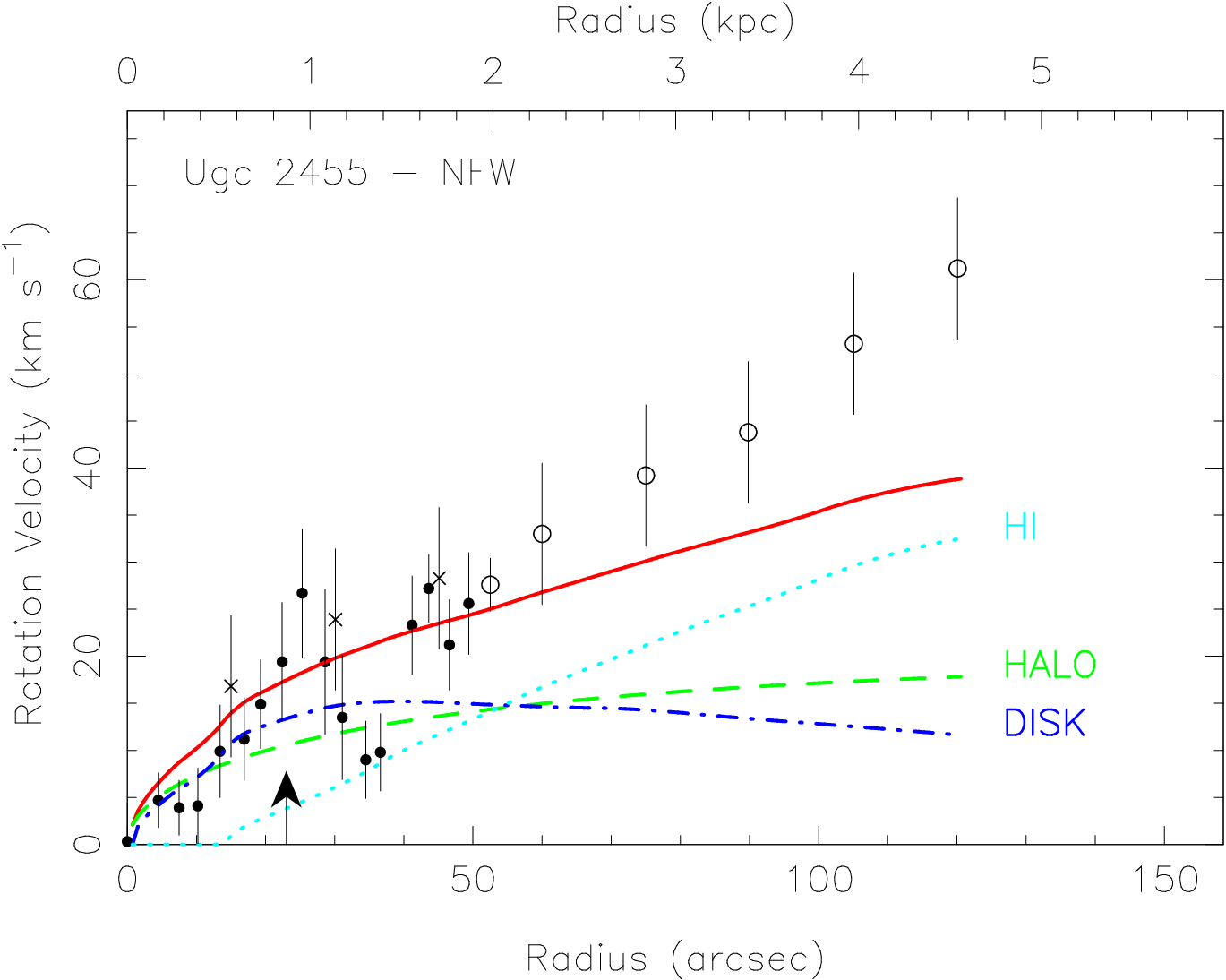} \\
\includegraphics[width=5.5cm,angle=-90]{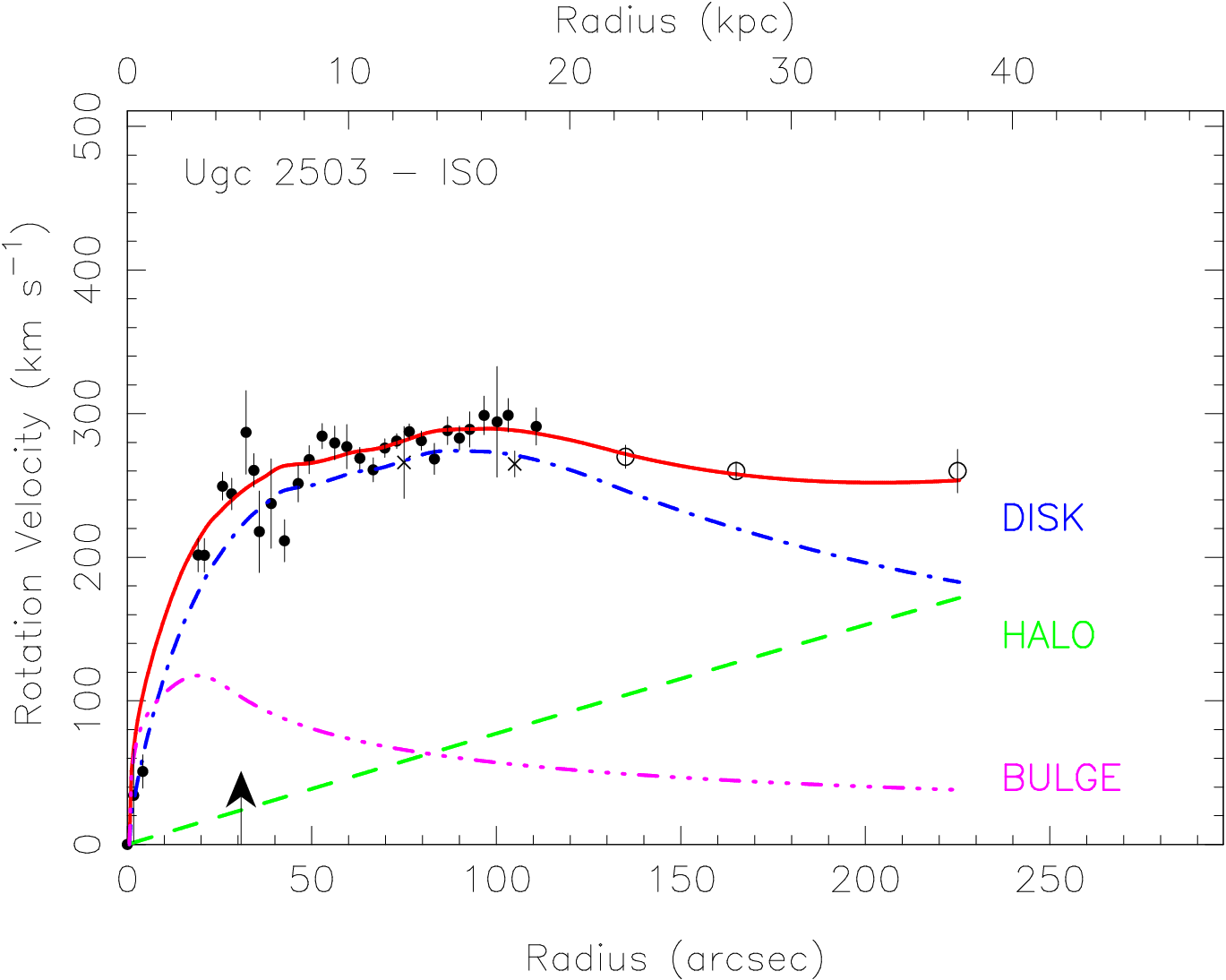}  \includegraphics[width=5.5cm,angle=-90]{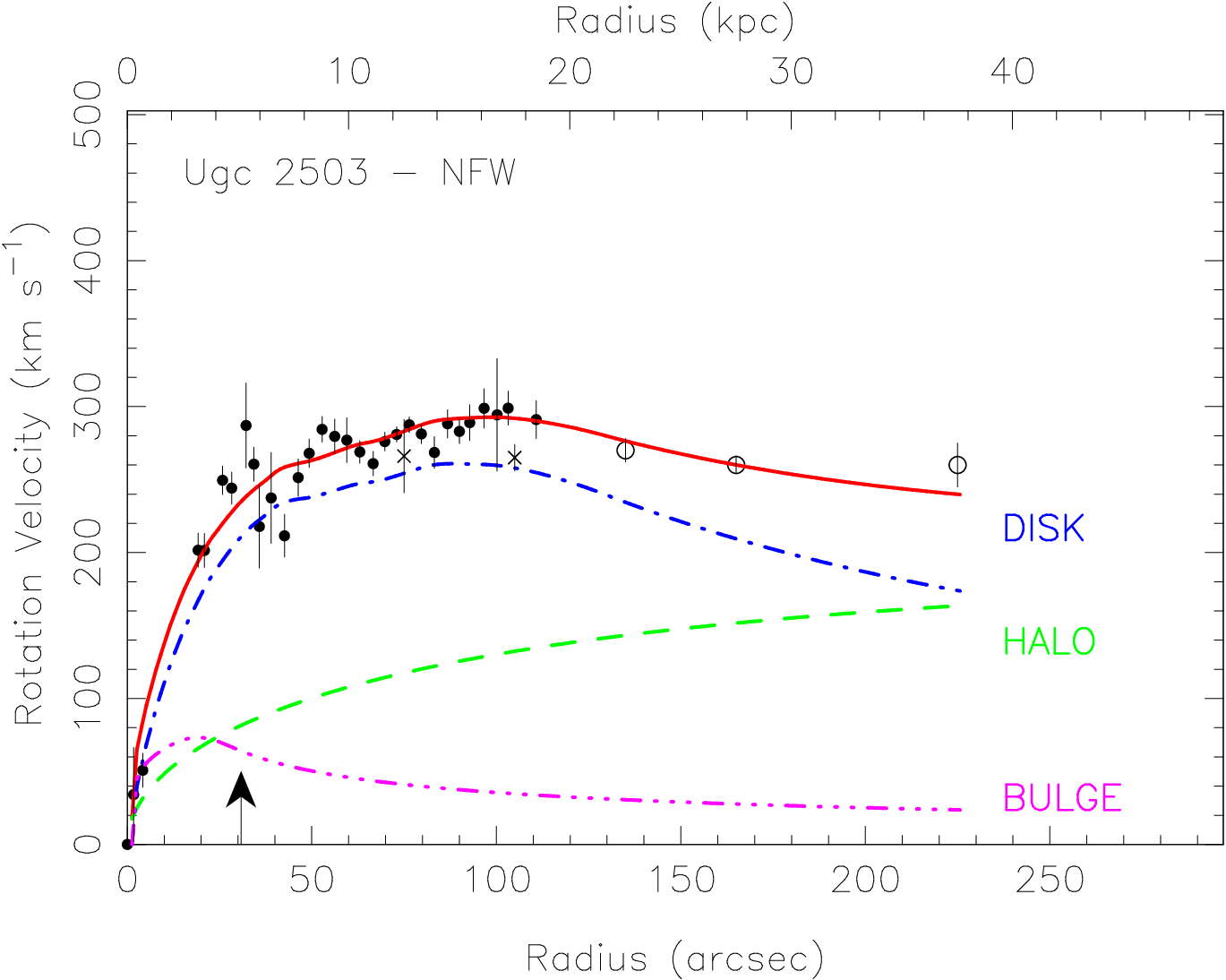} \\
\includegraphics[width=5.5cm,angle=-90]{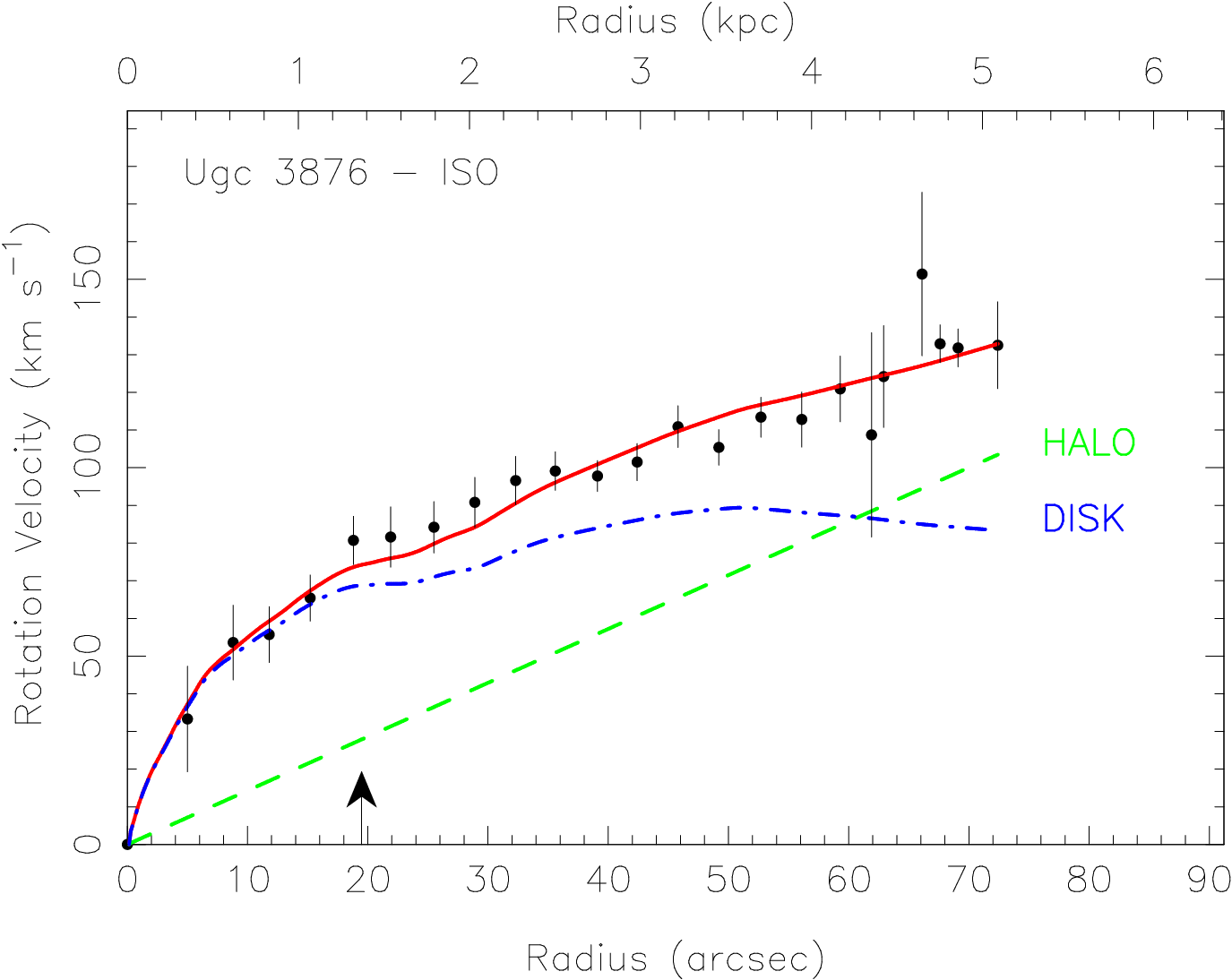}  \includegraphics[width=5.5cm,angle=-90]{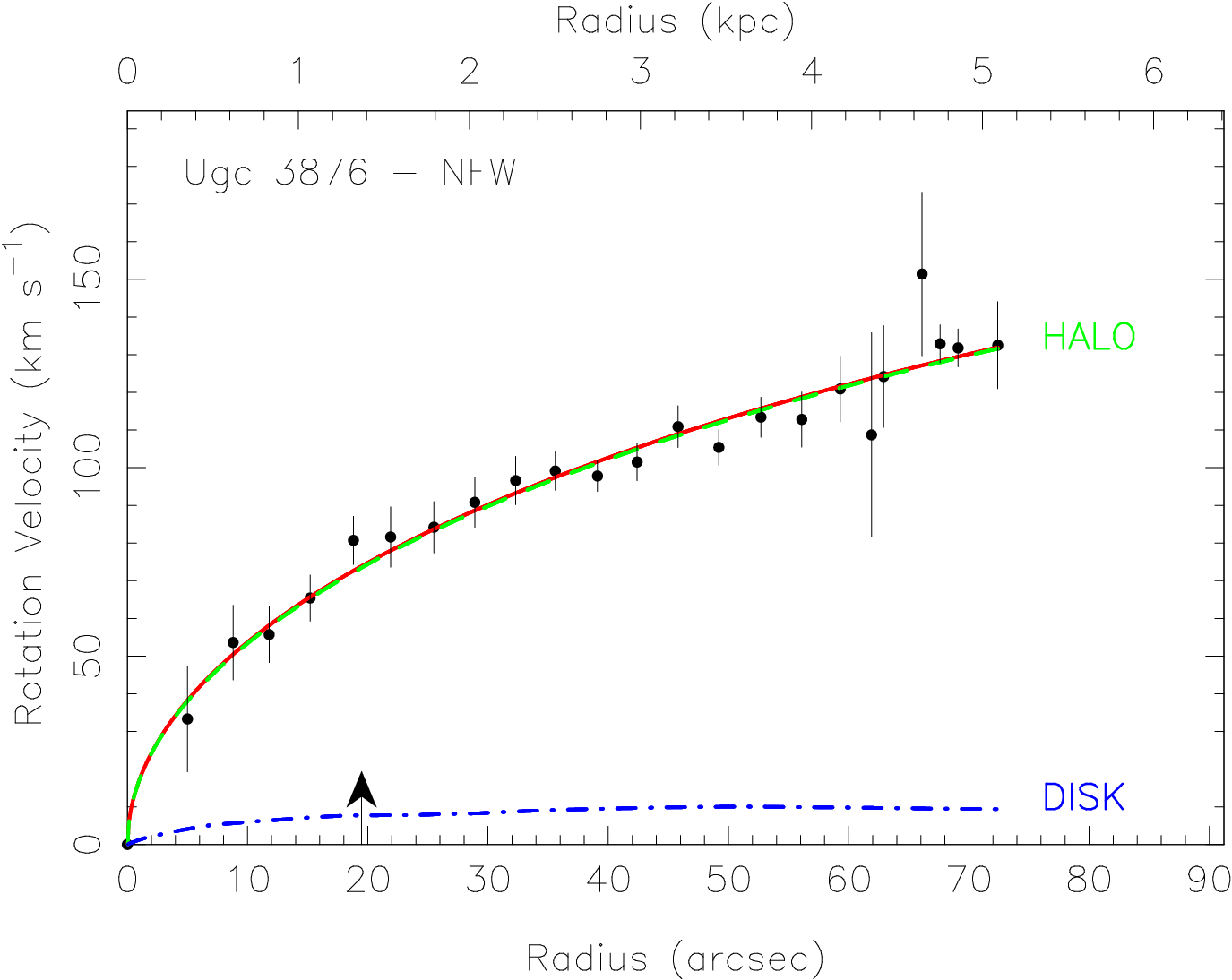} \\
\end{center}
\caption{Best fit model for the rotation curve of UGC2034,
UGC2455, UGC2503 and UGC3876 with isothermal sphere profile (left)
and NFW profile (right). Dots are for optical velocities, open
circles for HI velocities and crosses for HI in the central part,
not taken into account. The arrow on the X axis indicates the disk scale-length.}

\end{figure*}

\begin{figure*}
\begin{center}
\includegraphics[width=5.5cm,angle=-90]{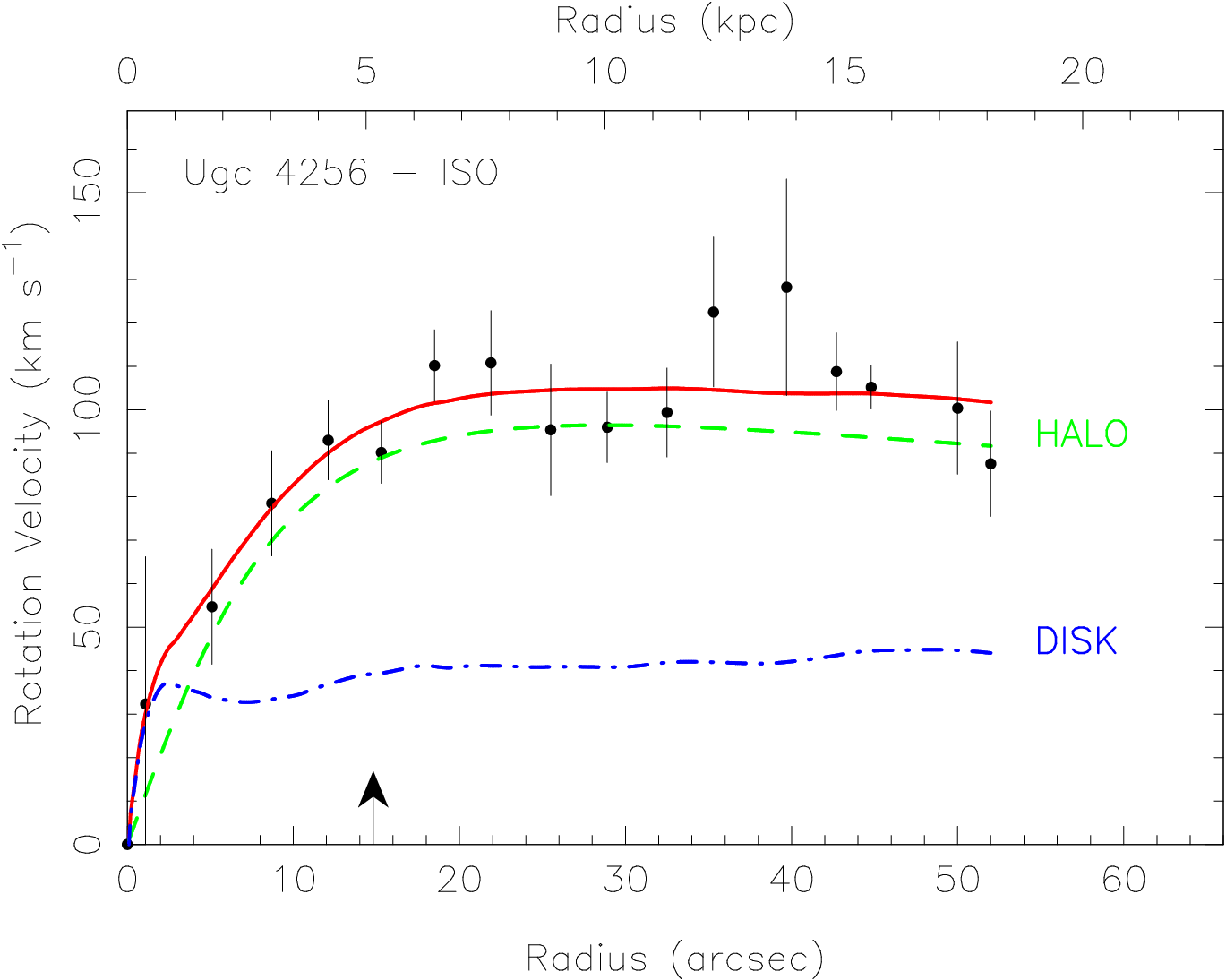} \includegraphics[width=5.5cm,angle=-90]{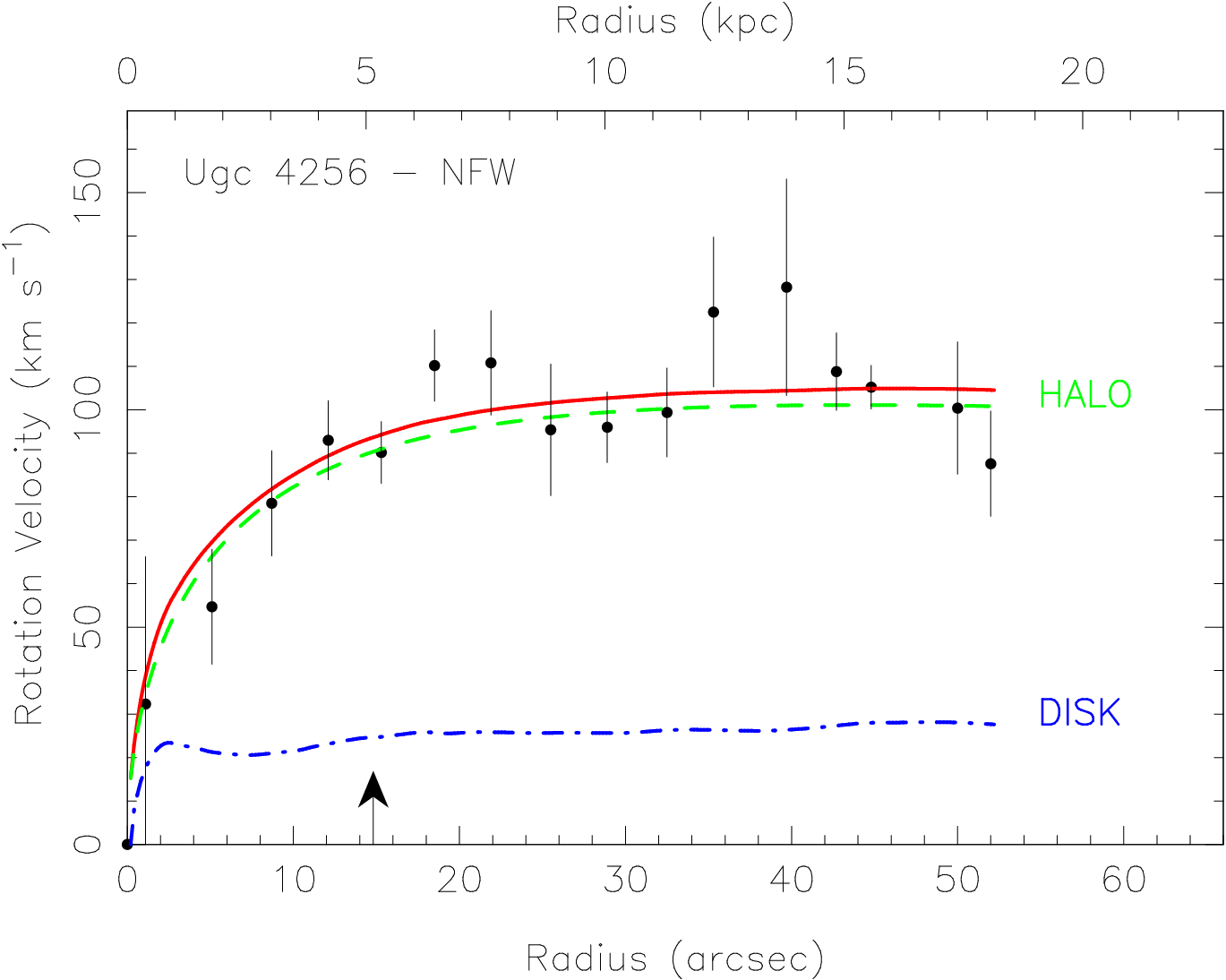} \\
\includegraphics[width=5.5cm,angle=-90]{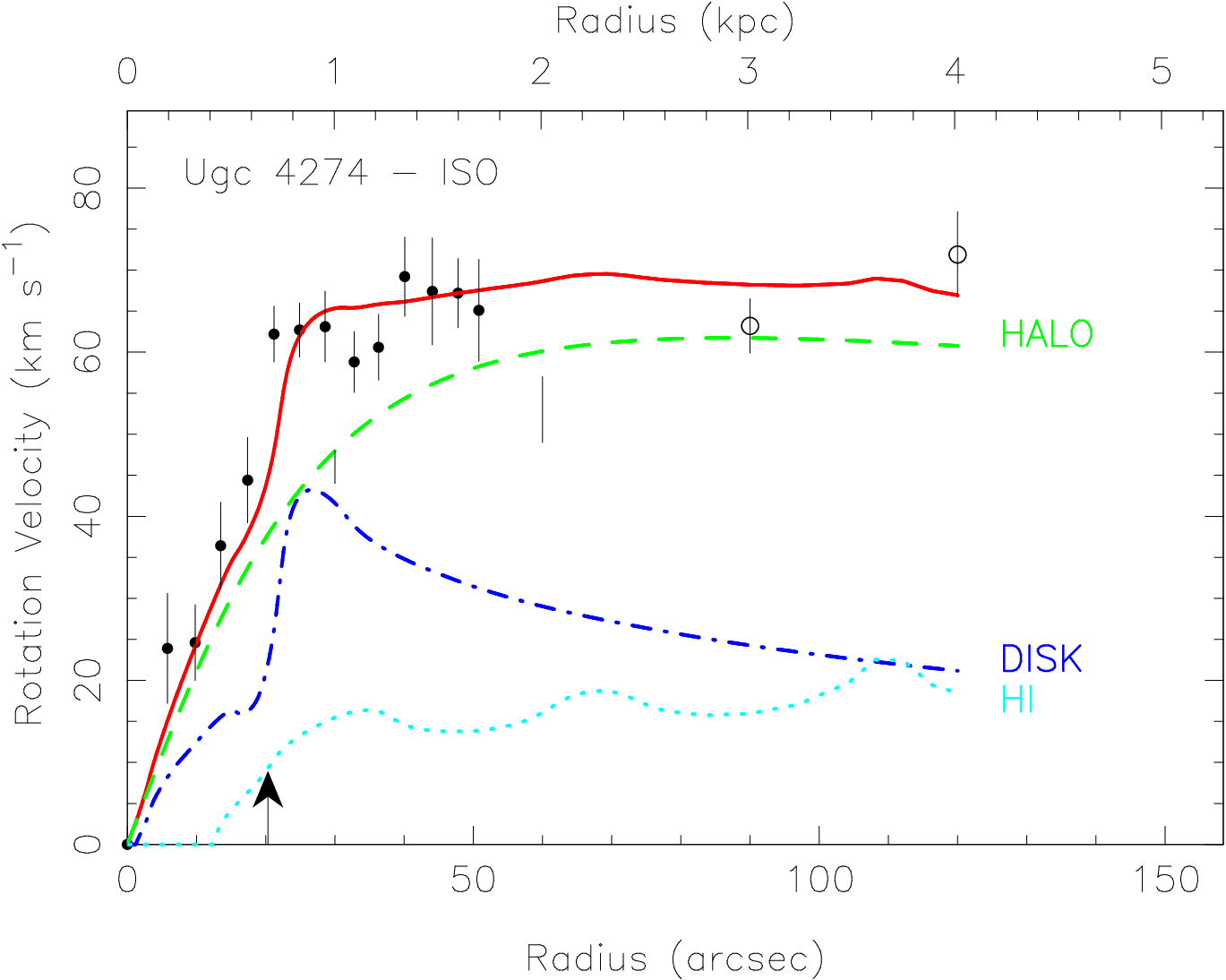} \includegraphics[width=5.5cm,angle=-90]{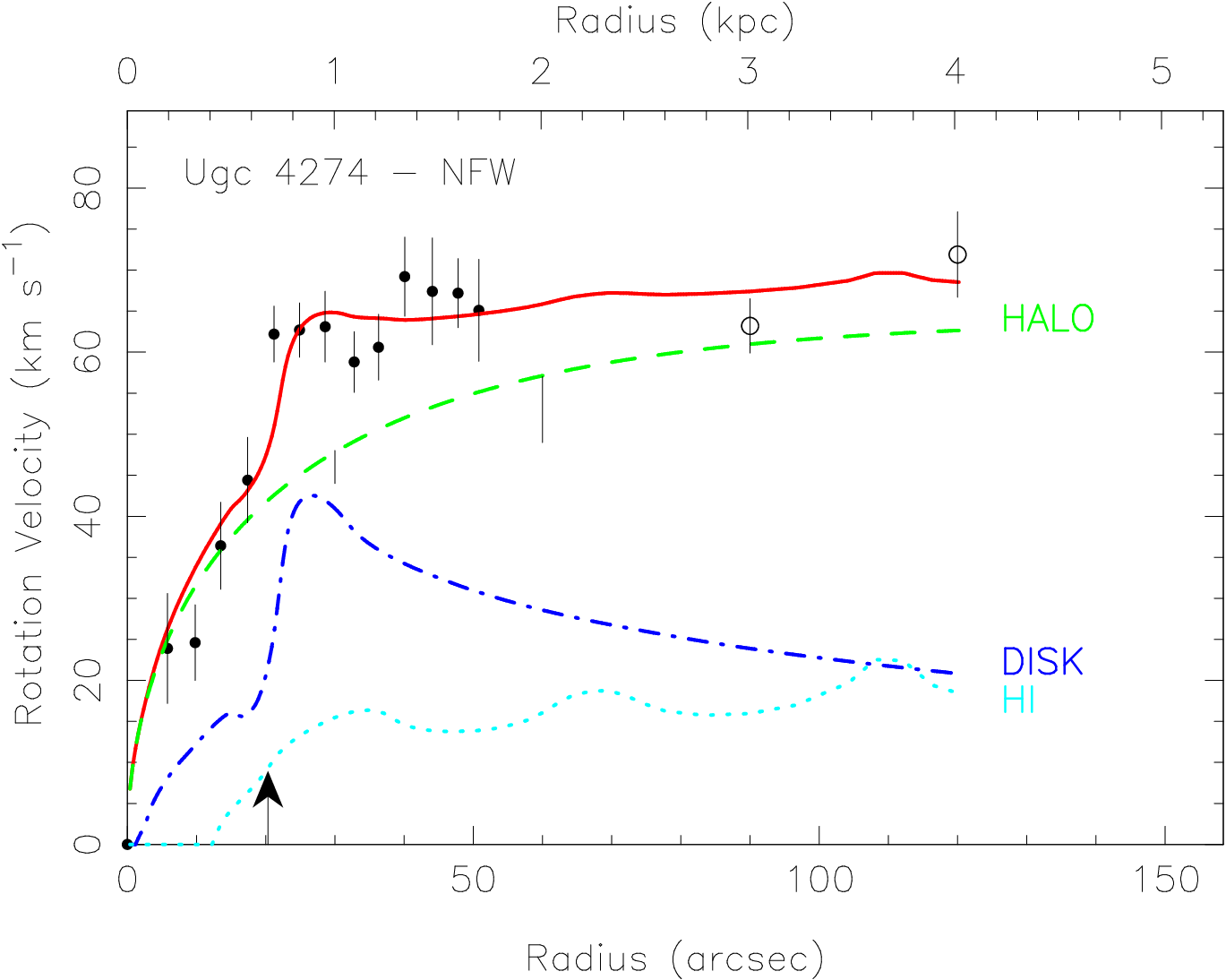} \\
\includegraphics[width=5.5cm,angle=-90]{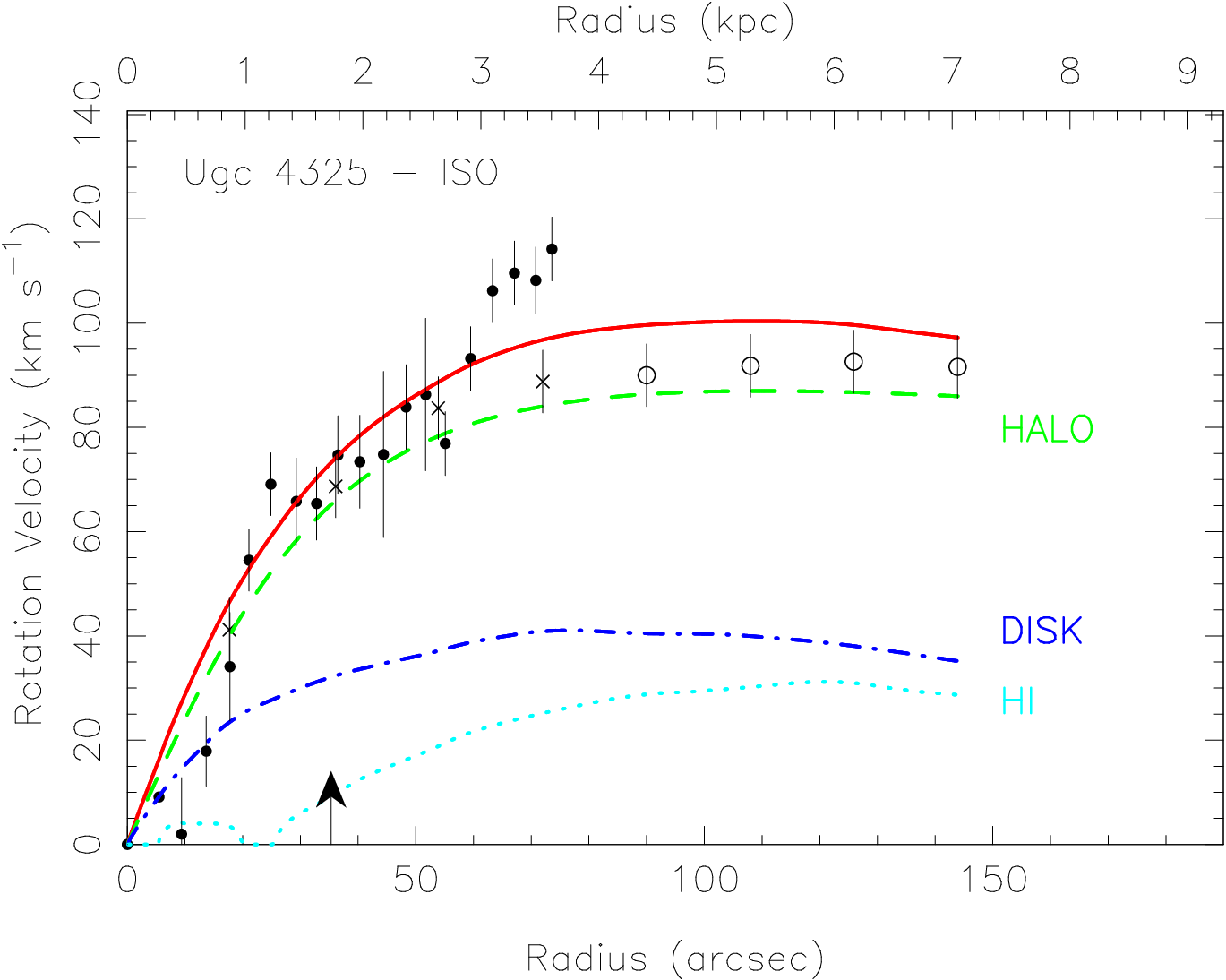} \includegraphics[width=5.5cm,angle=-90]{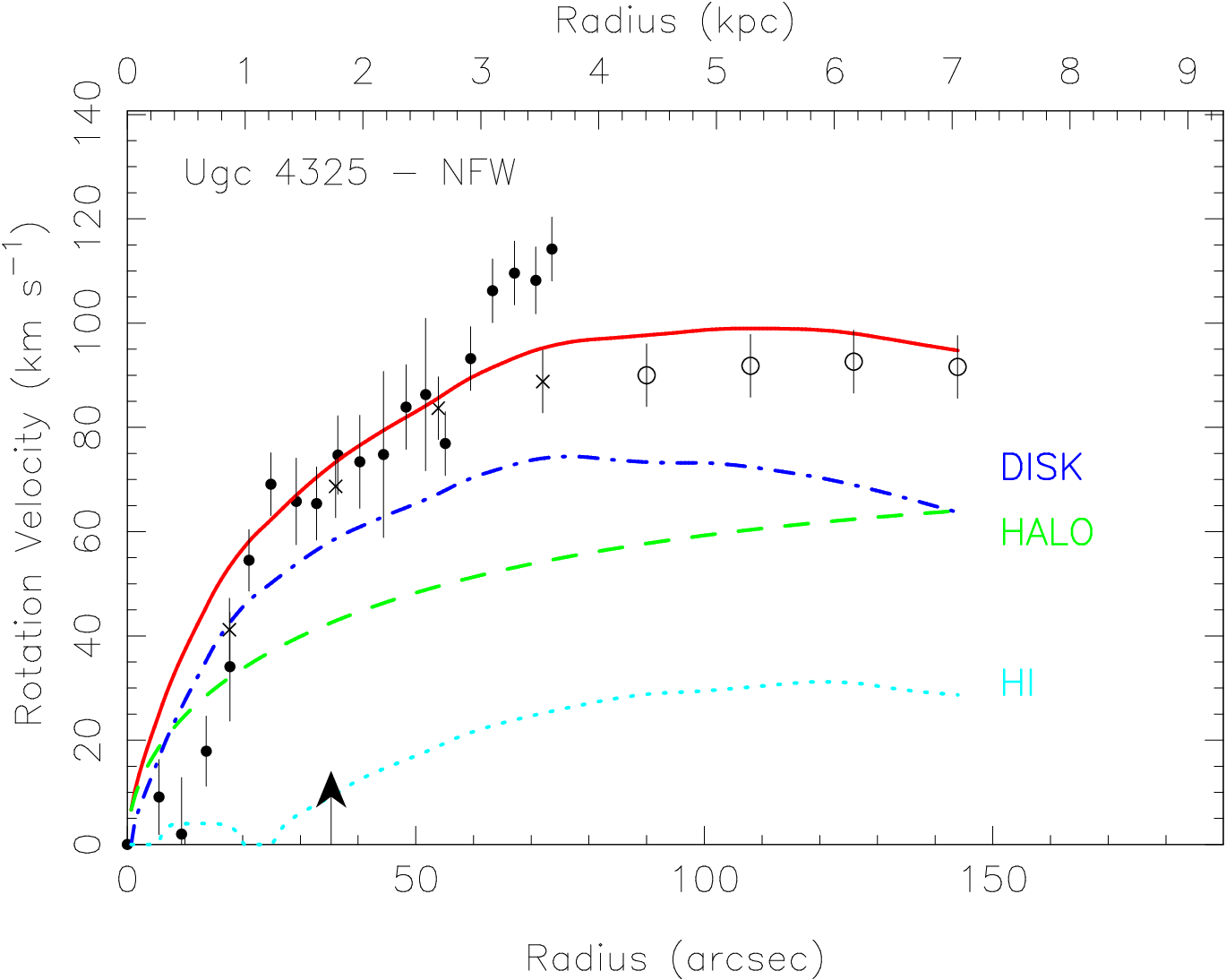} \\
\includegraphics[width=5.5cm,angle=-90]{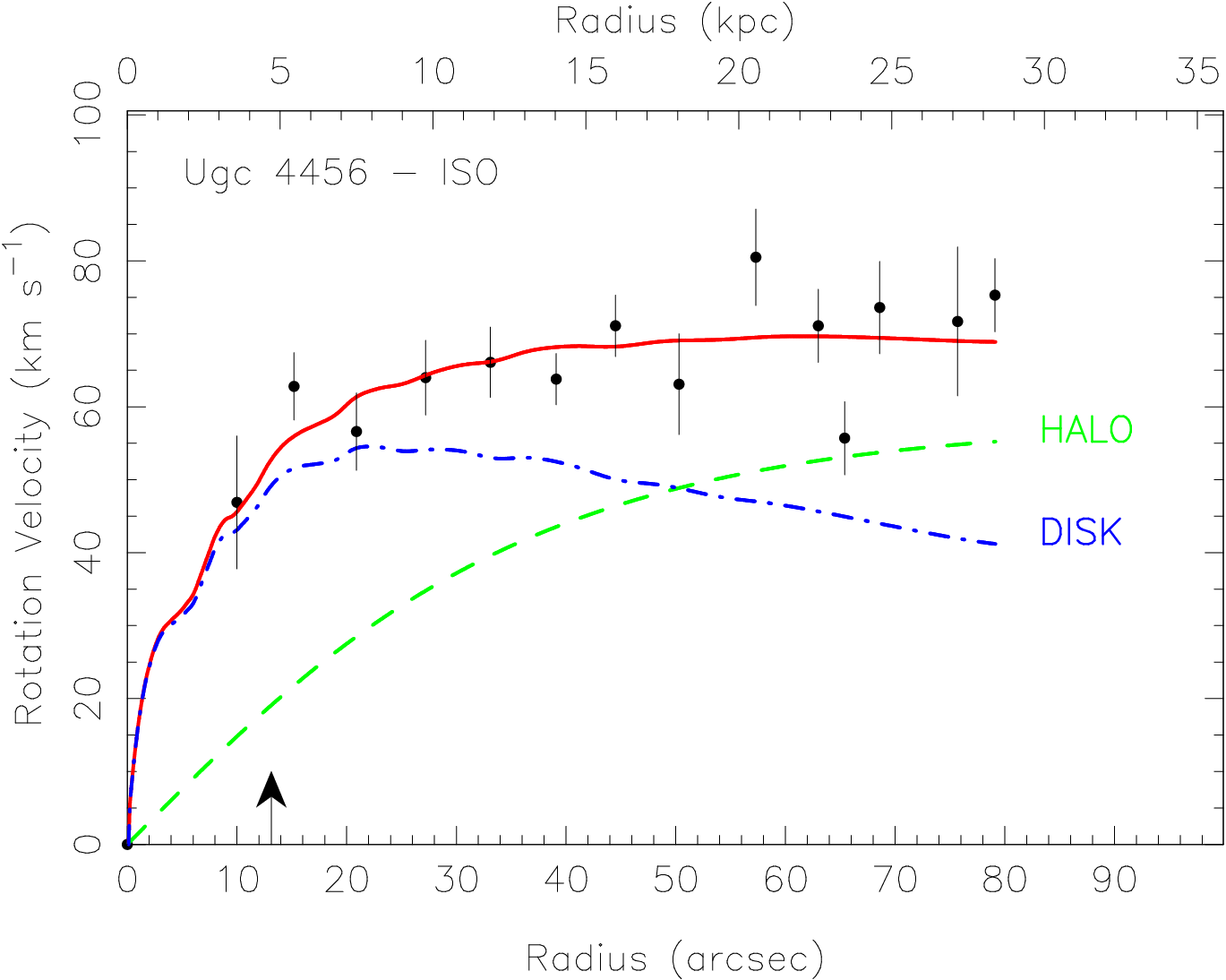} \includegraphics[width=5.5cm,angle=-90]{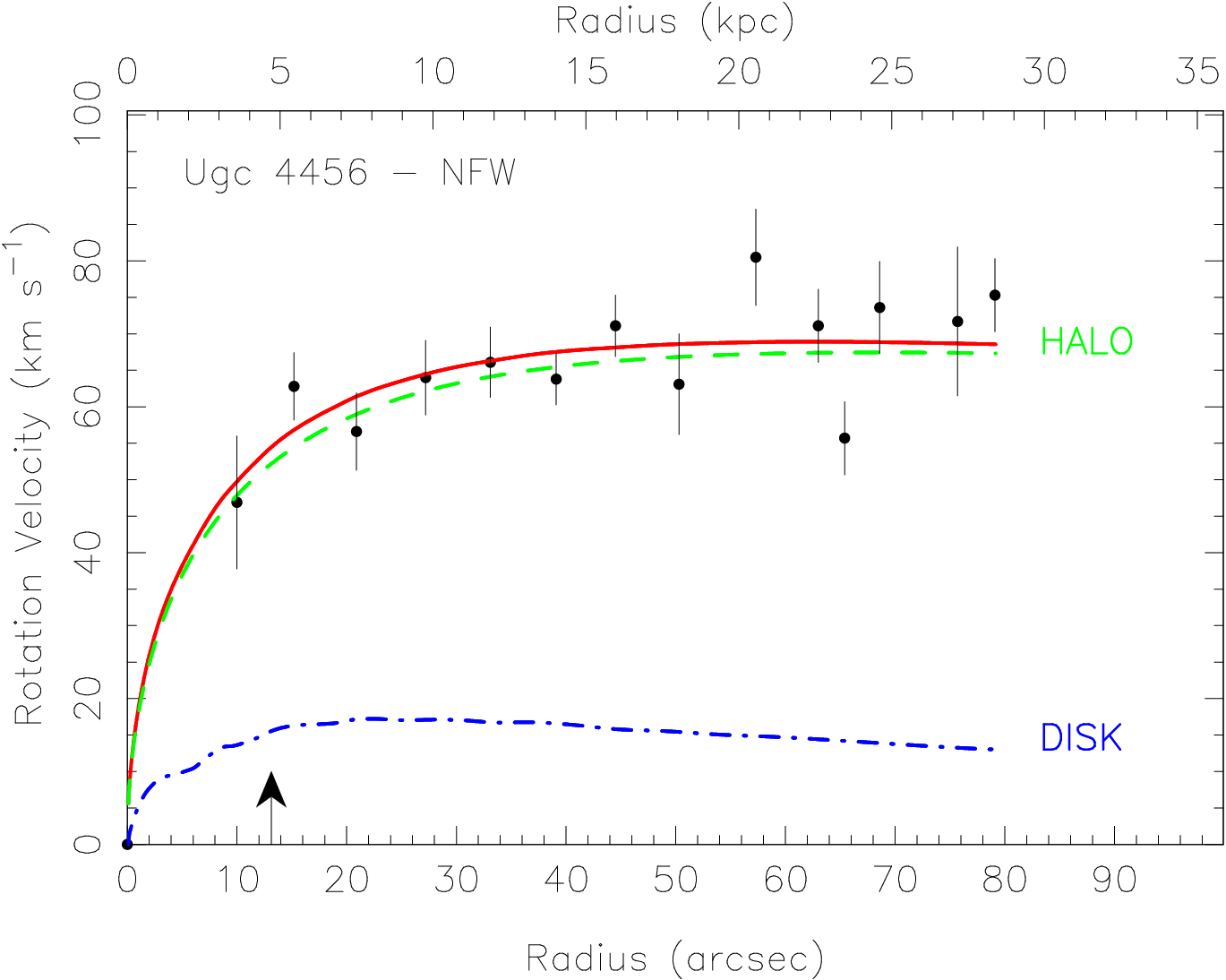}
\end{center}
\caption{Best fit model for the rotation curve of UGC4256,
UGC4274, UGC4325 and UGC4456 with isothermal sphere profile (left)
and NFW profile (right). Dots are for optical velocities, open
circles for HI velocities and crosses for HI in the central part,
not taken into account.The arrow on the X axis indicates the disk scale-length.}
\end{figure*}

\begin{figure*}
\begin{center}
\includegraphics[width=5.5cm,angle=-90]{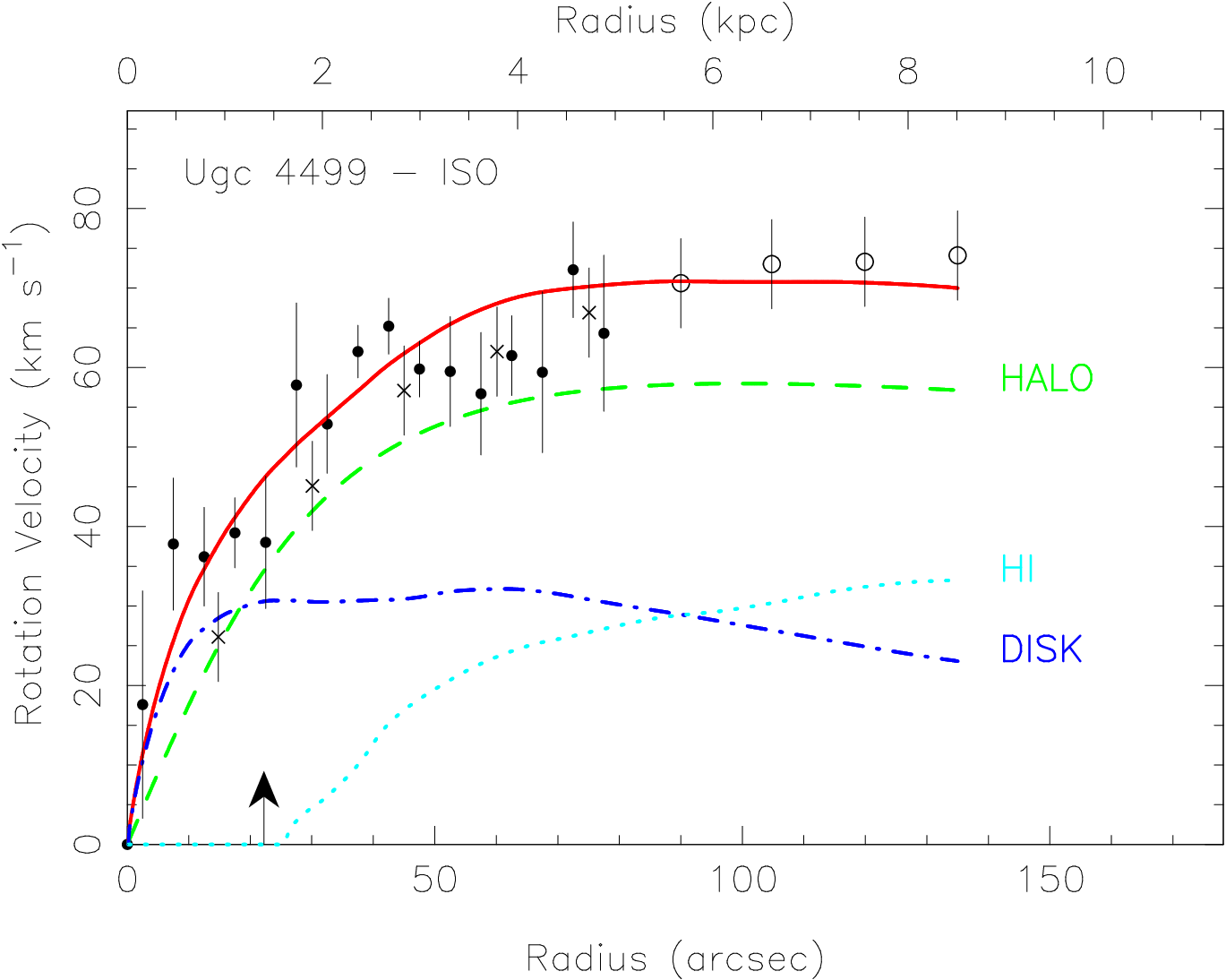}   \includegraphics[width=5.5cm,angle=-90]{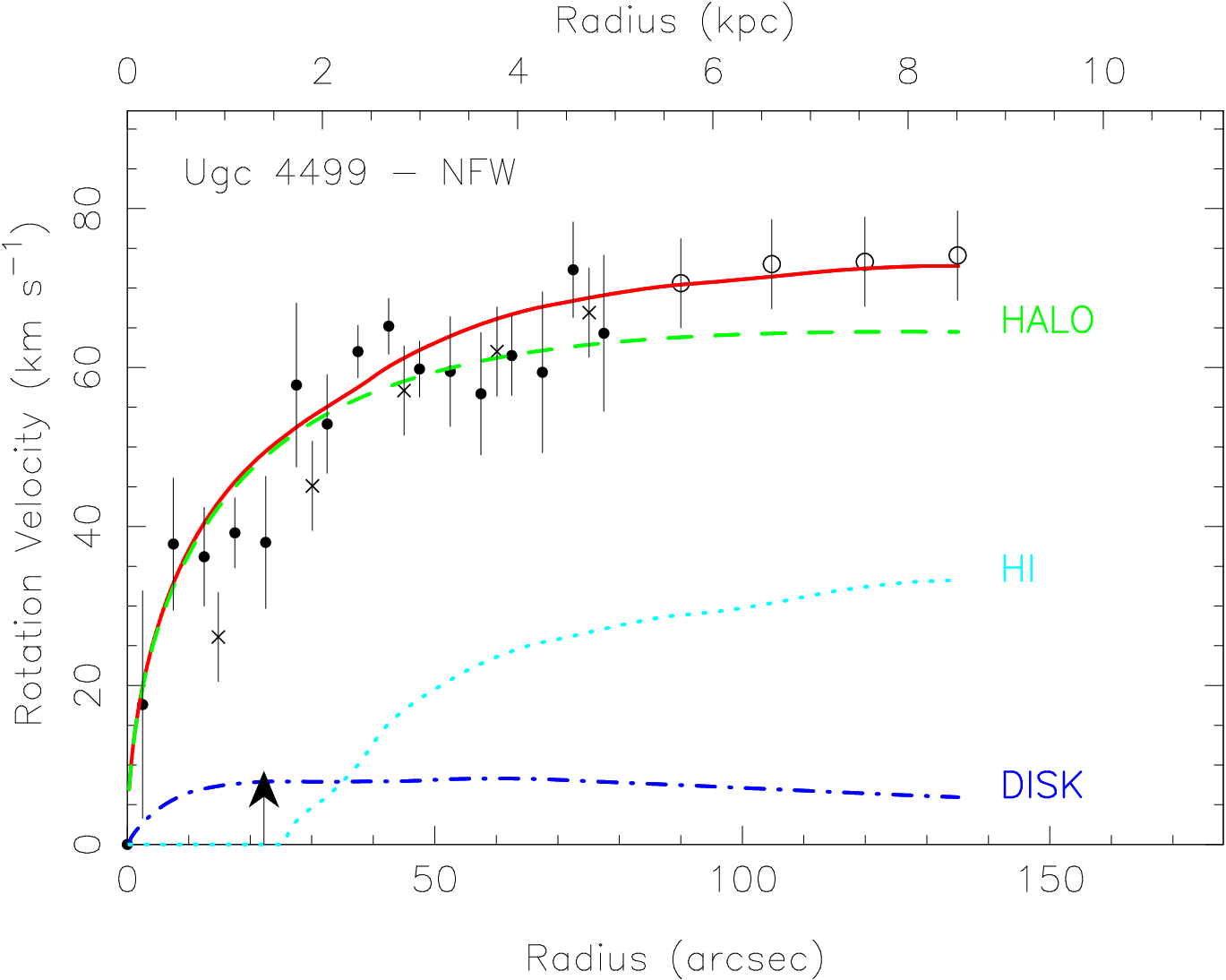}  \\
 \includegraphics[width=5.5cm,angle=-90]{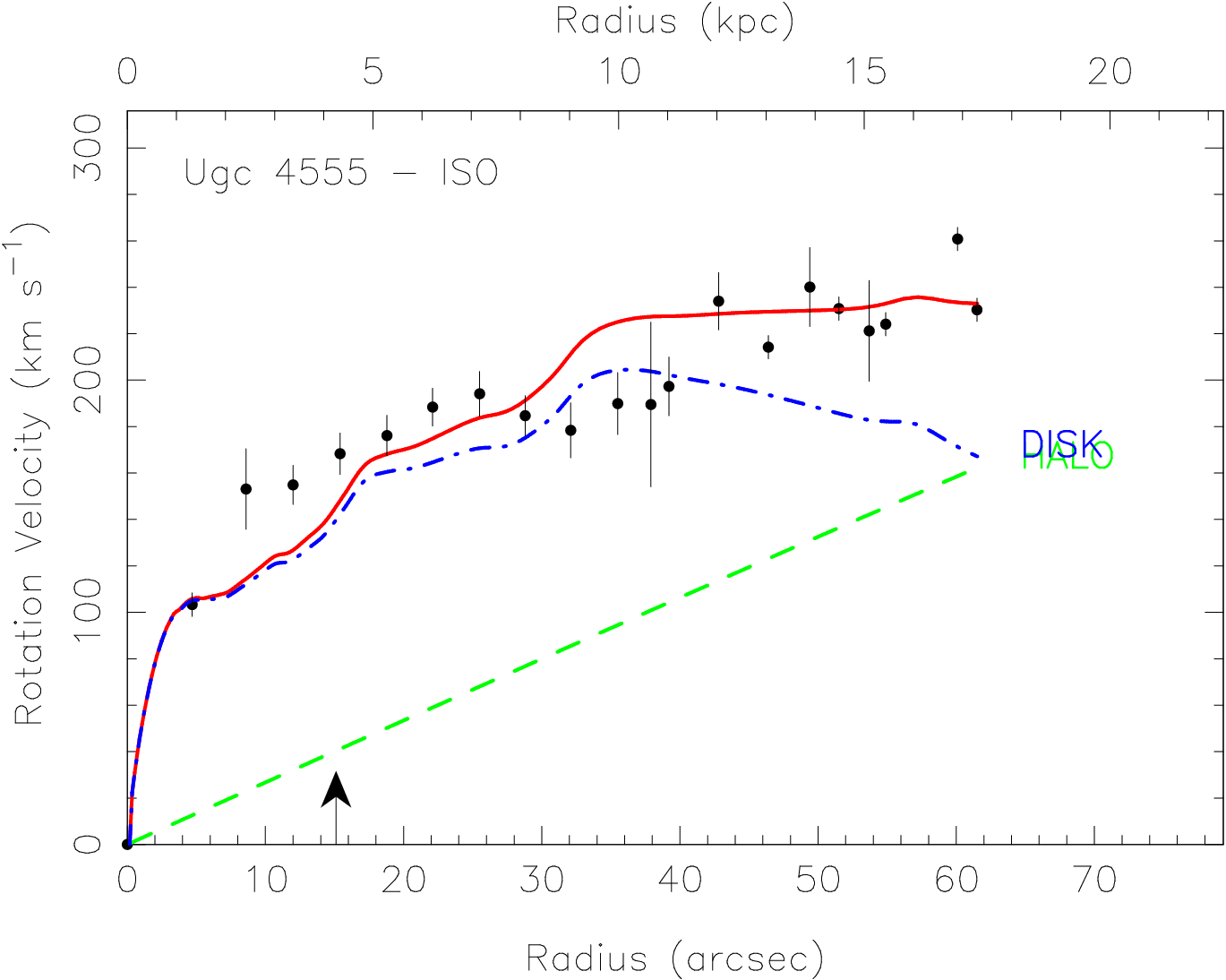}   \includegraphics[width=5.5cm,angle=-90]{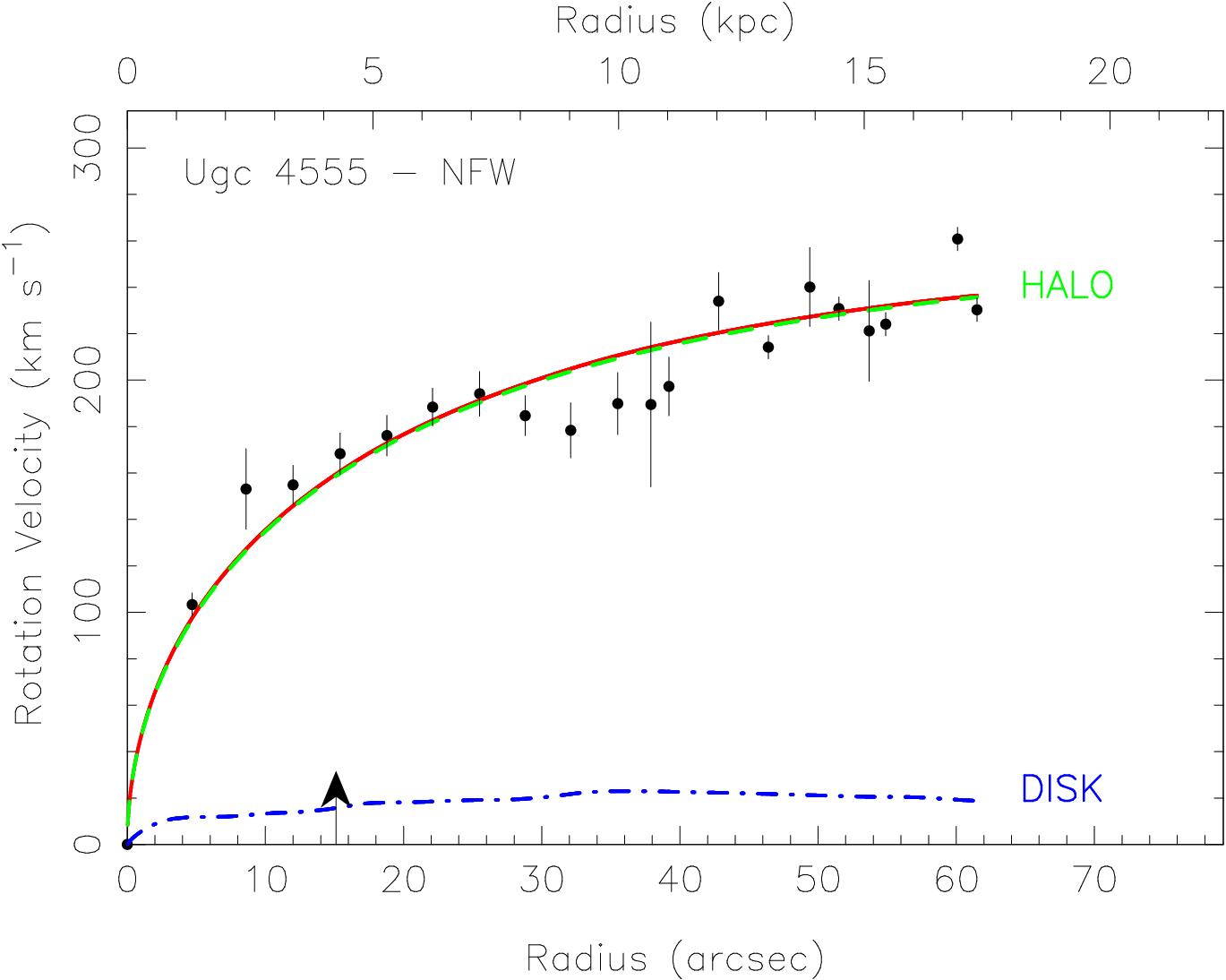} \\
\includegraphics[width=5.5cm,angle=-90]{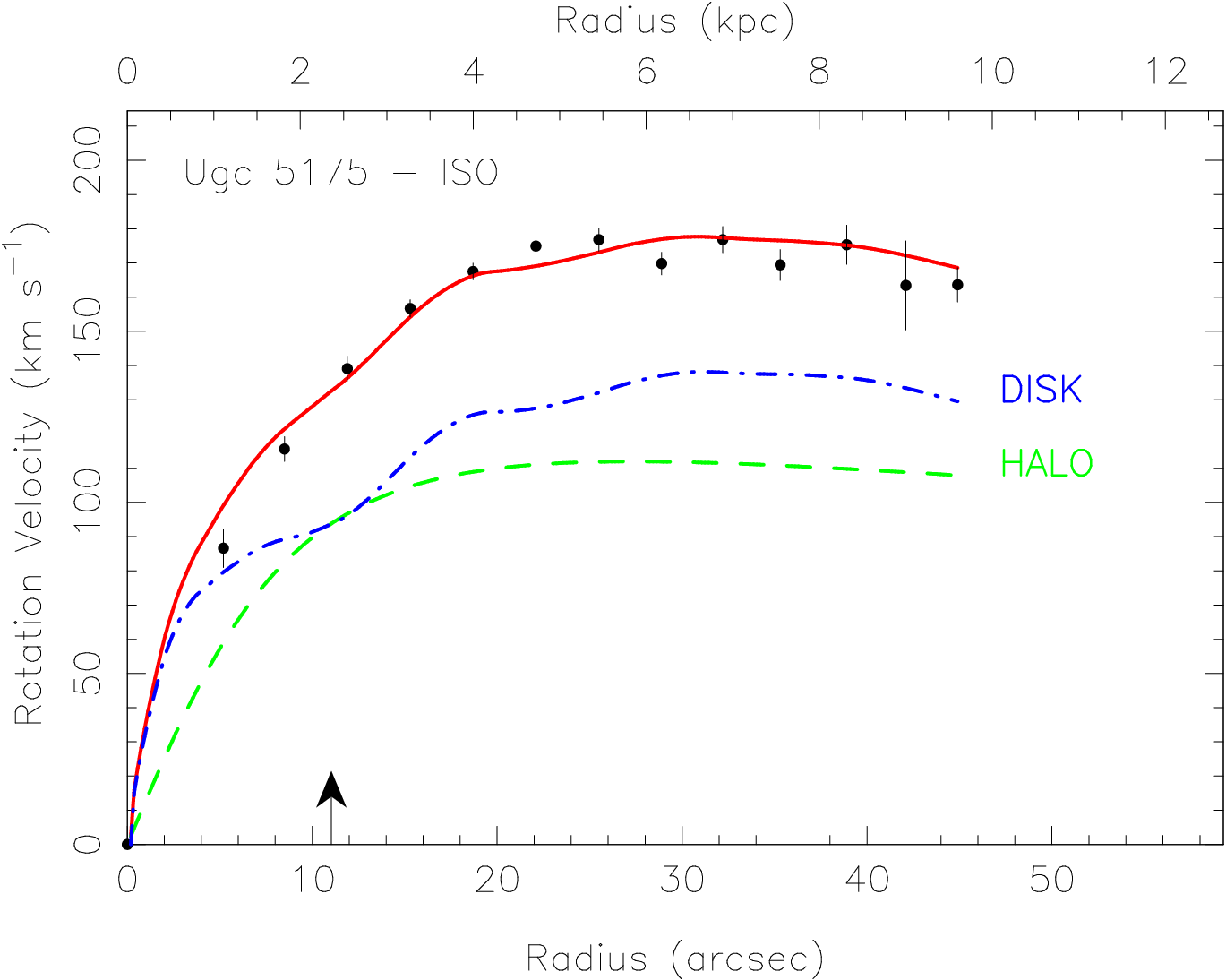}   \includegraphics[width=5.5cm,angle=-90]{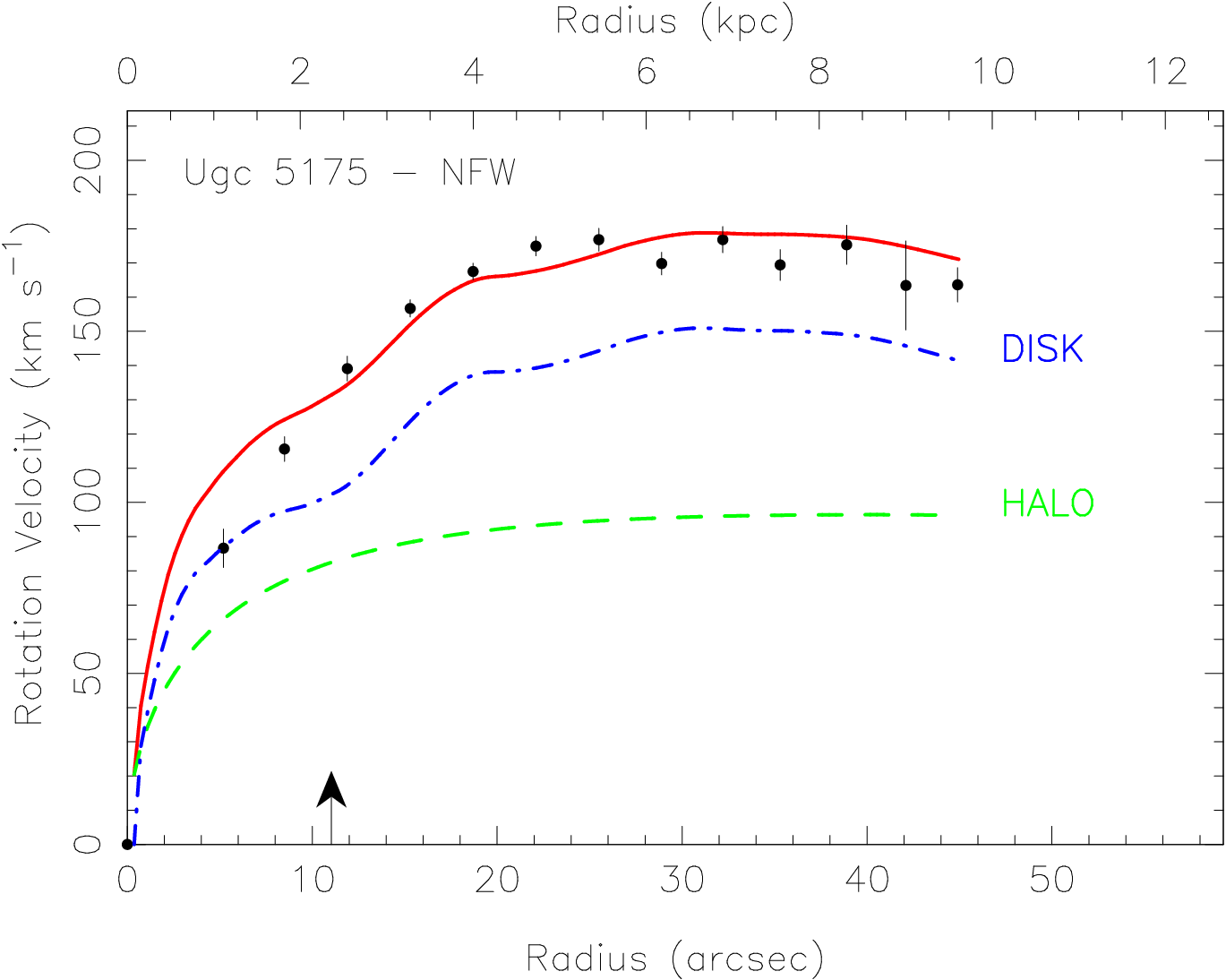}  \\
\includegraphics[width=5.5cm,angle=-90]{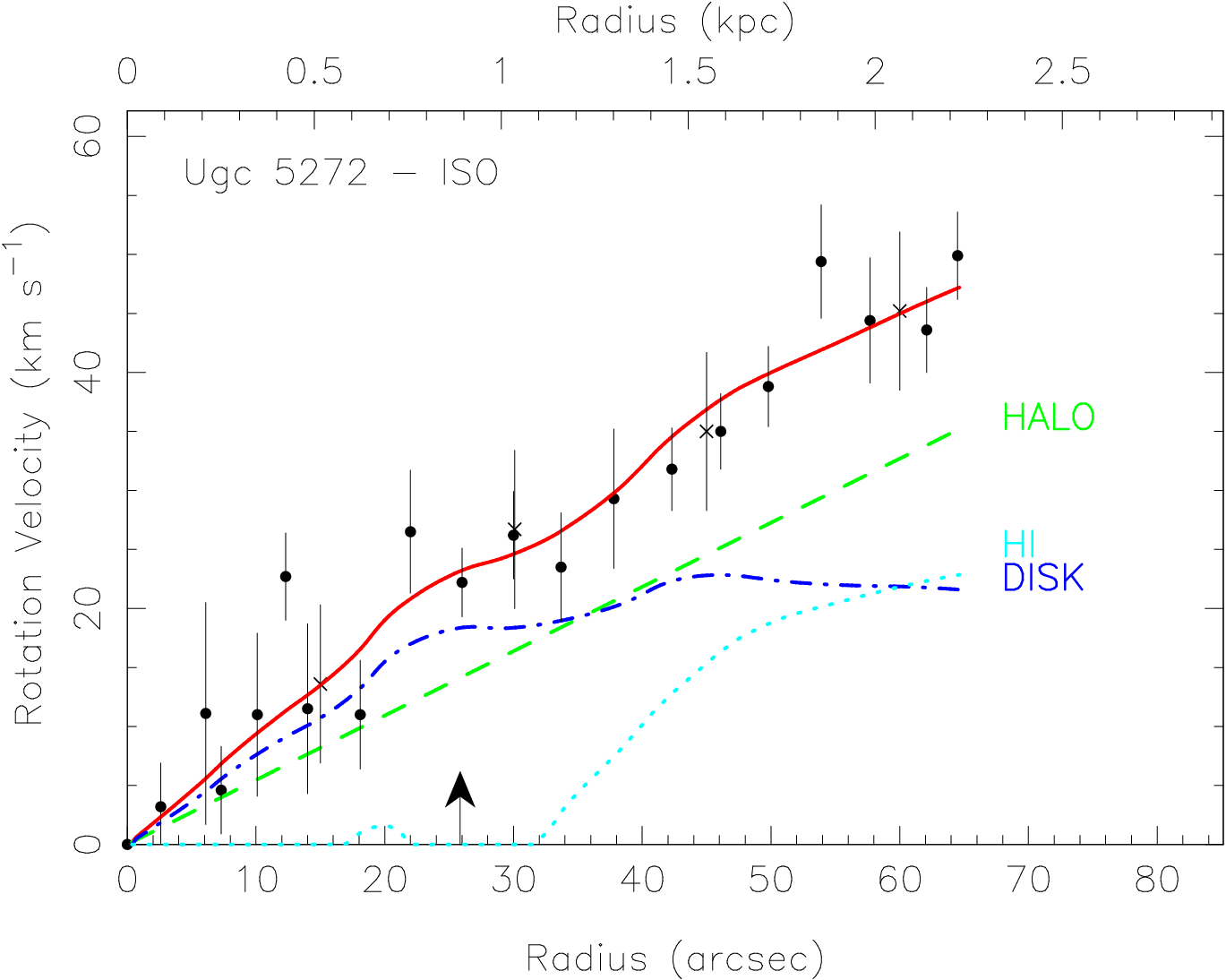}   \includegraphics[width=5.5cm,angle=-90]{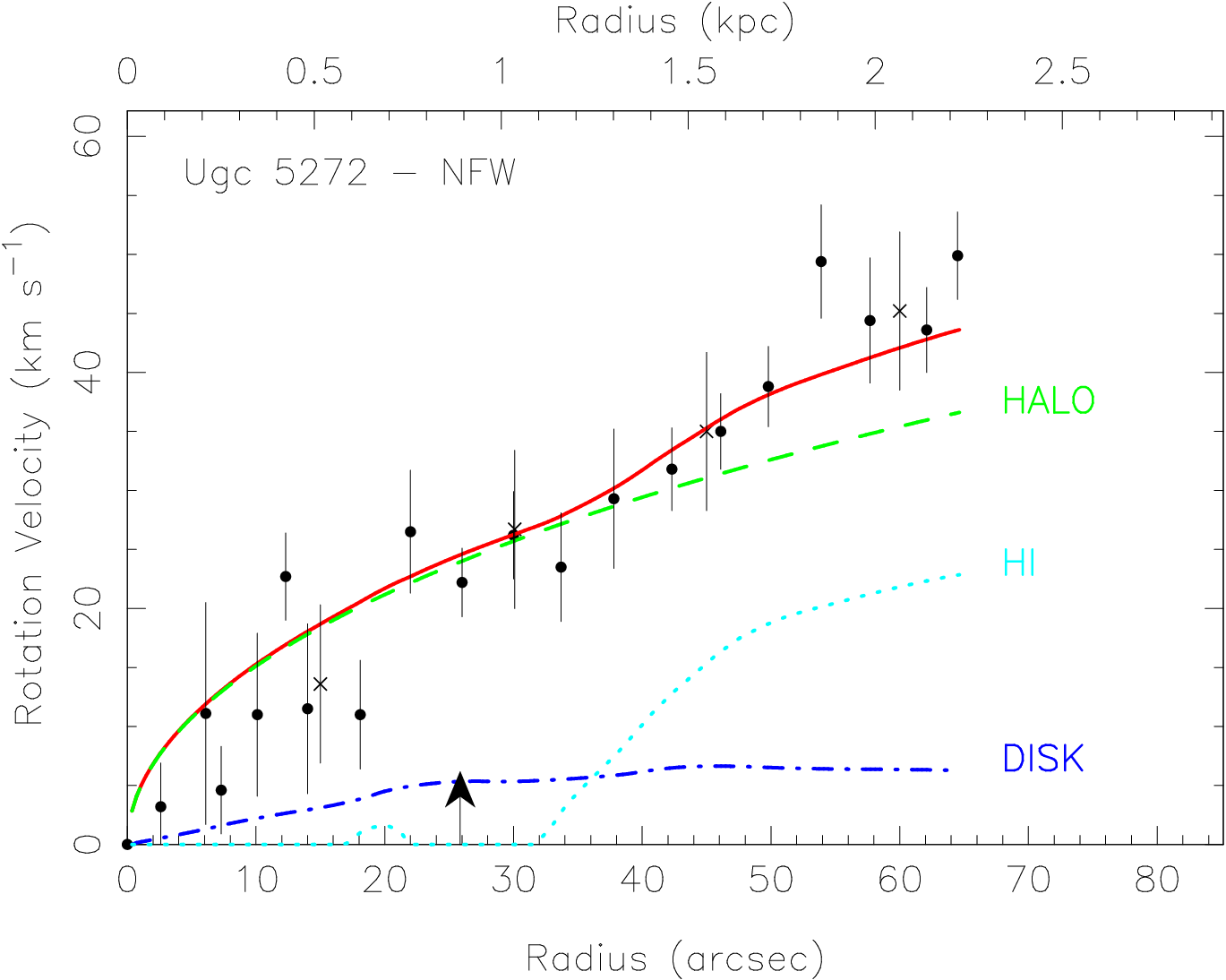}
\end{center}
\caption{Best fit model for the rotation curve of UGC4499,
UGC4555, UGC5175 and UGC5272 with isothermal sphere profile (left)
and NFW profile (right). Dots are for optical velocities, open
circles for HI velocities and crosses for HI in the central part,
not taken into account.The arrow on the X axis indicates the disk scale-length.}
\end{figure*}

\begin{figure*}
\begin{center}
\includegraphics[width=5.5cm,angle=-90]{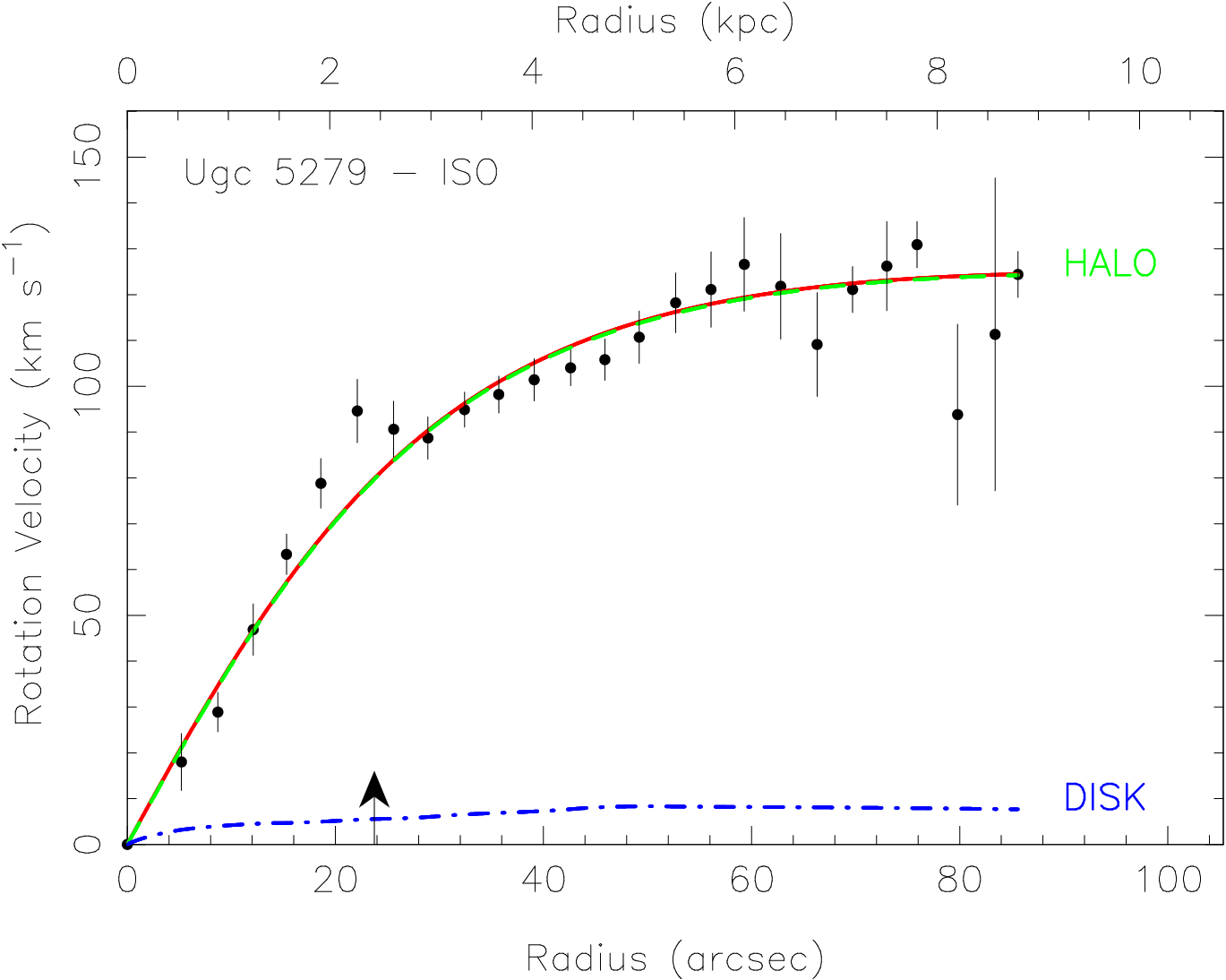}   \includegraphics[width=5.5cm,angle=-90]{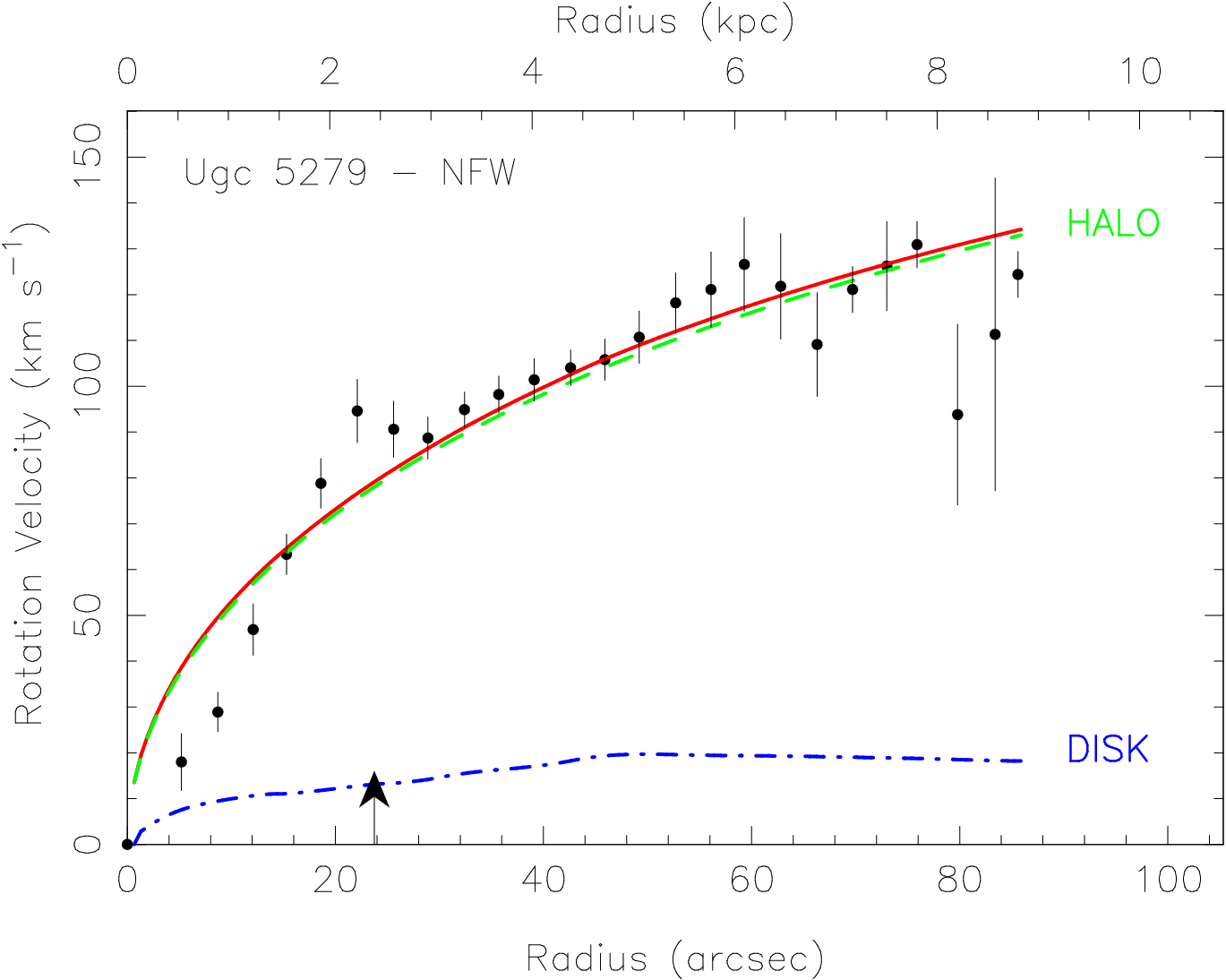}  \\
 \includegraphics[width=5.5cm,angle=-90]{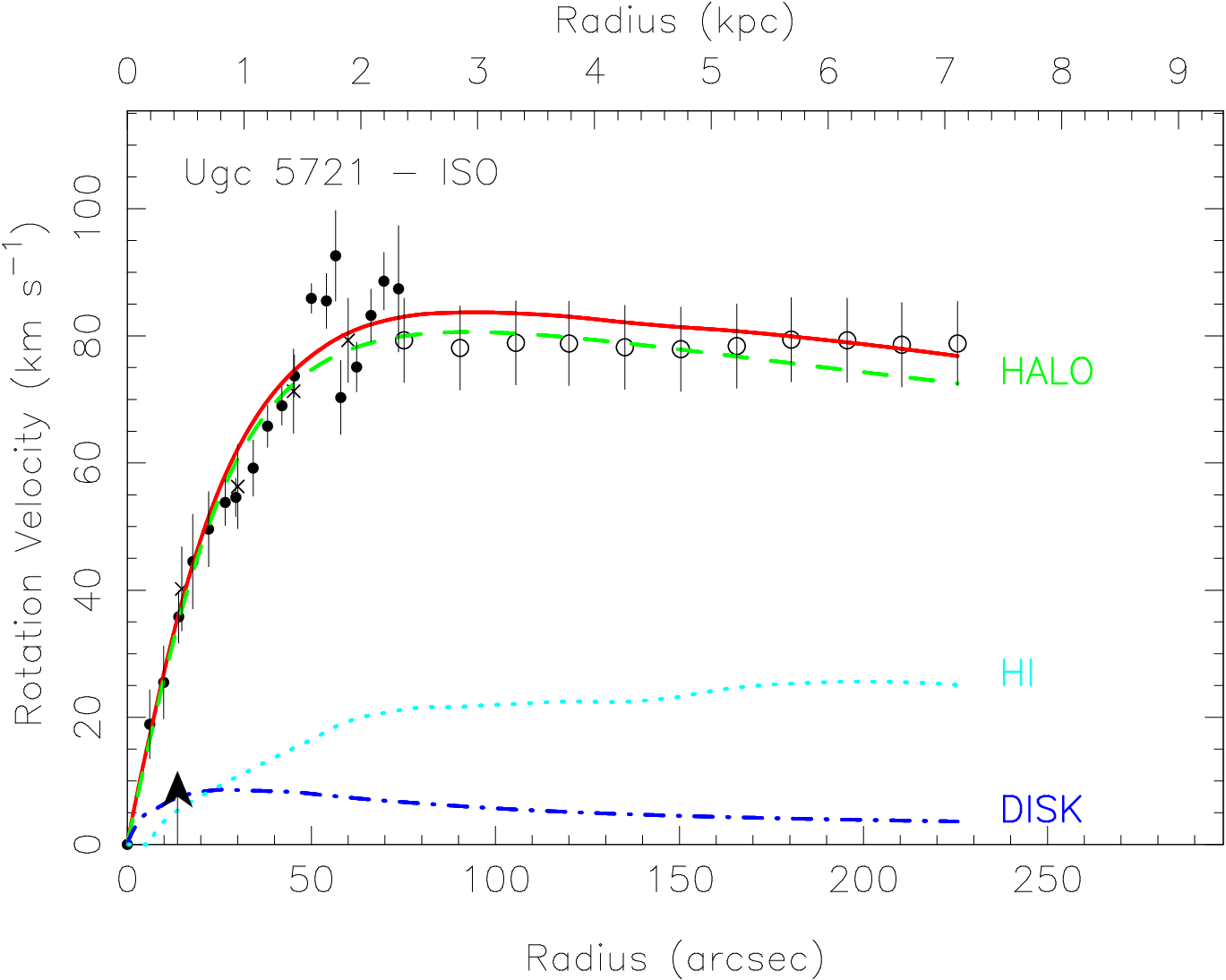}   \includegraphics[width=5.5cm,angle=-90]{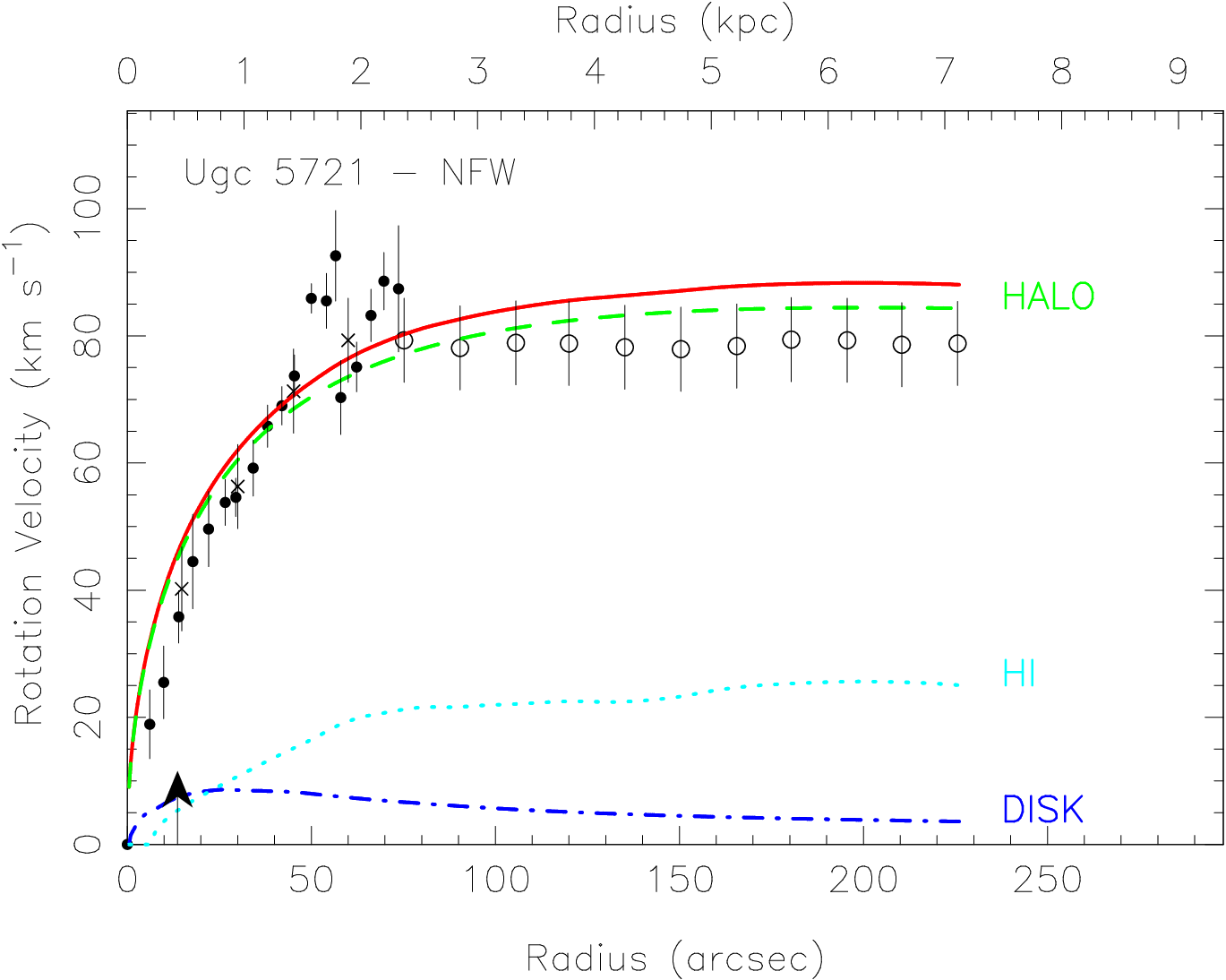} \\
\includegraphics[width=5.5cm,angle=-90]{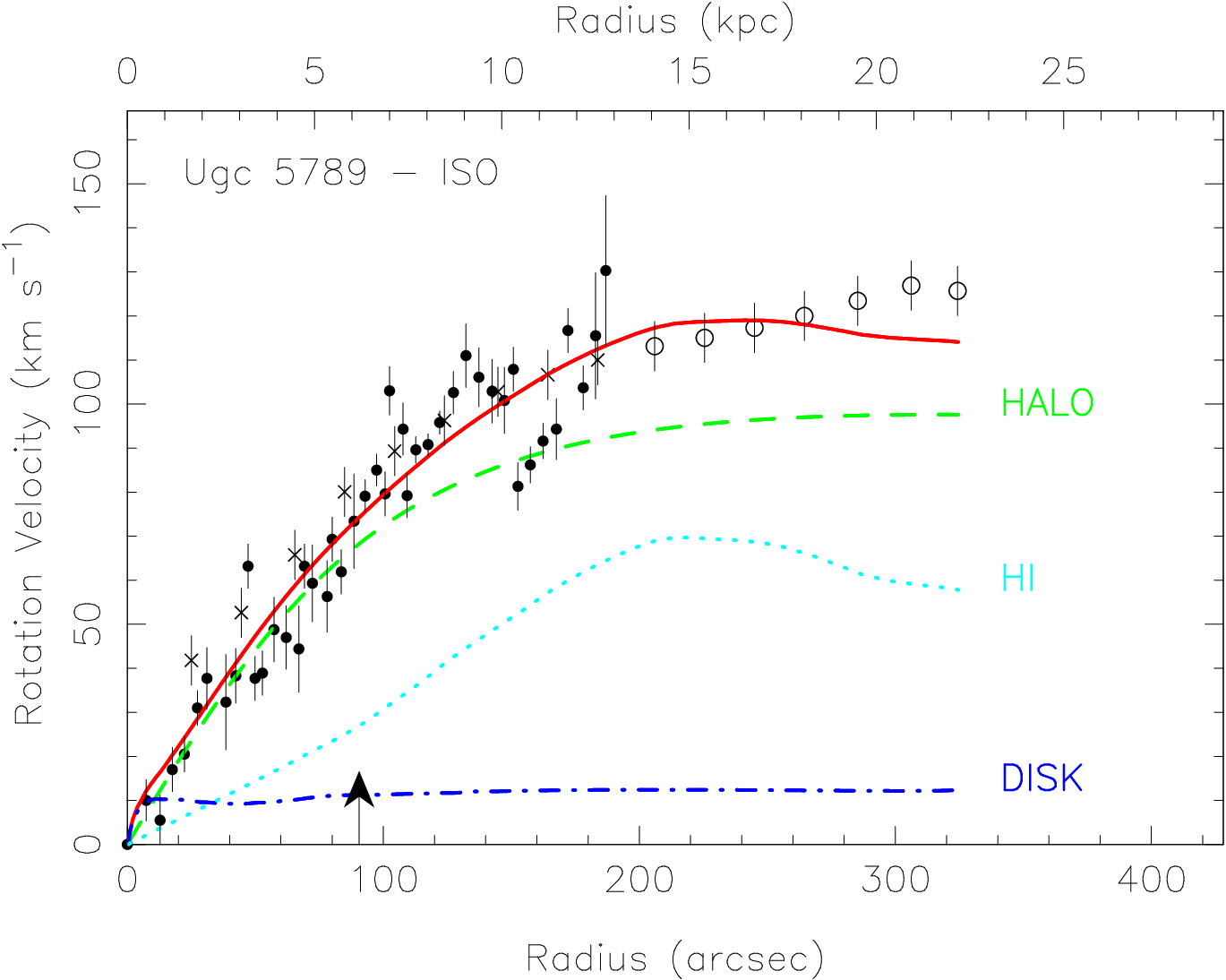}   \includegraphics[width=5.5cm,angle=-90]{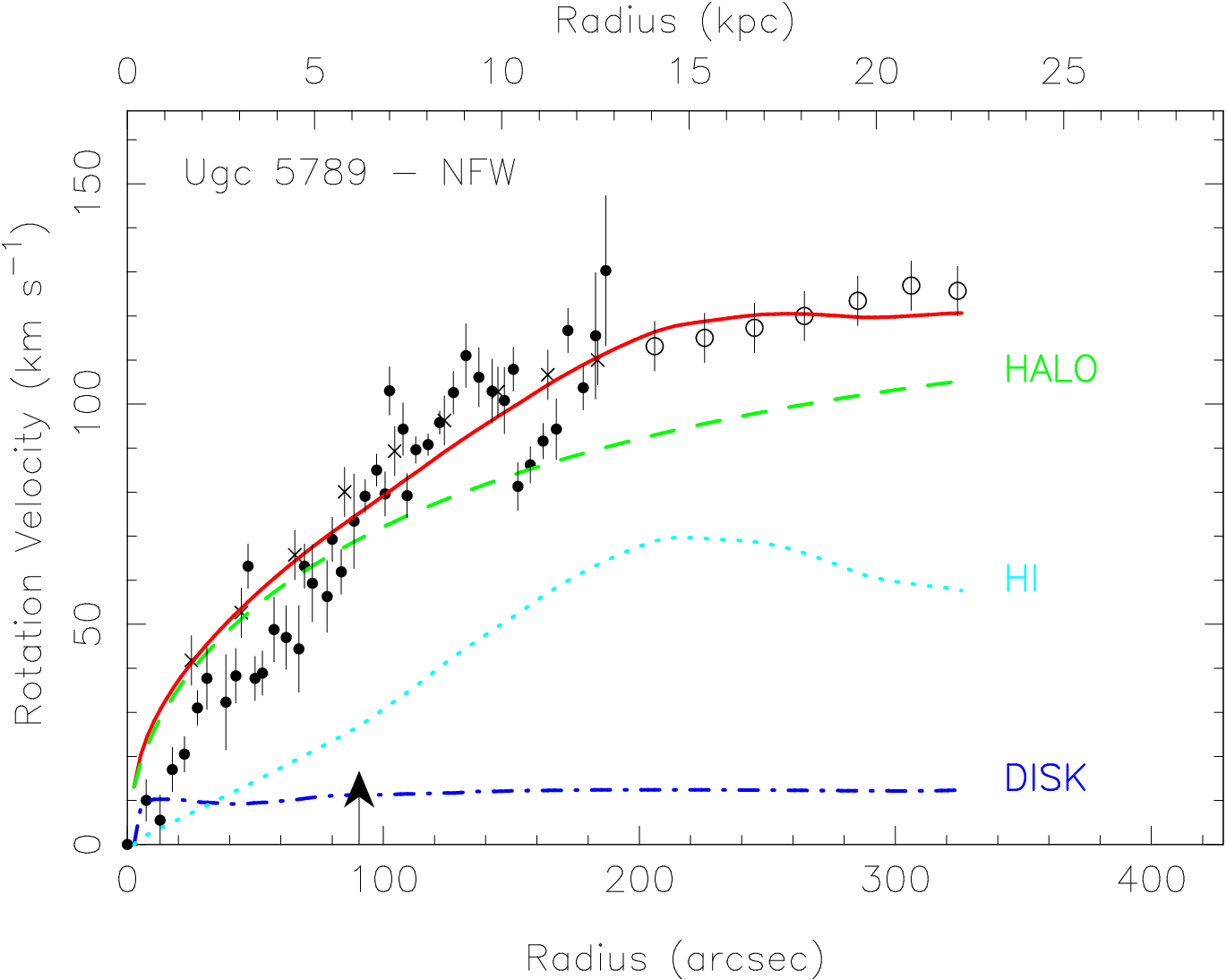}  \\
\includegraphics[width=5.5cm,angle=-90]{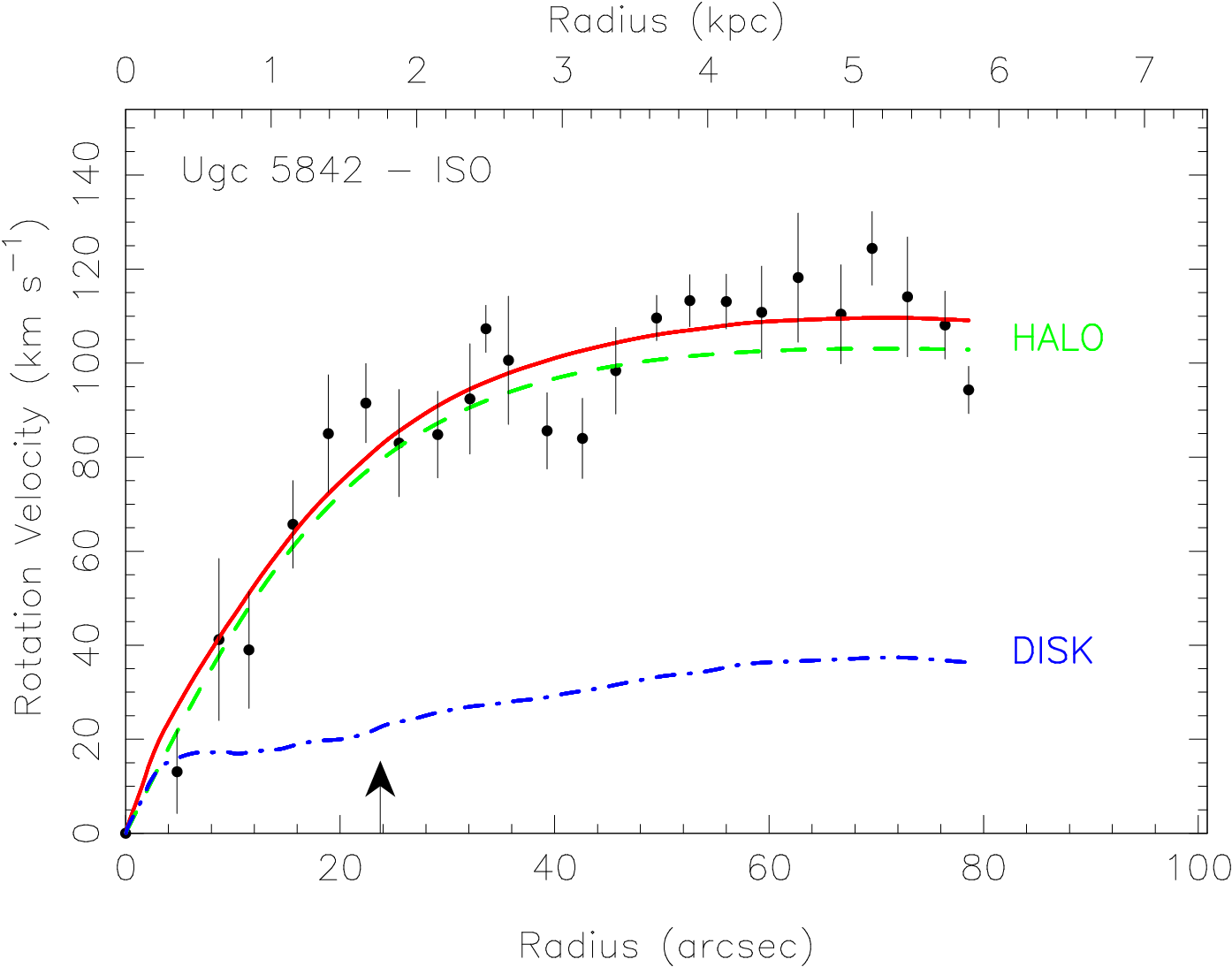}   \includegraphics[width=5.5cm,angle=-90]{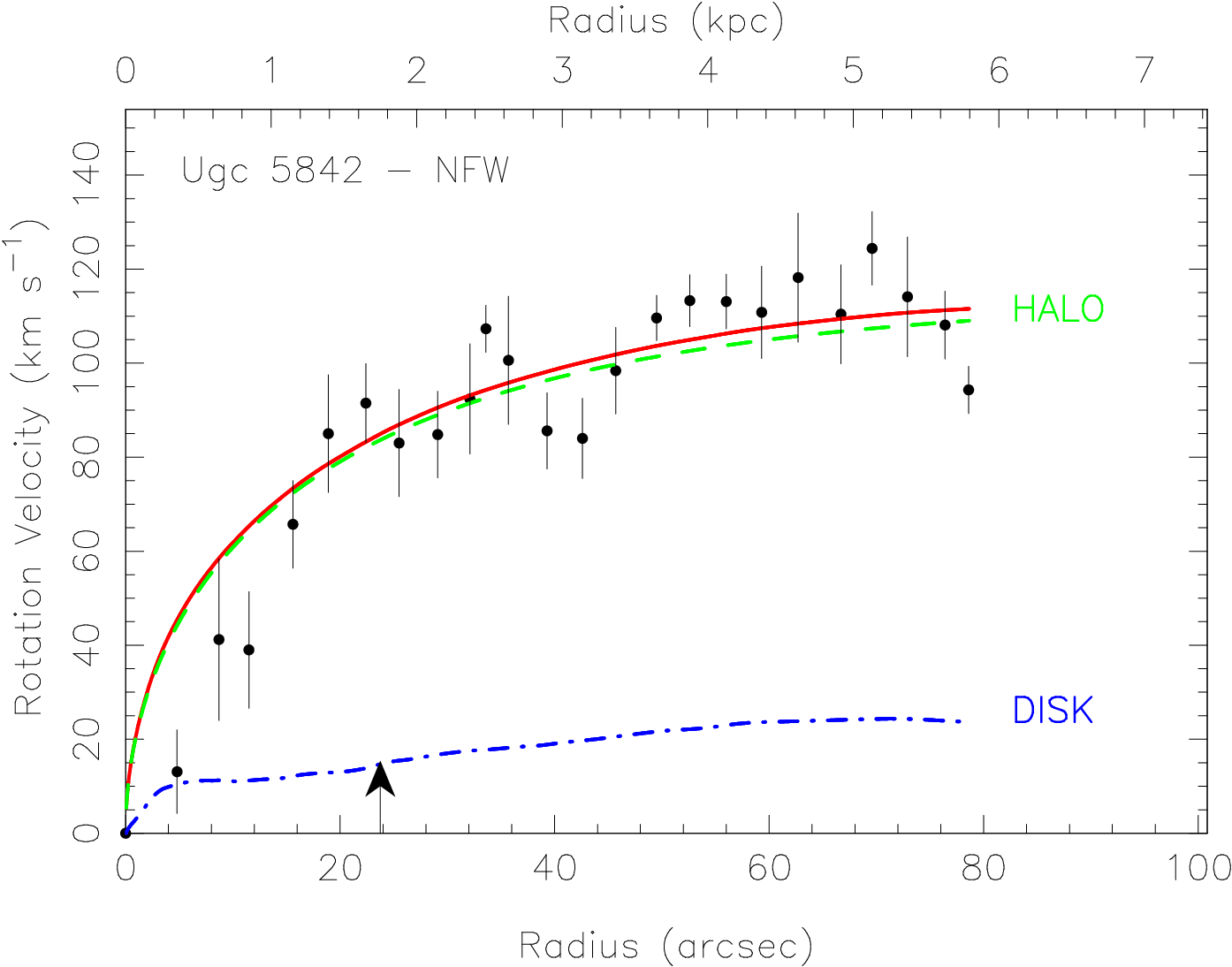}
\end{center}
\caption{Best fit model for the rotation curve of UGC5279,
UGC5721, UGC5789 and UGC5842 with isothermal sphere profile (left)
and NFW profile (right). Dots are for optical velocities, open
circles for HI velocities and crosses for HI in the central part,
not taken into account.The arrow on the X axis indicates the disk scale-length.}
\end{figure*}

\begin{figure*}
\begin{center}
\includegraphics[width=5.5cm,angle=-90]{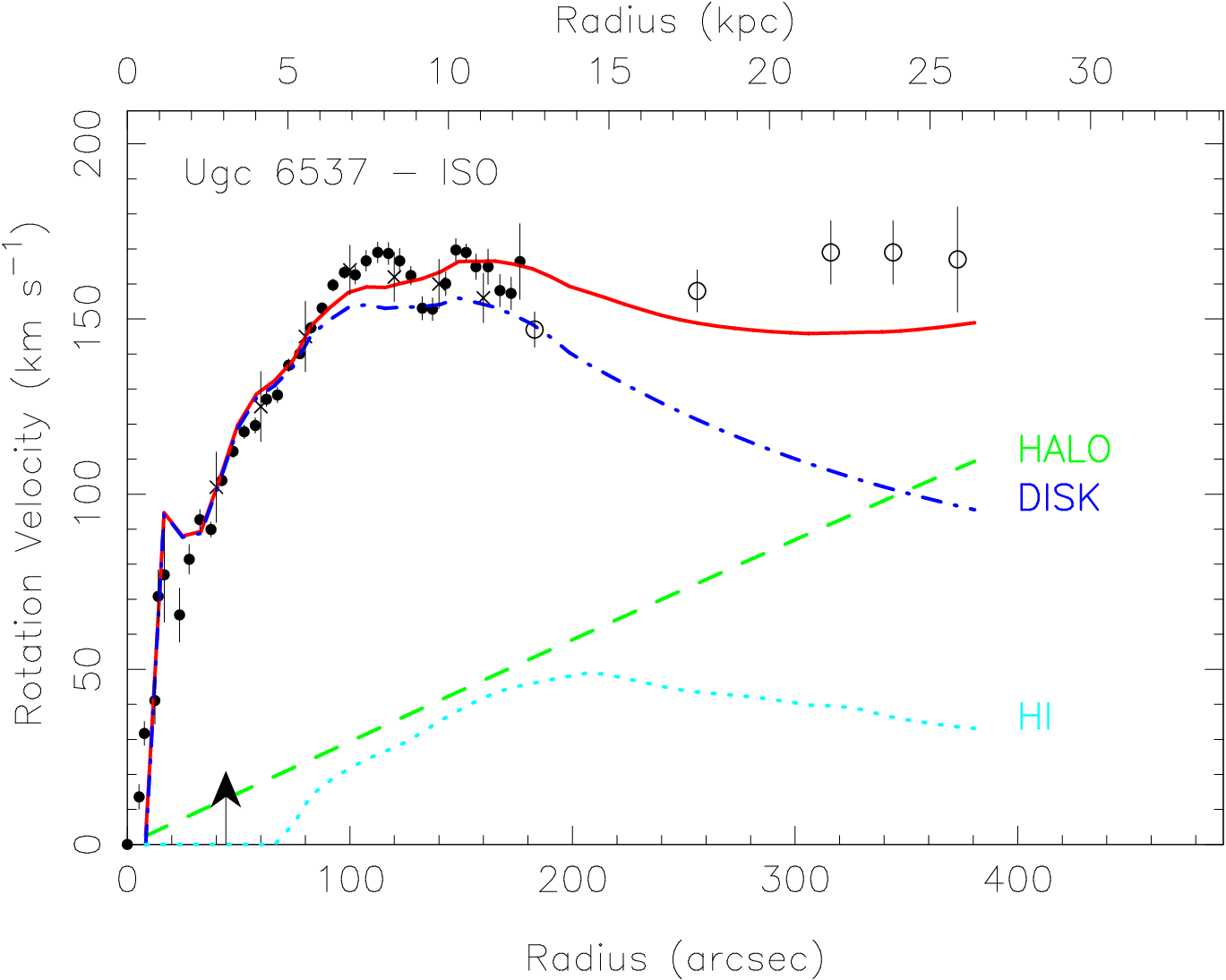}   \includegraphics[width=5.5cm,angle=-90]{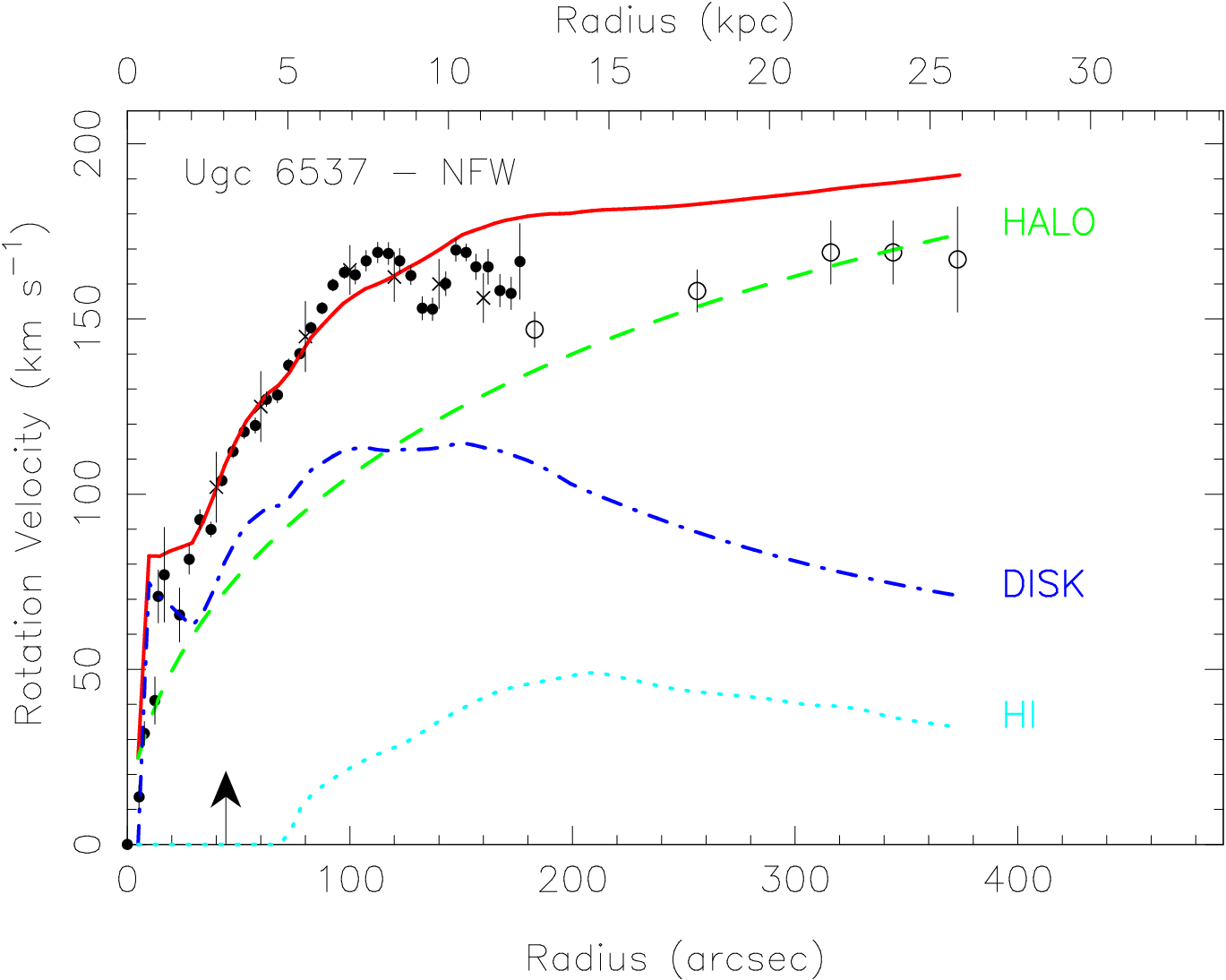}  \\
\includegraphics[width=5.5cm,angle=-90]{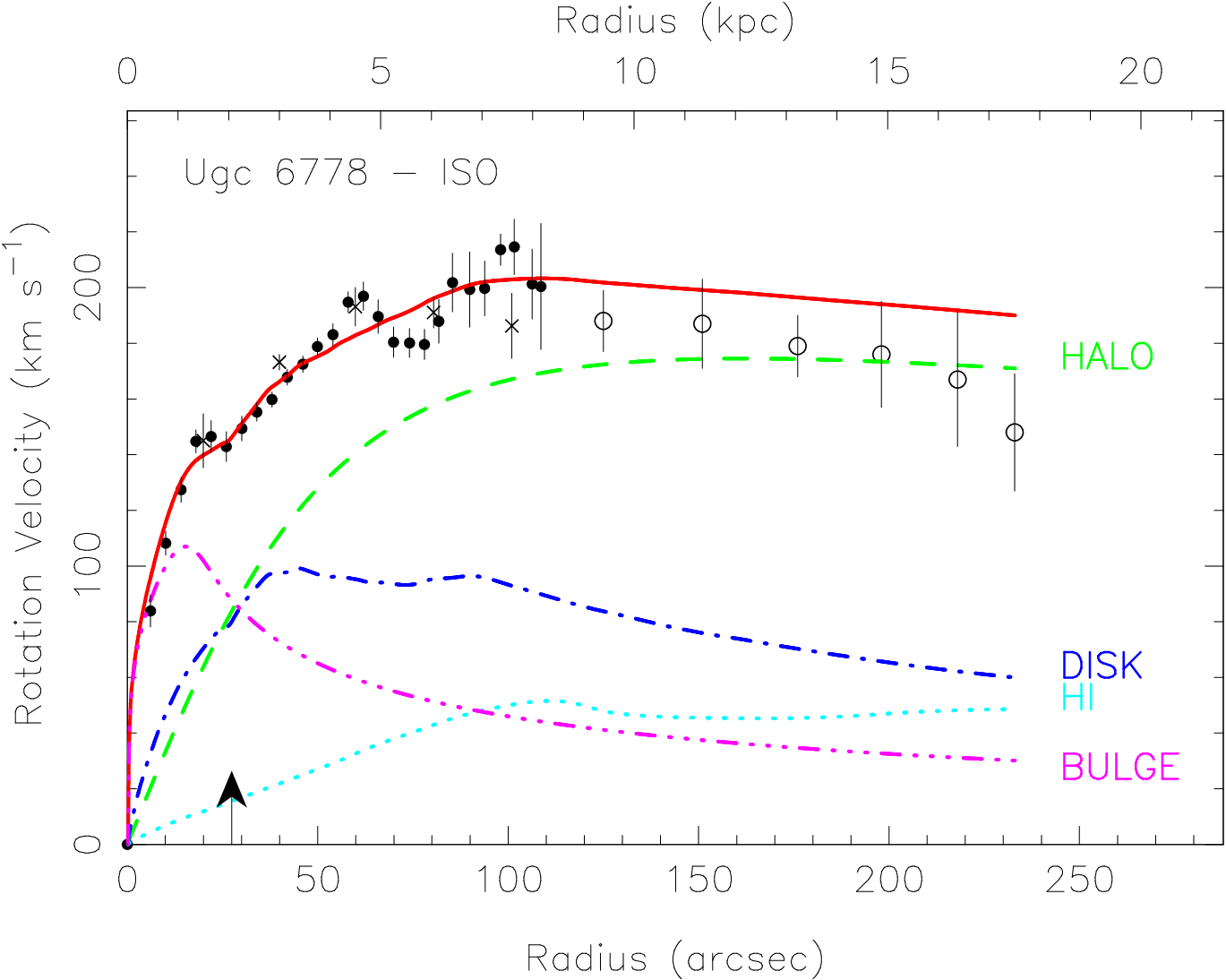}   \includegraphics[width=5.5cm,angle=-90]{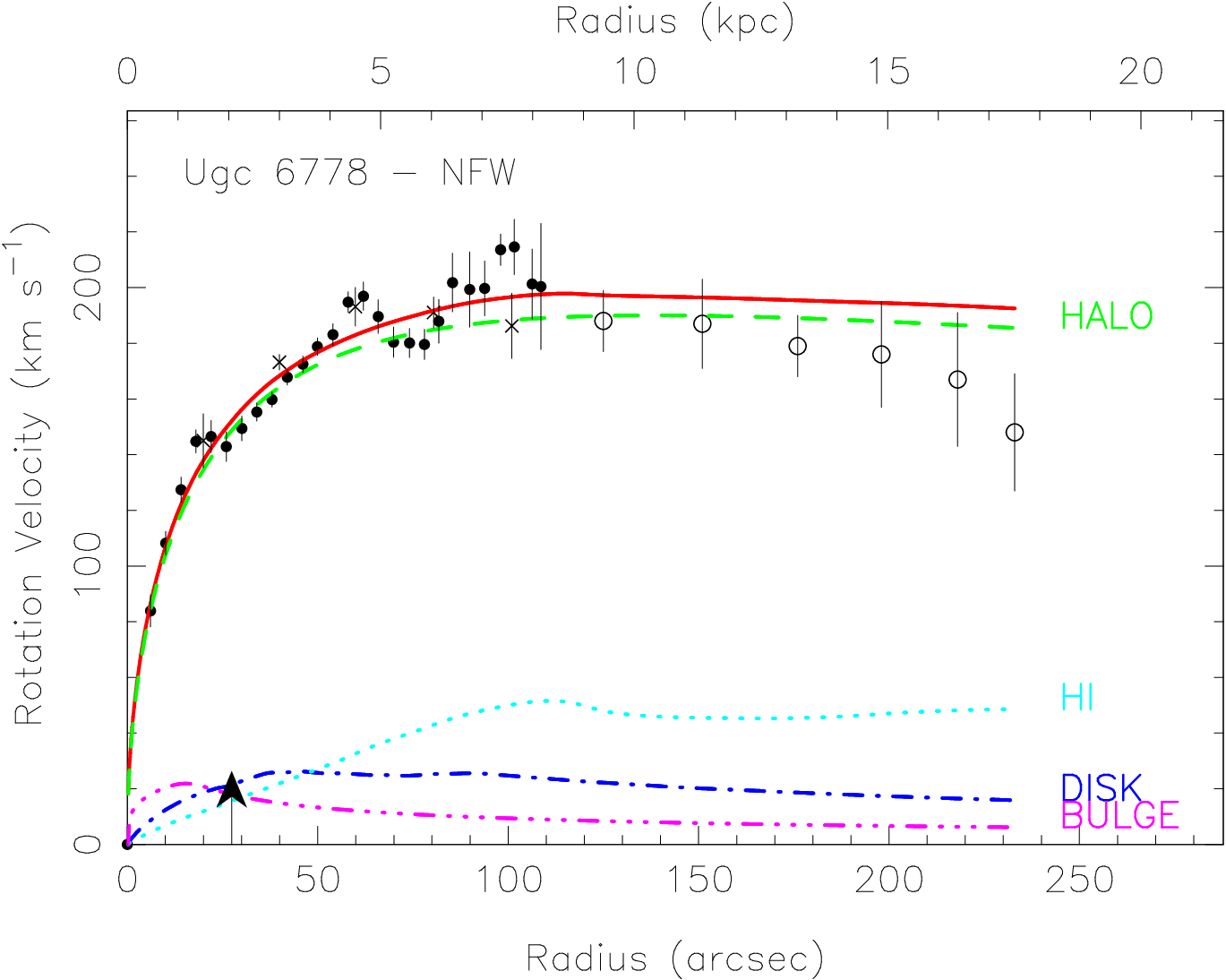} \\
\includegraphics[width=5.5cm,angle=-90]{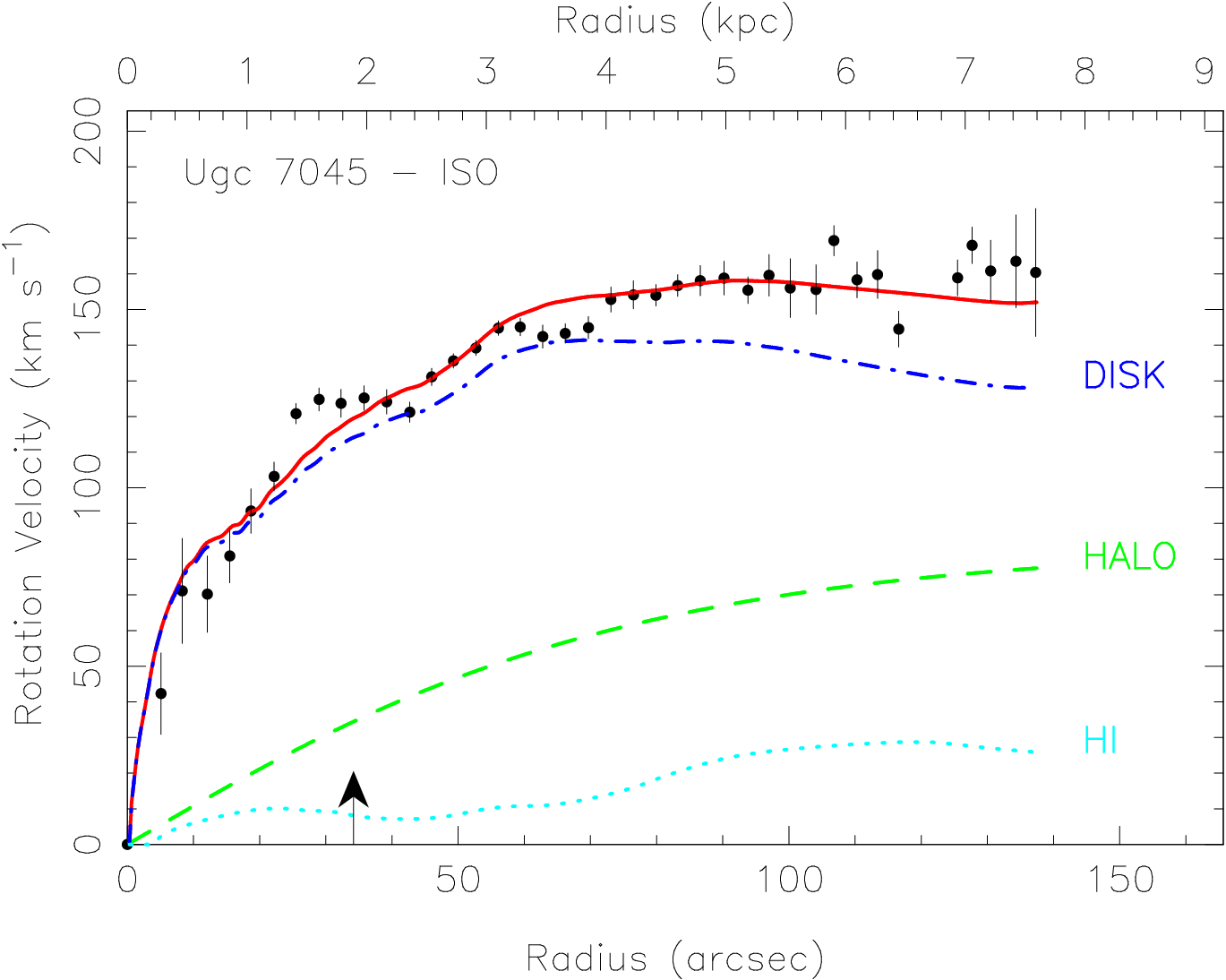}   \includegraphics[width=5.5cm,angle=-90]{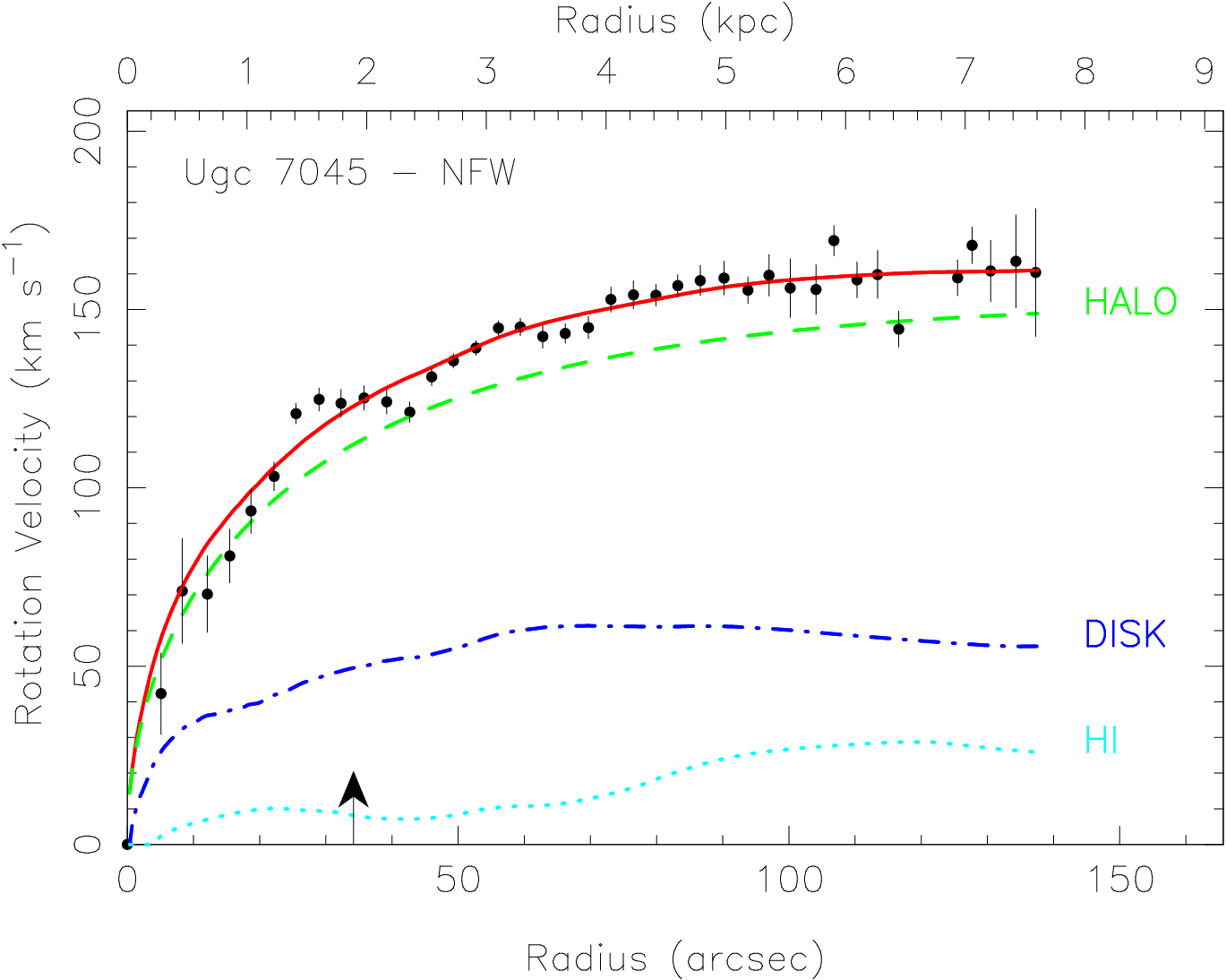}  \\
\includegraphics[width=5.5cm,angle=-90]{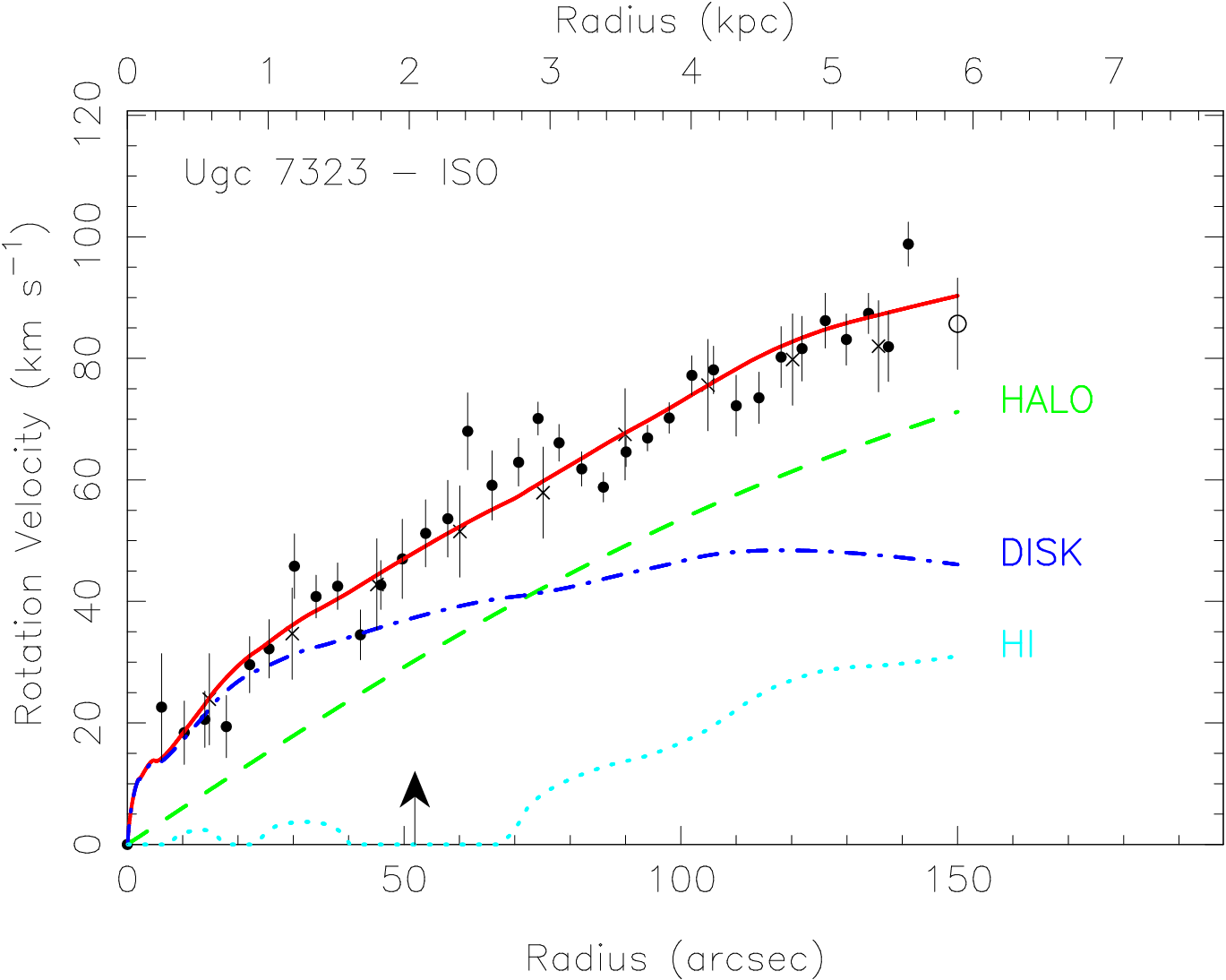}   \includegraphics[width=5.5cm,angle=-90]{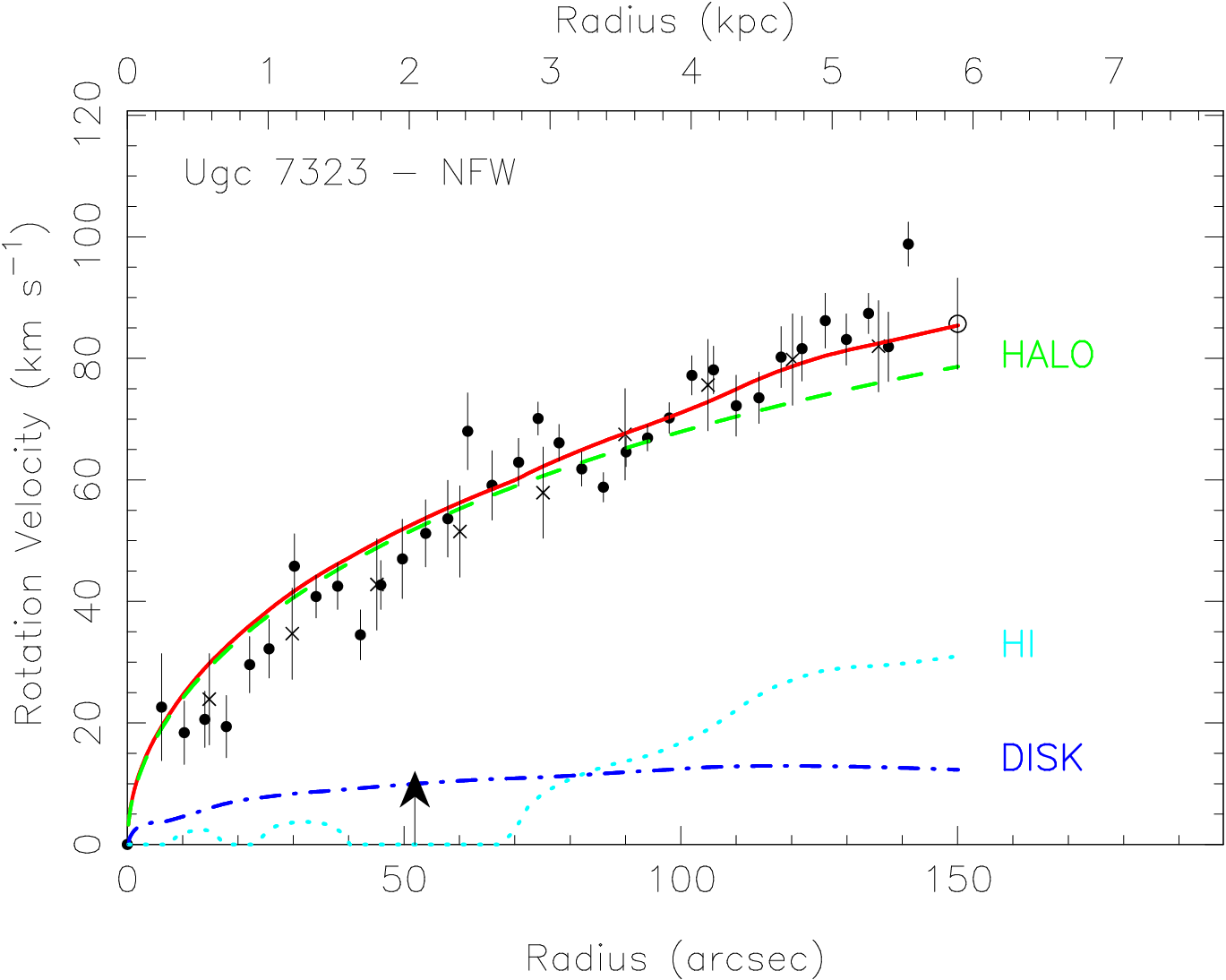}
\end{center}
\caption{Best fit model for the rotation curve of UGC6537,
UGC6778, UGC7045 and UGC7323 with isothermal sphere profile (left)
and NFW profile (right). Dots are for optical velocities, open
circles for HI velocities and crosses for HI in the central part,
not taken into account.The arrow on the X axis indicates the disk scale-length.}
\end{figure*}

\begin{figure*}
\begin{center}
\includegraphics[width=5.5cm,angle=-90]{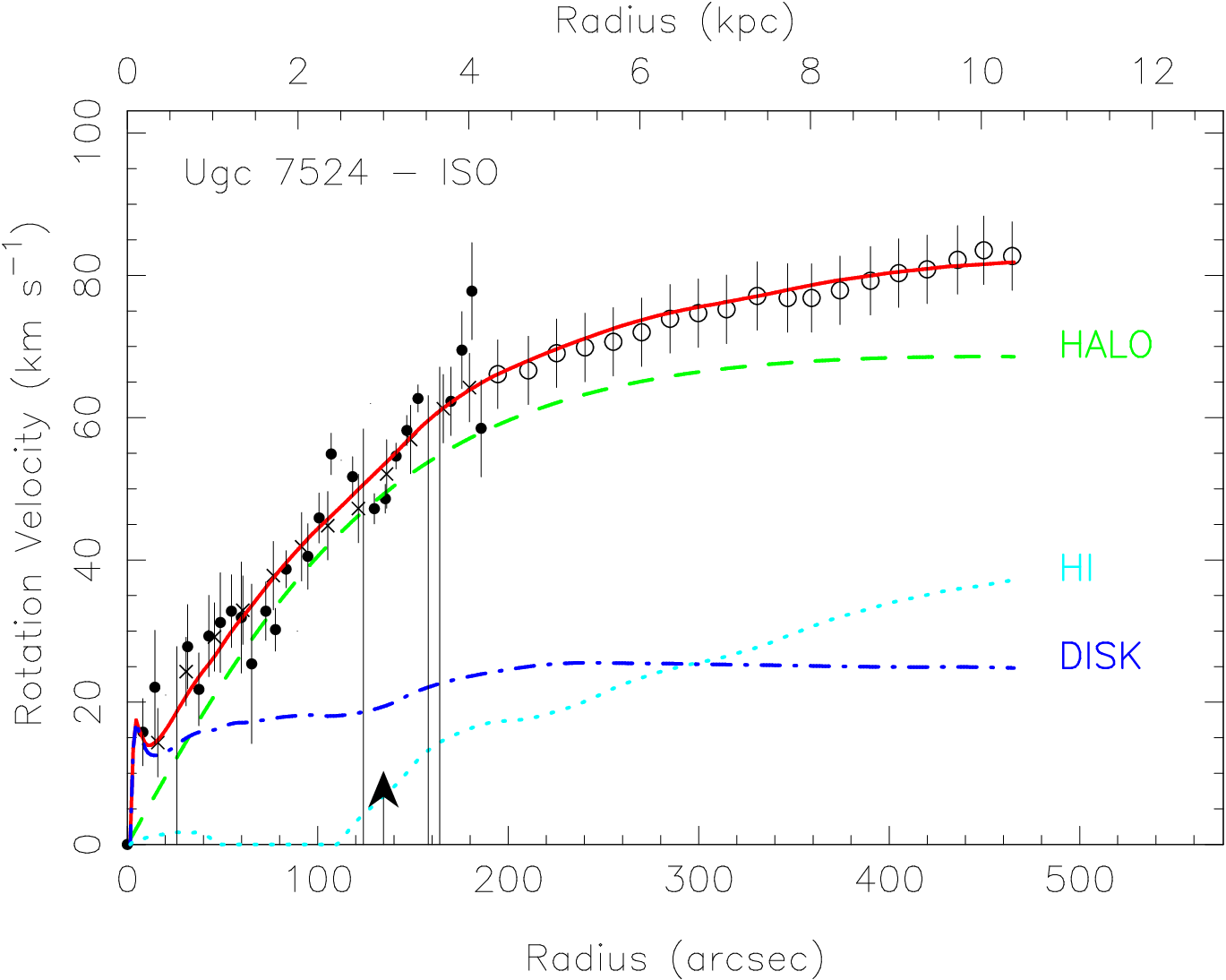}   \includegraphics[width=5.5cm,angle=-90]{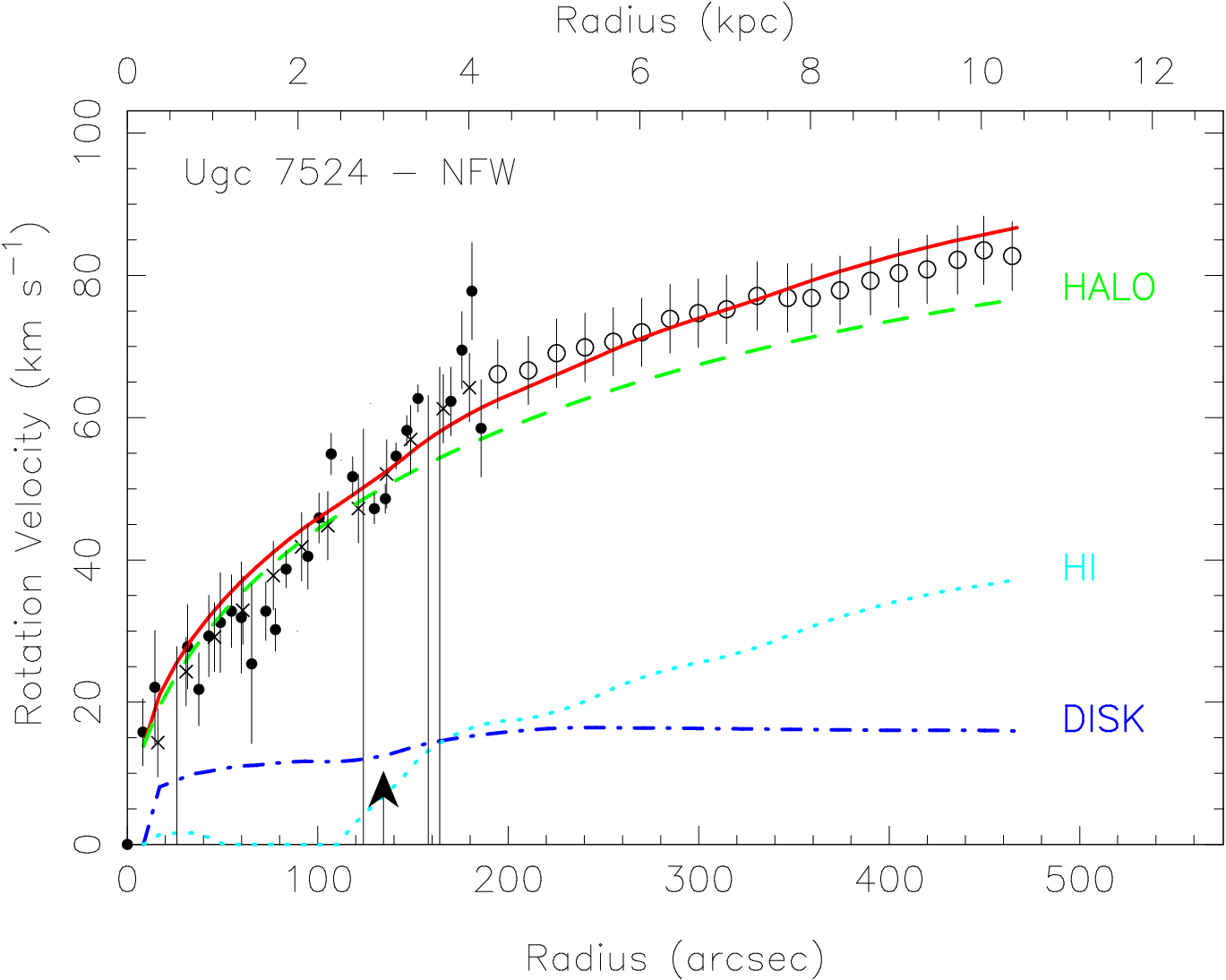}  \\
\includegraphics[width=5.5cm,angle=-90]{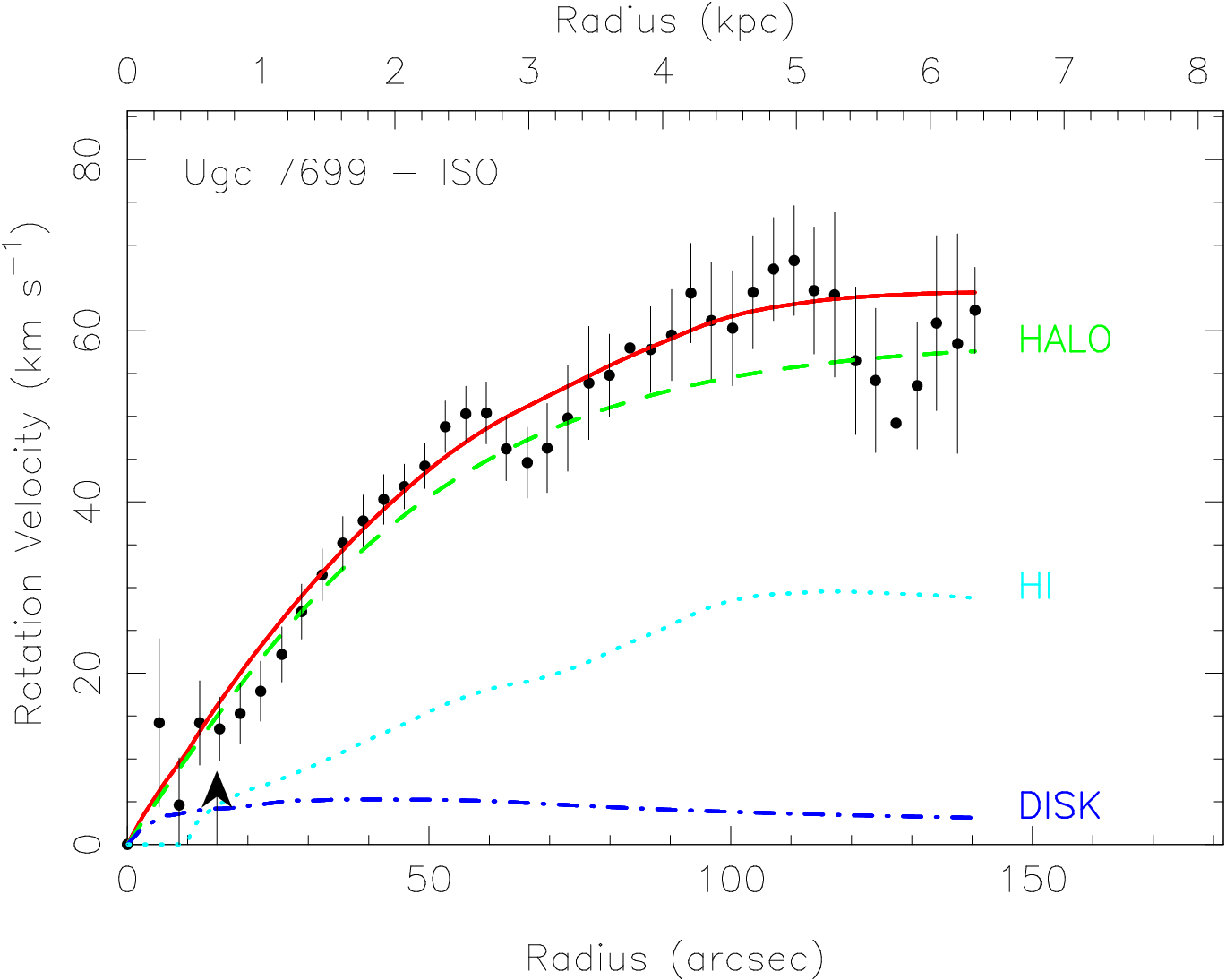}   \includegraphics[width=5.5cm,angle=-90]{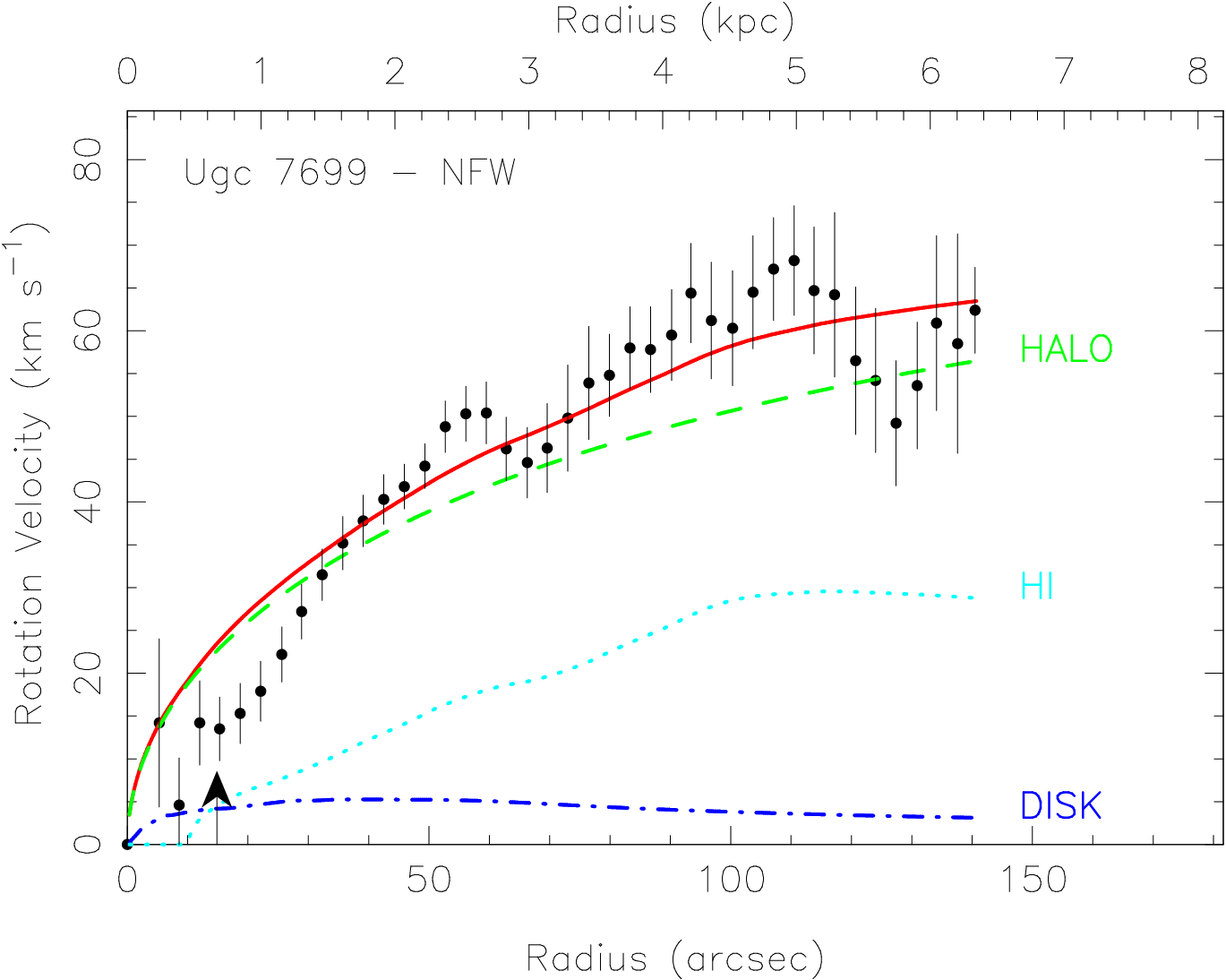} \\
\includegraphics[width=5.5cm,angle=-90]{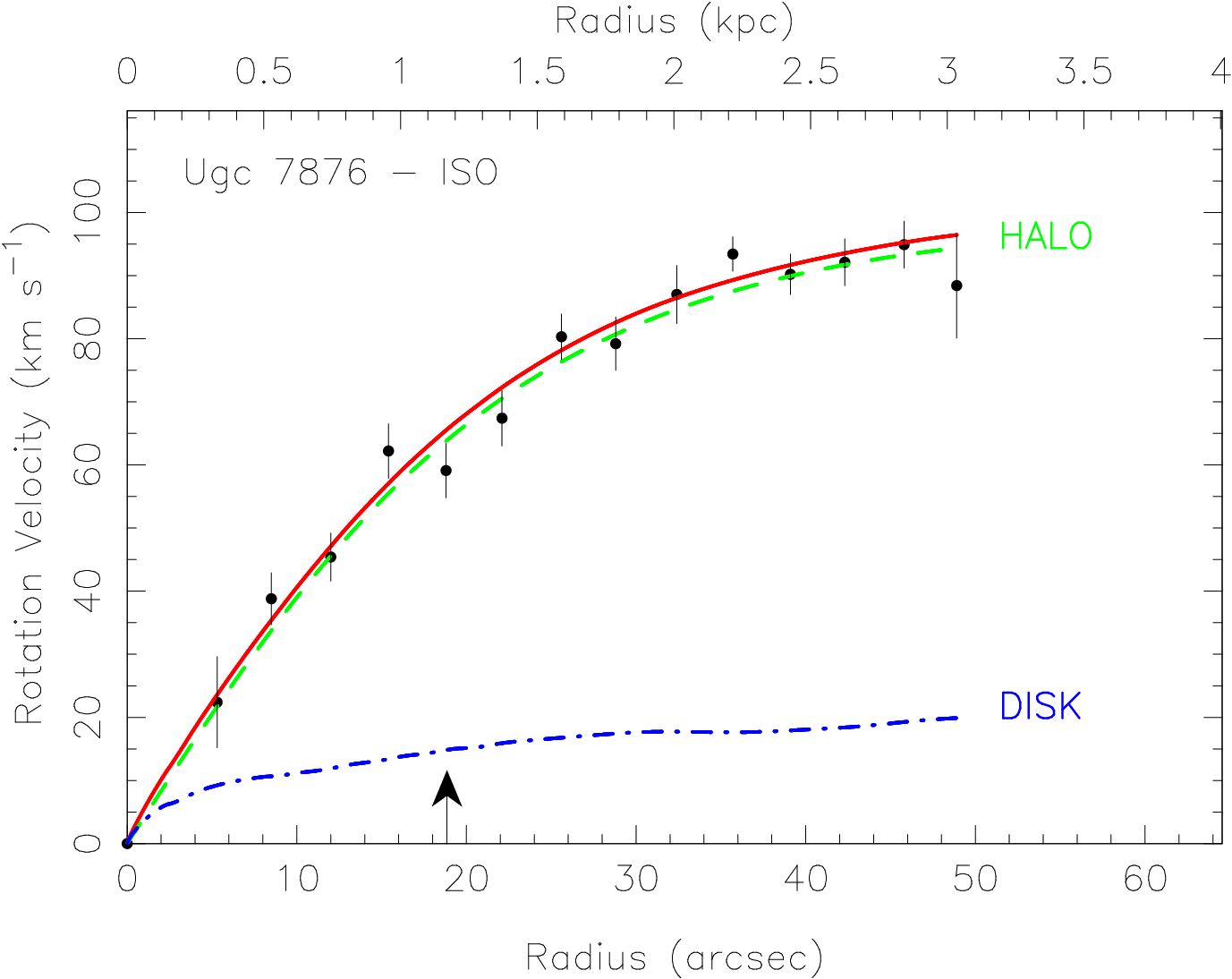}   \includegraphics[width=5.5cm,angle=-90]{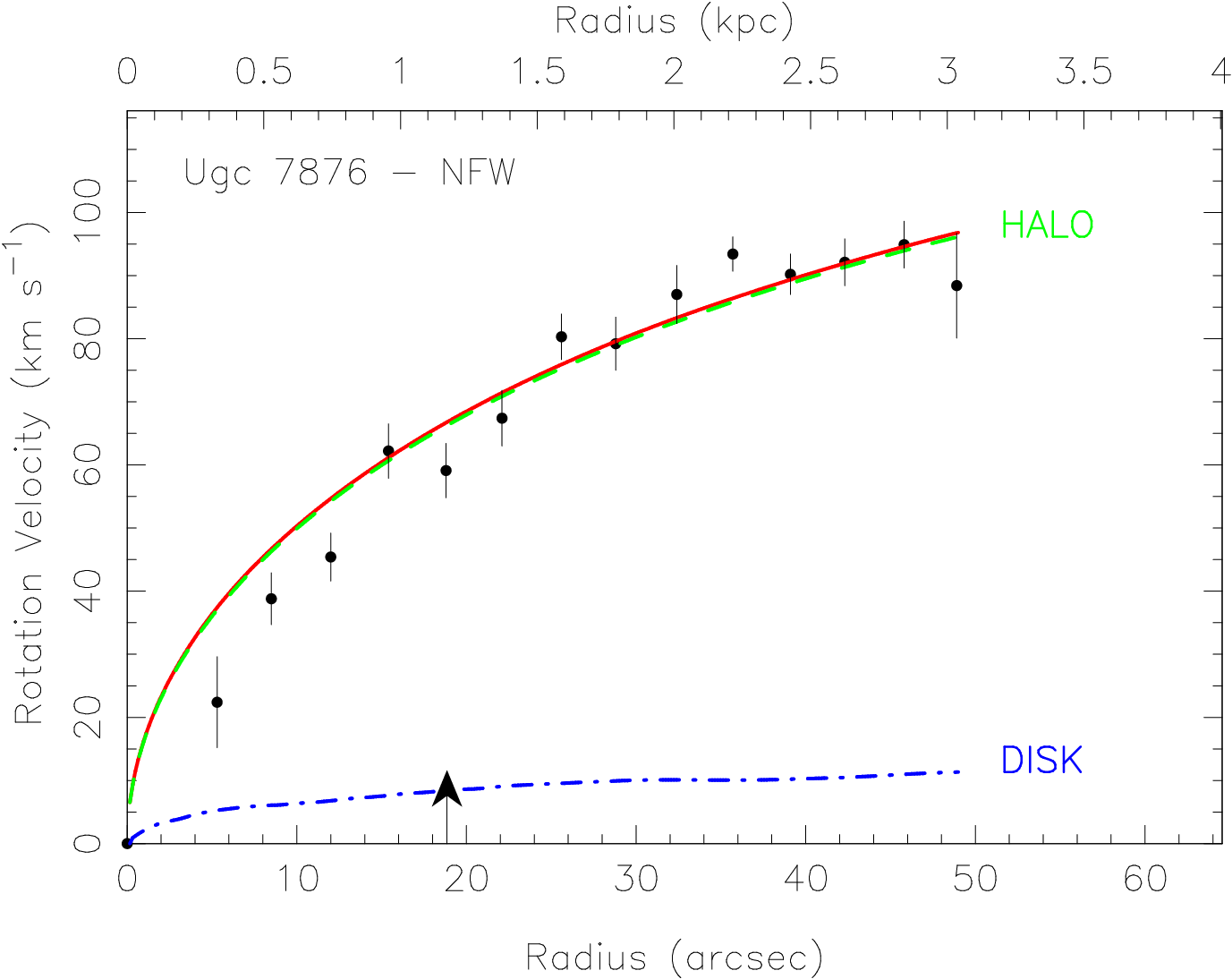}  \\
\includegraphics[width=5.5cm,angle=-90]{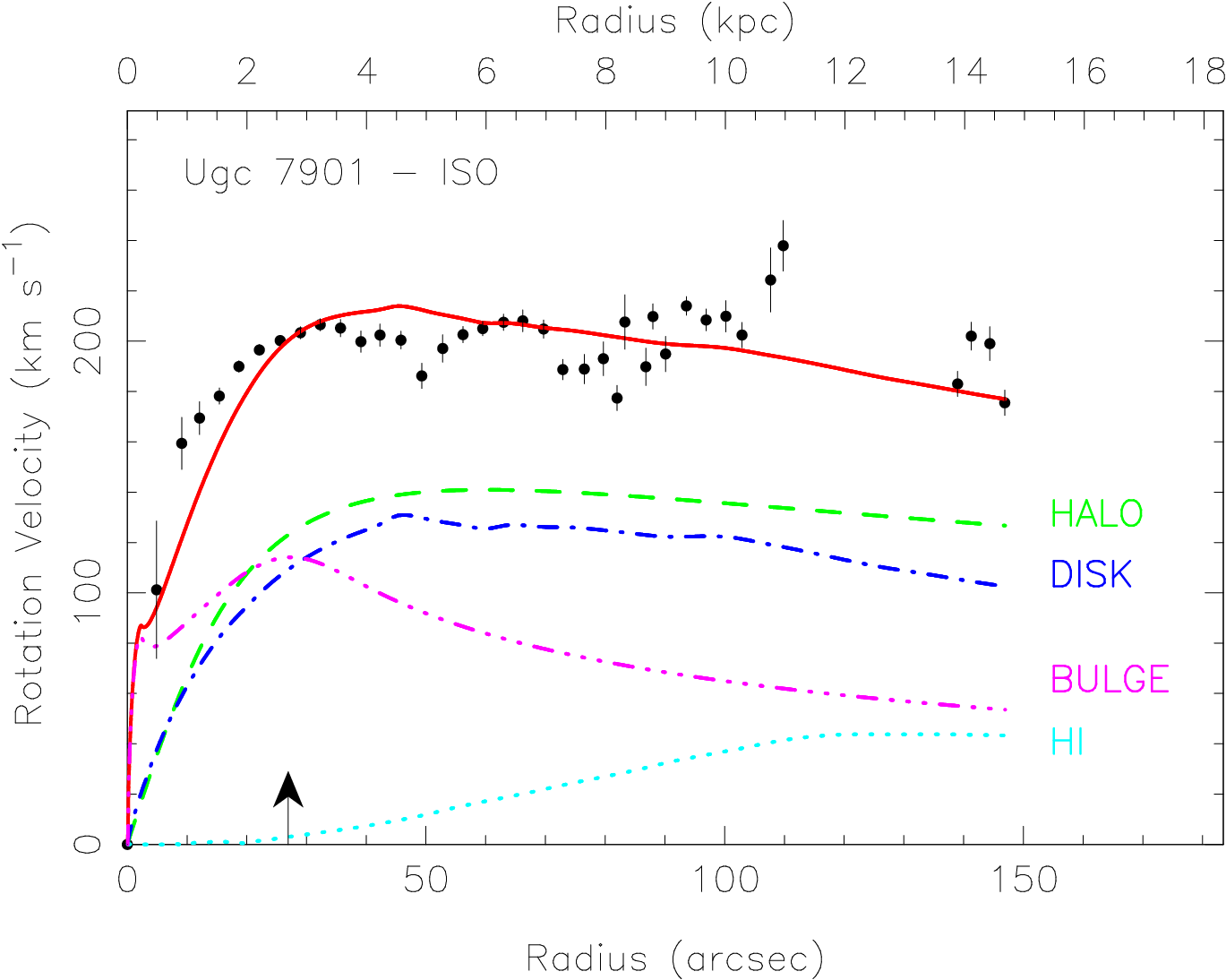}   \includegraphics[width=5.5cm,angle=-90]{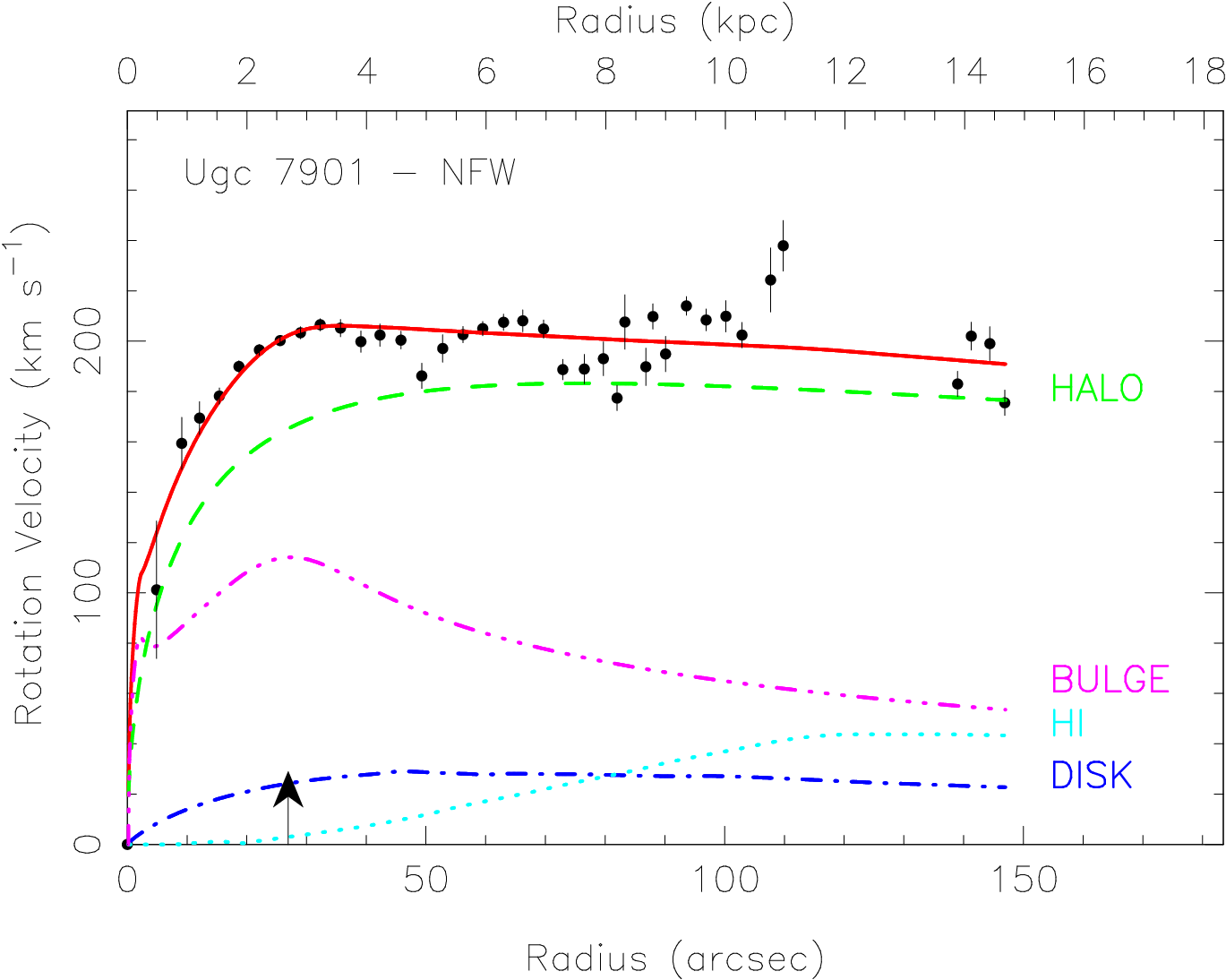}
\end{center}
\caption{Best fit model for the rotation curve of UGC7524,
UGC7699, UGC7876 and UGC7901 with isothermal sphere profile (left)
and NFW profile (right). Dots are for optical velocities, open
circles for HI velocities and crosses for HI in the central part,
not taken into account.The arrow on the X axis indicates the disk scale-length.}
\end{figure*}

\clearpage

\begin{figure*}
\begin{center}
\includegraphics[width=5.5cm,angle=-90]{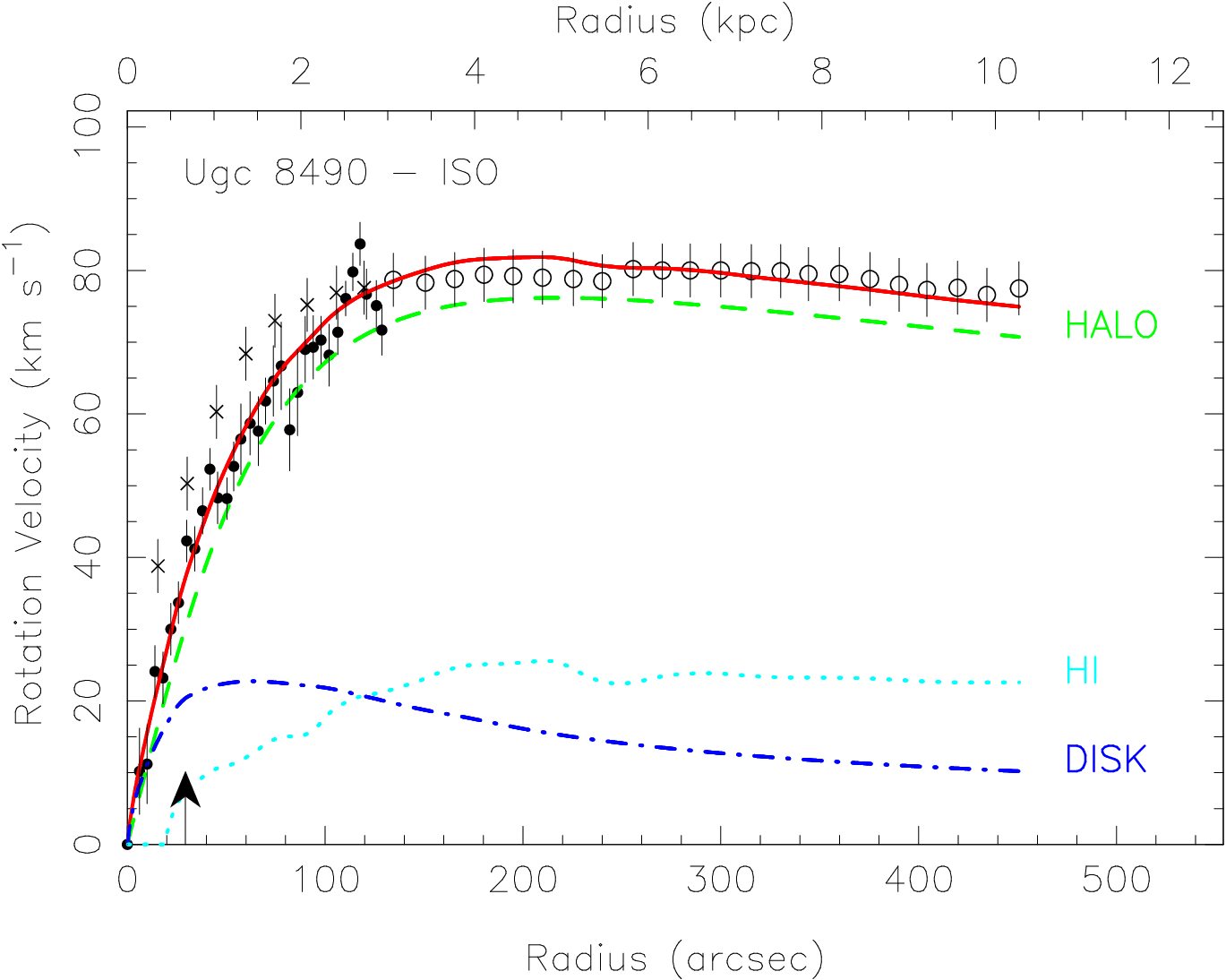}   \includegraphics[width=5.5cm,angle=-90]{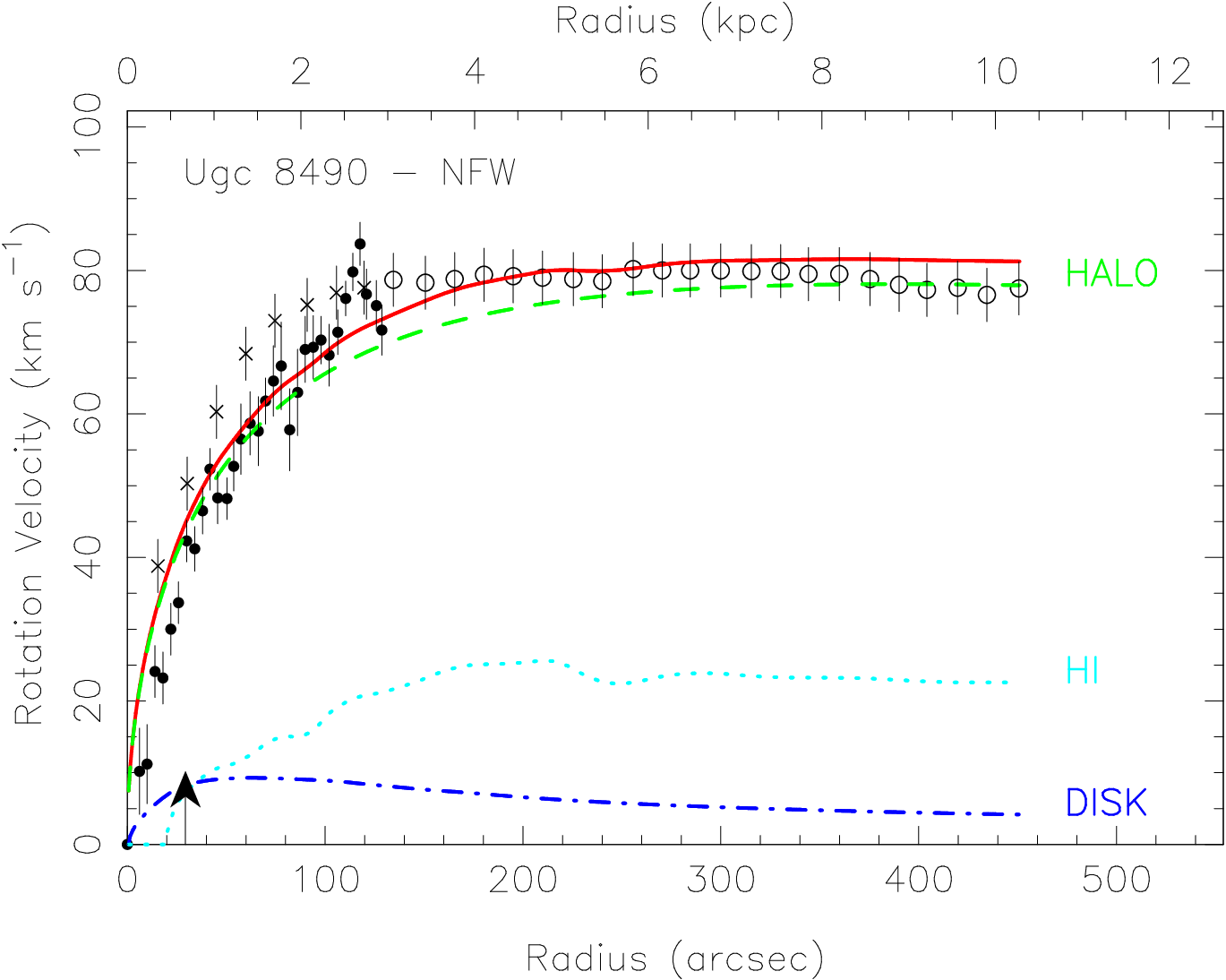}  \\
\includegraphics[width=5.5cm,angle=-90]{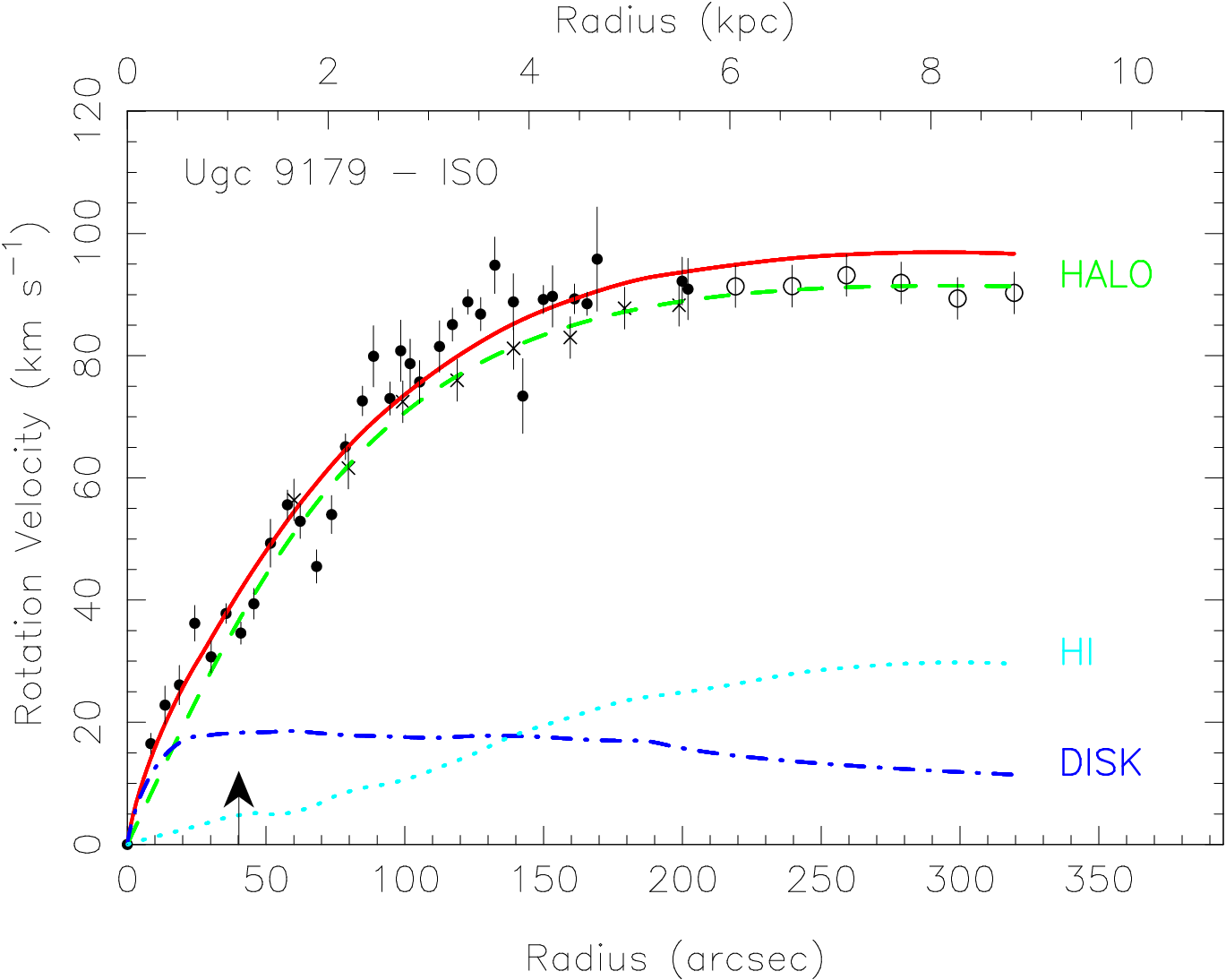}   \includegraphics[width=5.5cm,angle=-90]{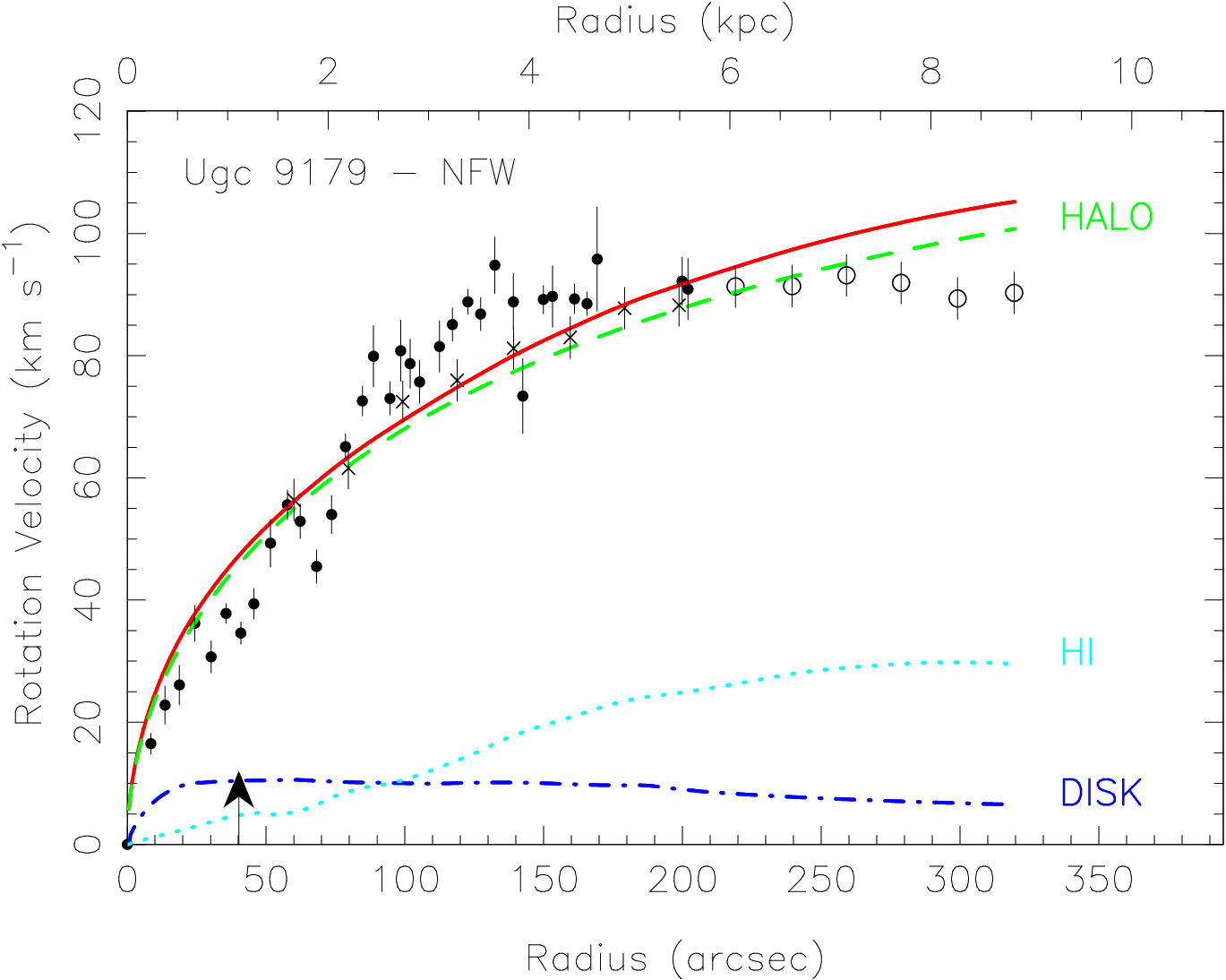} \\
\includegraphics[width=5.5cm,angle=-90]{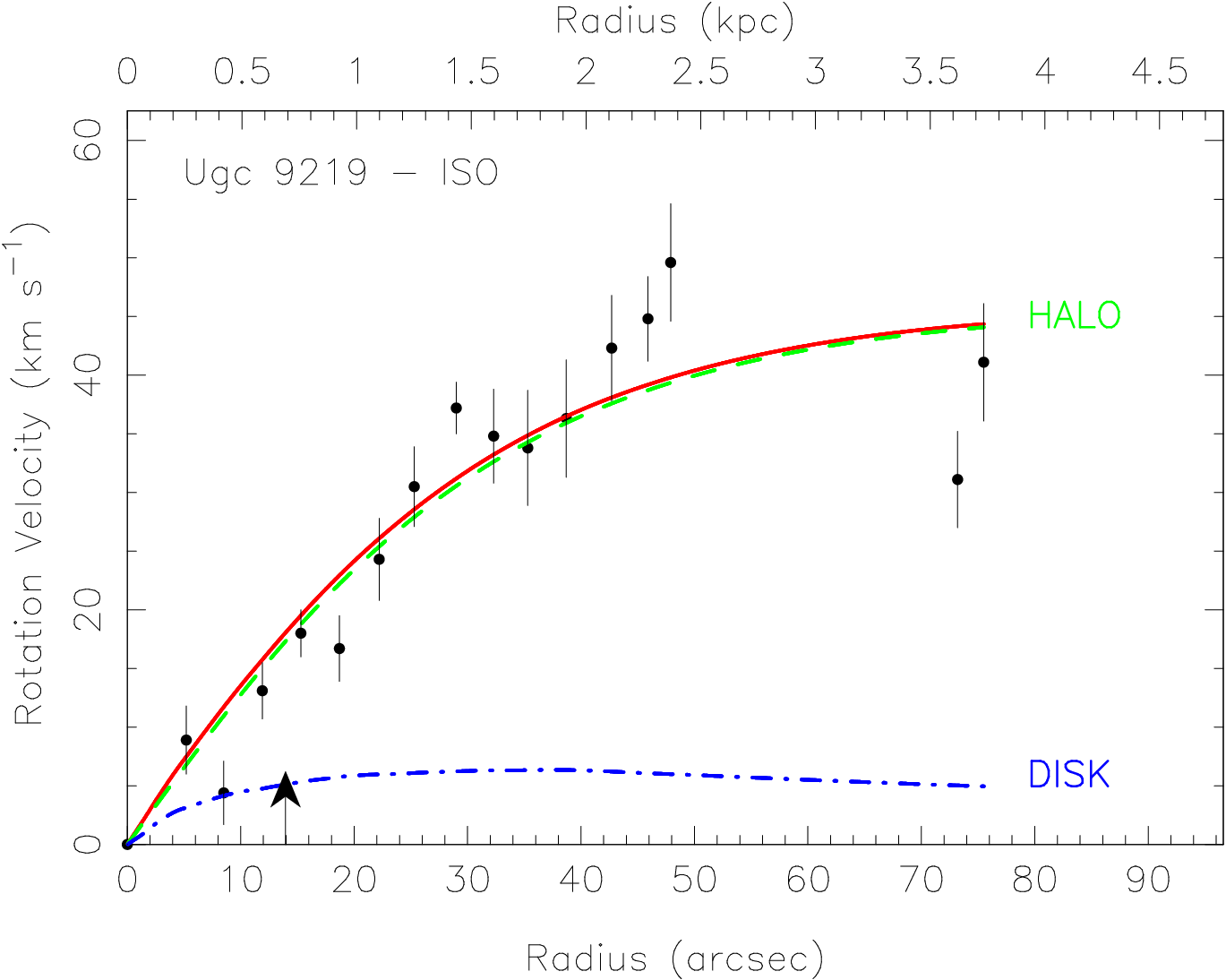}   \includegraphics[width=5.5cm,angle=-90]{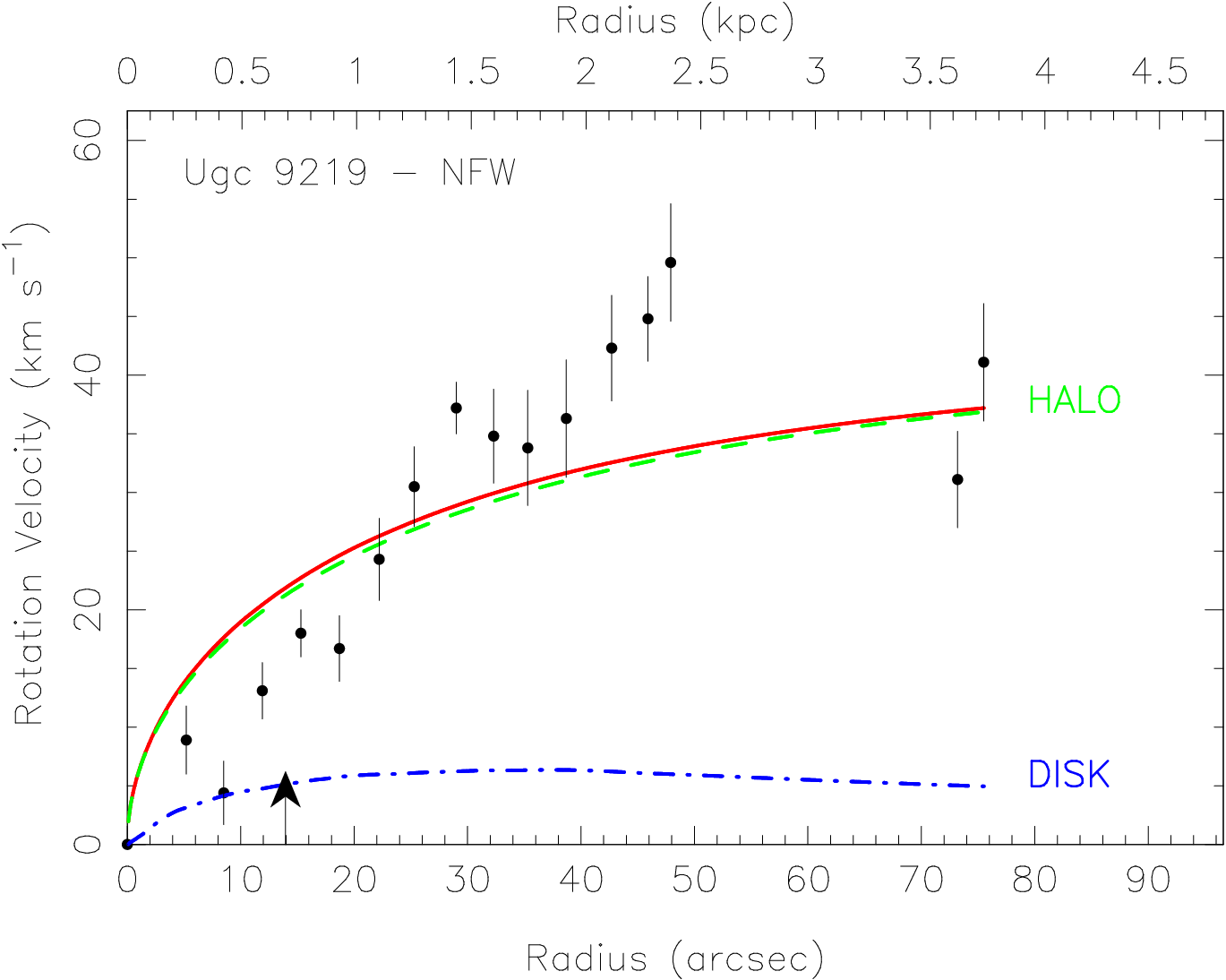}  \\
 \includegraphics[width=5.5cm,angle=-90]{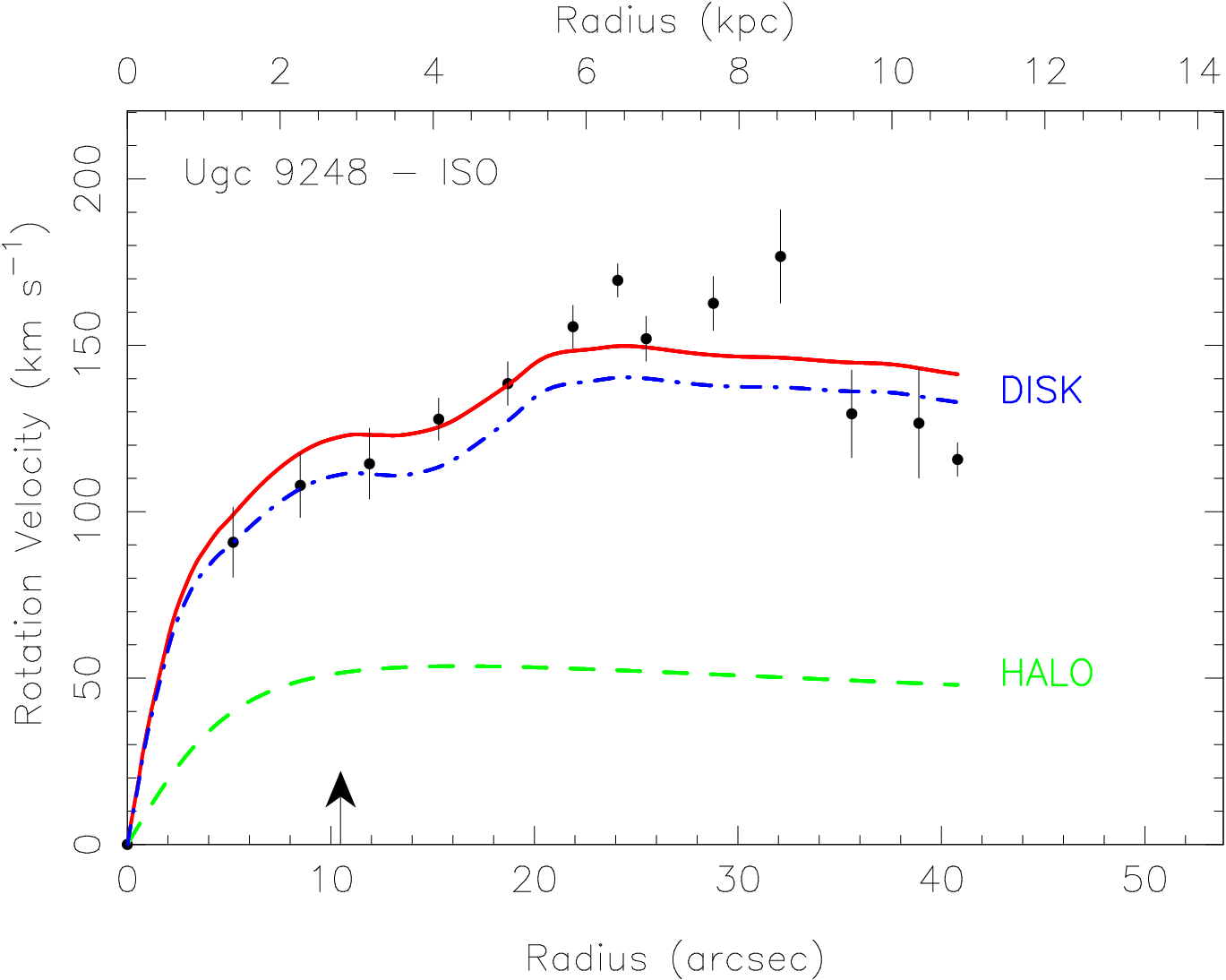}   \includegraphics[width=5.5cm,angle=-90]{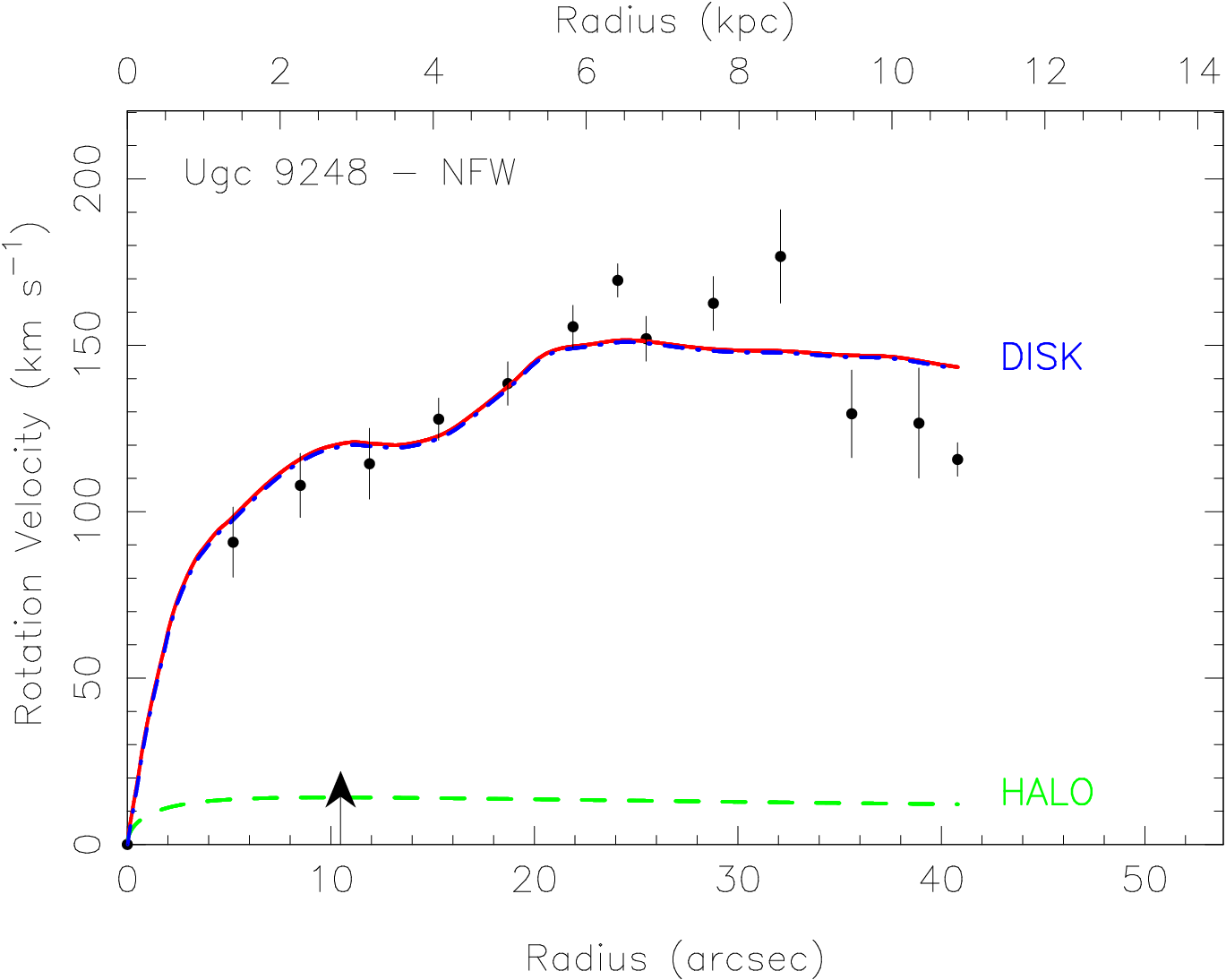}
\end{center}
\caption{Best fit model for the rotation curve of UGC8490,
UGC9179, UGC9219 and UGC9248 with isothermal sphere profile (left)
and NFW profile (right). Dots are for optical velocities, open
circles for HI velocities and crosses for HI in the central part,
not taken into account.The arrow on the X axis indicates the disk scale-length.}
\end{figure*}

\begin{figure*}
\begin{center}
\includegraphics[width=5.5cm,angle=-90]{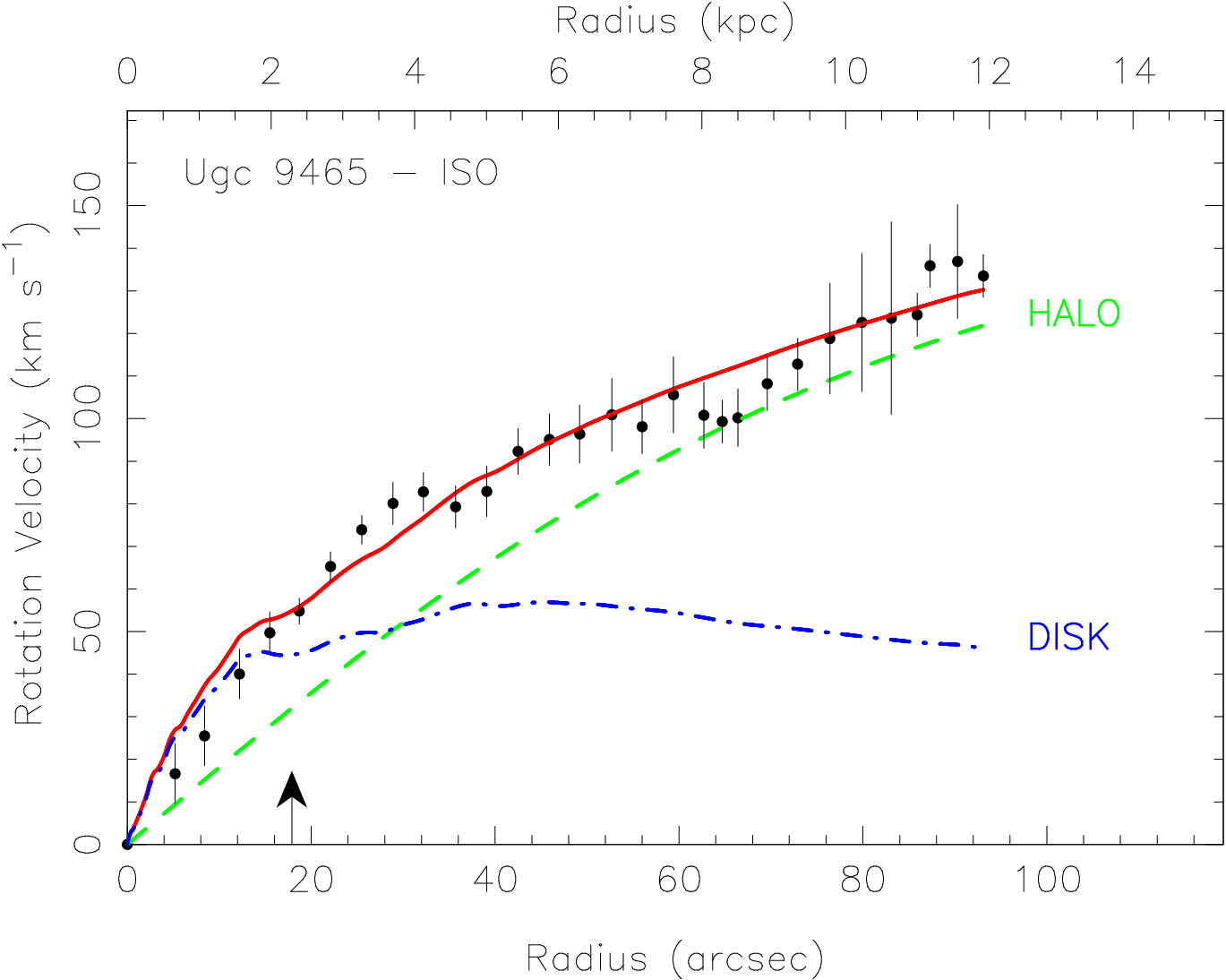}   \includegraphics[width=5.5cm,angle=-90]{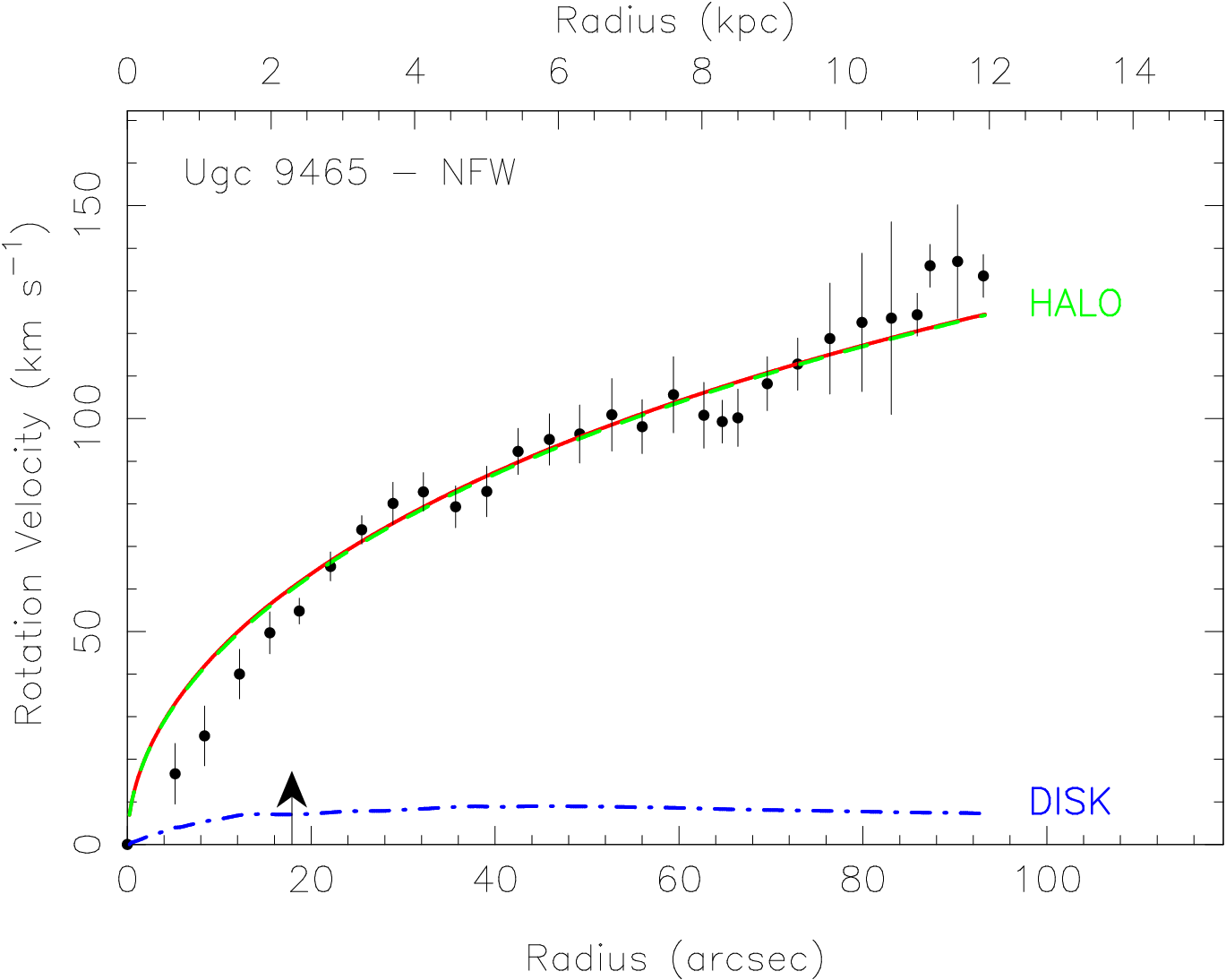}  \\
\includegraphics[width=5.5cm,angle=-90]{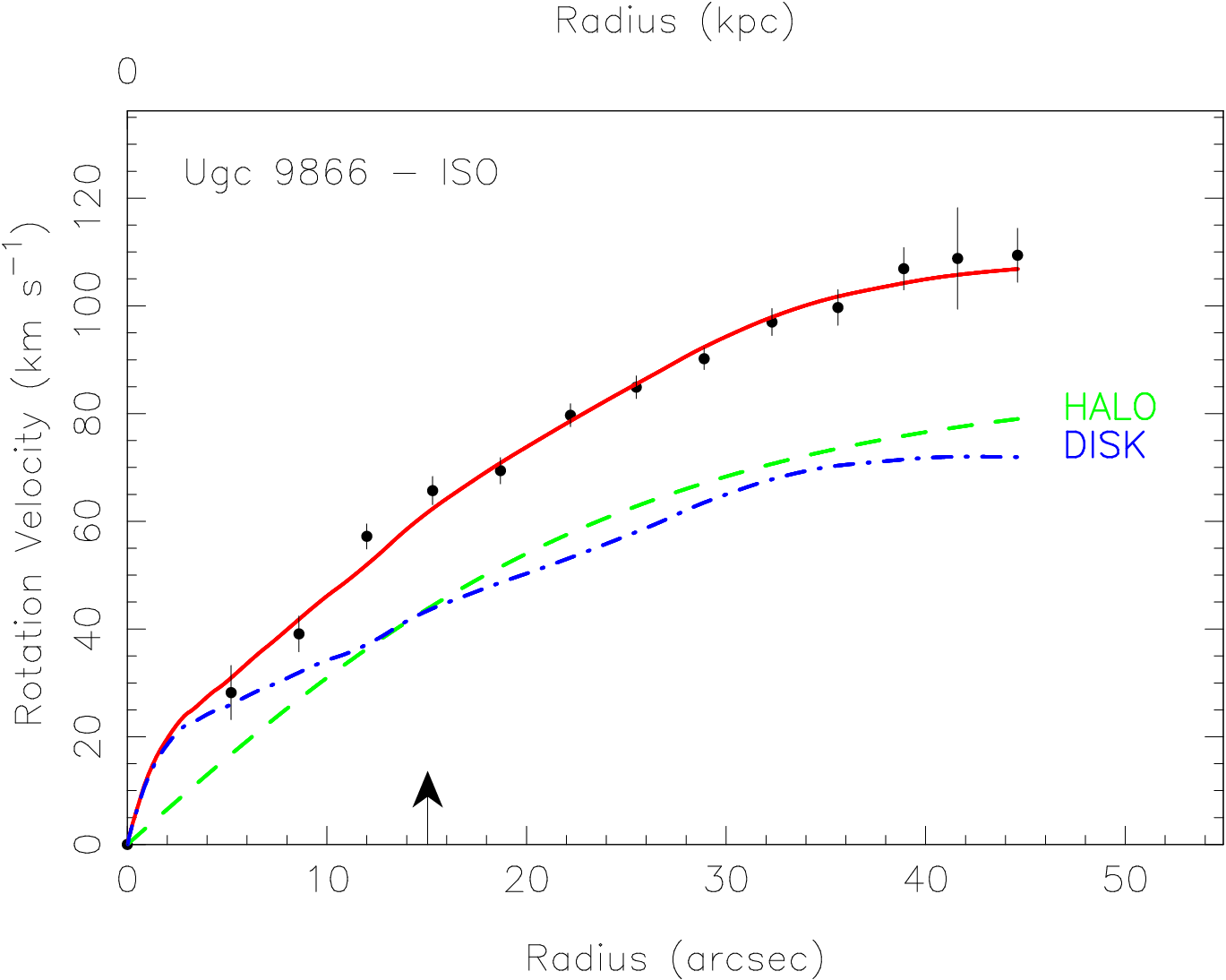}   \includegraphics[width=5.5cm,angle=-90]{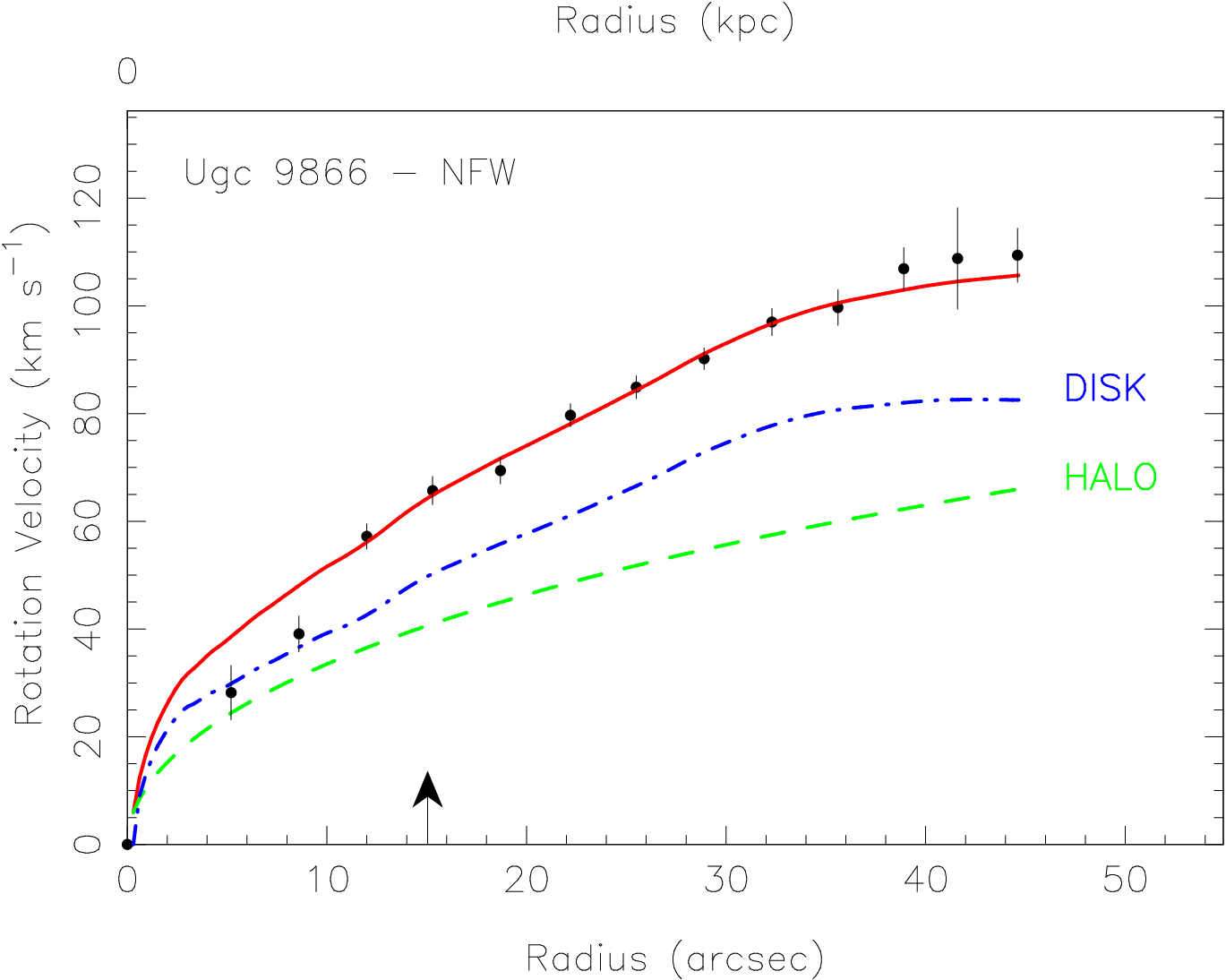} \\
\includegraphics[width=5.5cm,angle=-90]{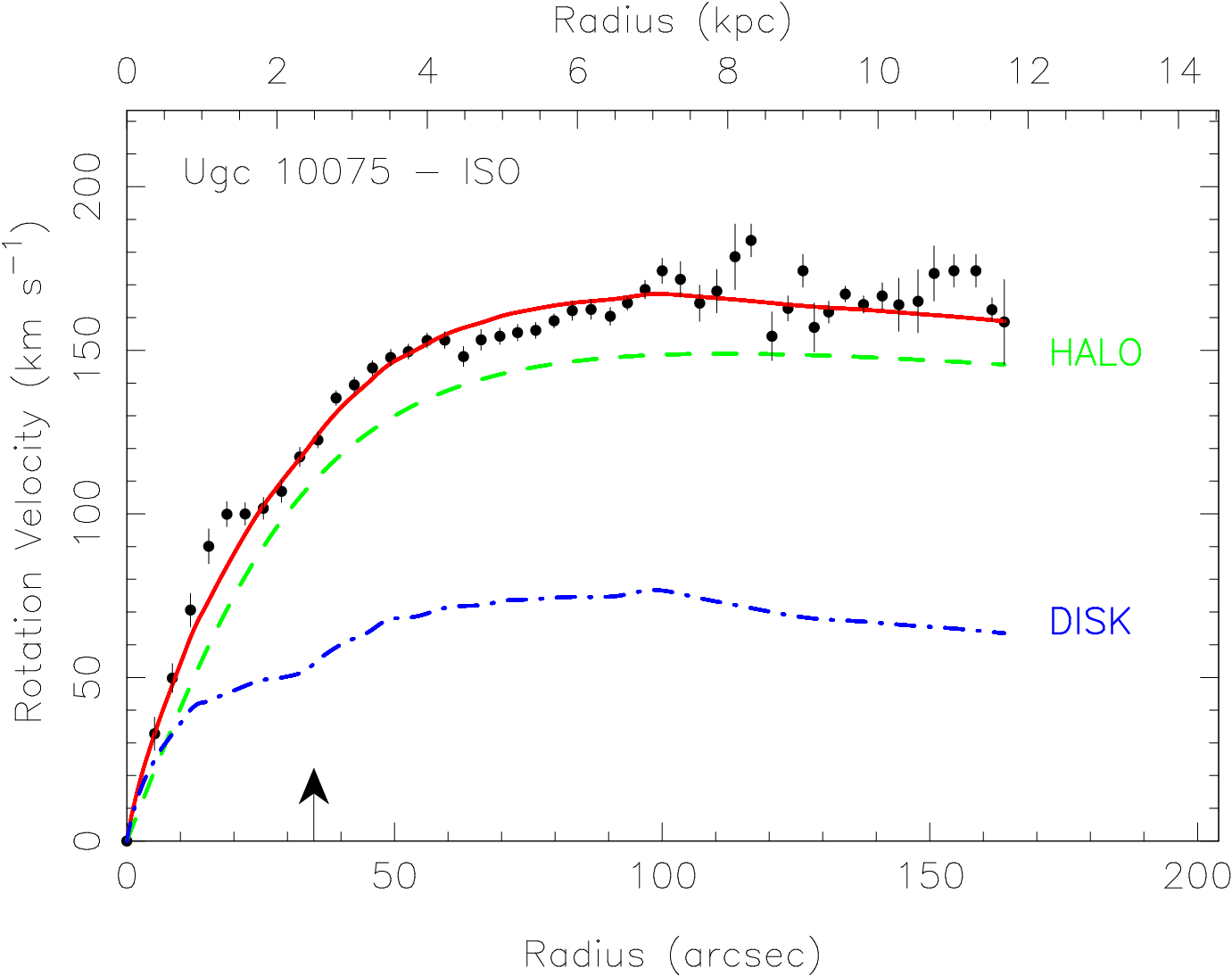}   \includegraphics[width=5.5cm,angle=-90]{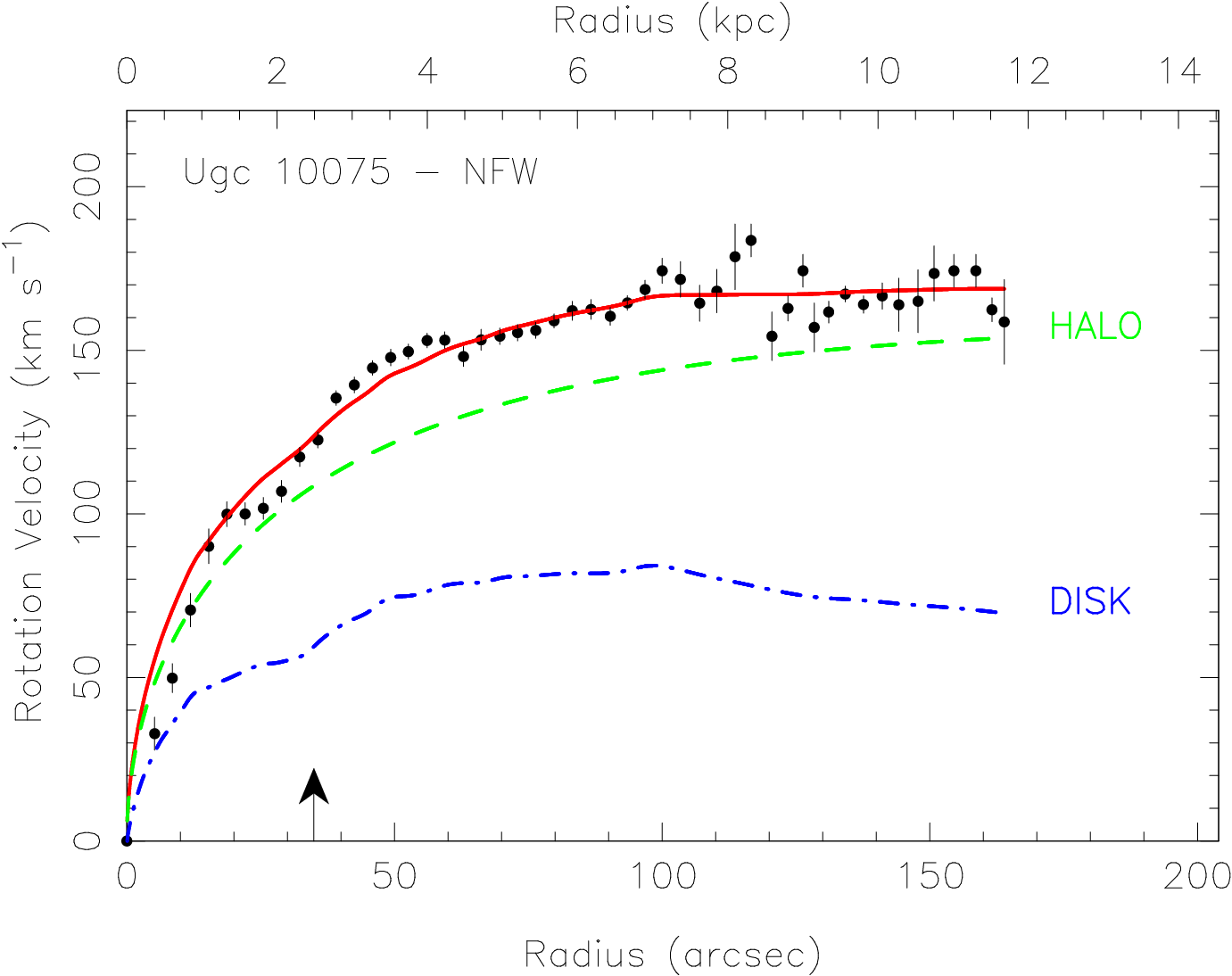}  \\
\includegraphics[width=5.5cm,angle=-90]{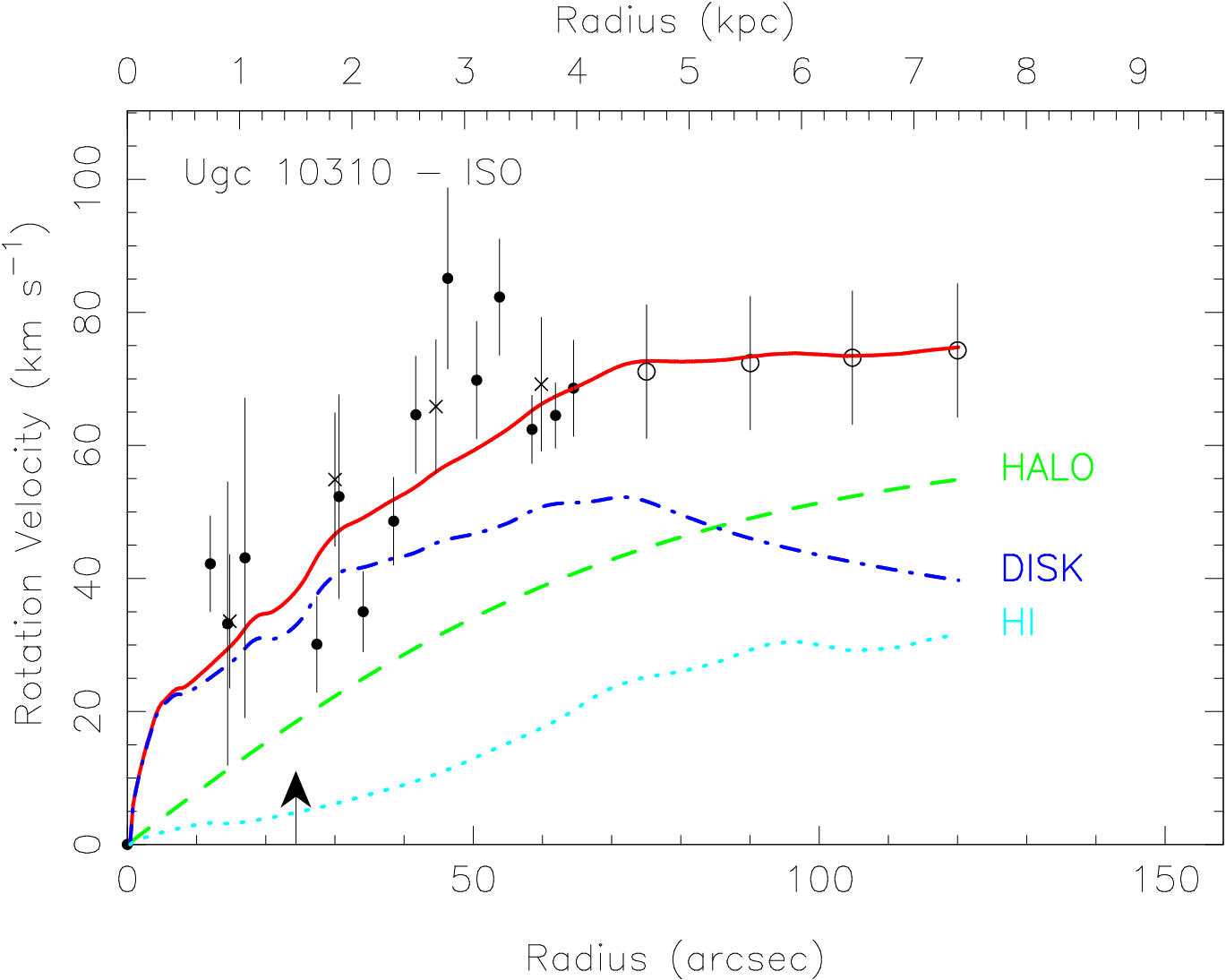}   \includegraphics[width=5.5cm,angle=-90]{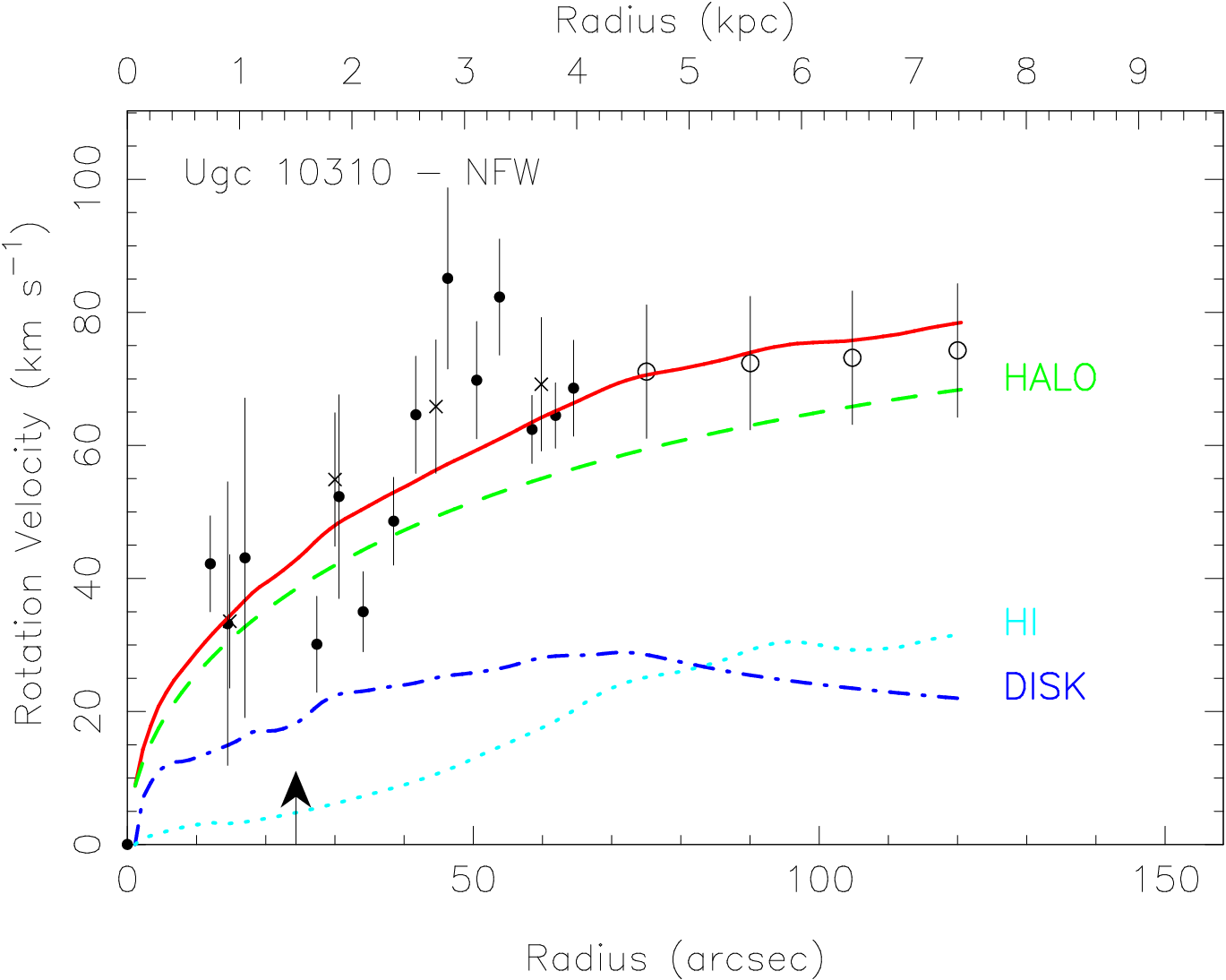}
\end{center}
\caption{Best fit model for the rotation curve of UGC9465,
UGC9866, UGC10075 and UGC10310 with isothermal sphere profile
(left) and NFW profile (right). Dots are for optical velocities,
open circles for HI velocities and crosses for HI in the central
part, not taken into account.The arrow on the X axis indicates the disk scale-length.}
\end{figure*}

\begin{figure*}
\begin{center}
\includegraphics[width=5.5cm,angle=-90]{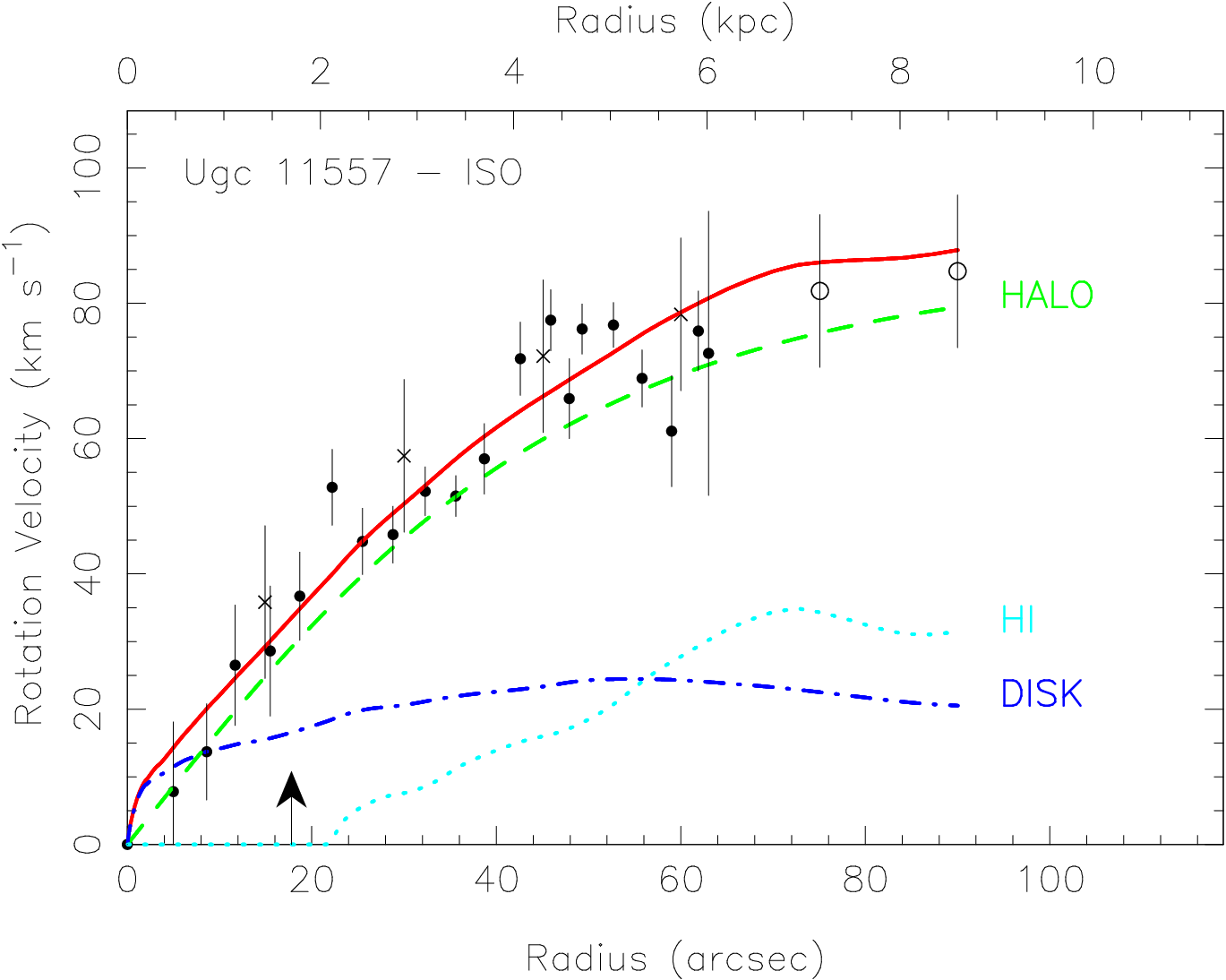}   \includegraphics[width=5.5cm,angle=-90]{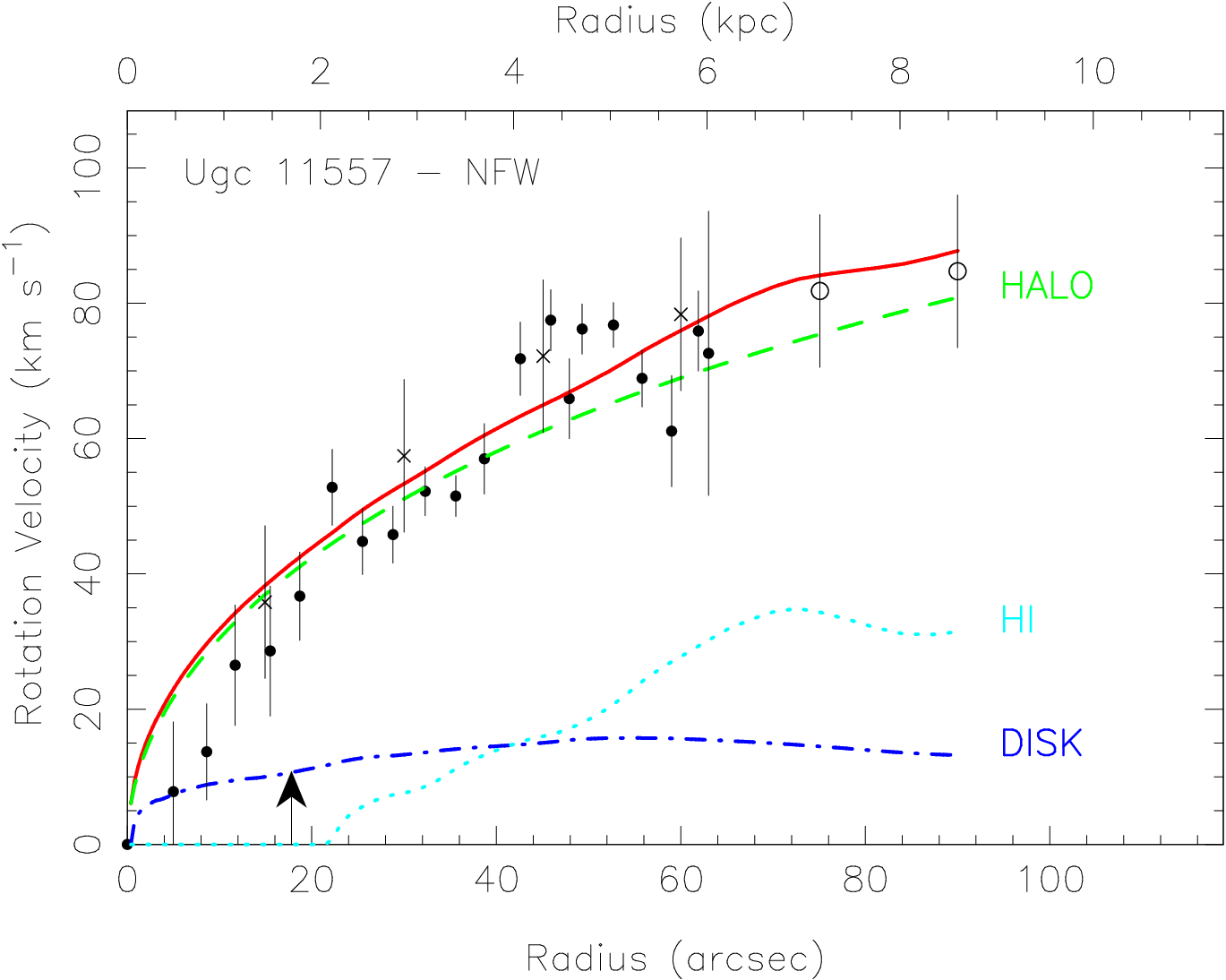}  \\
 \includegraphics[width=5.5cm,angle=-90]{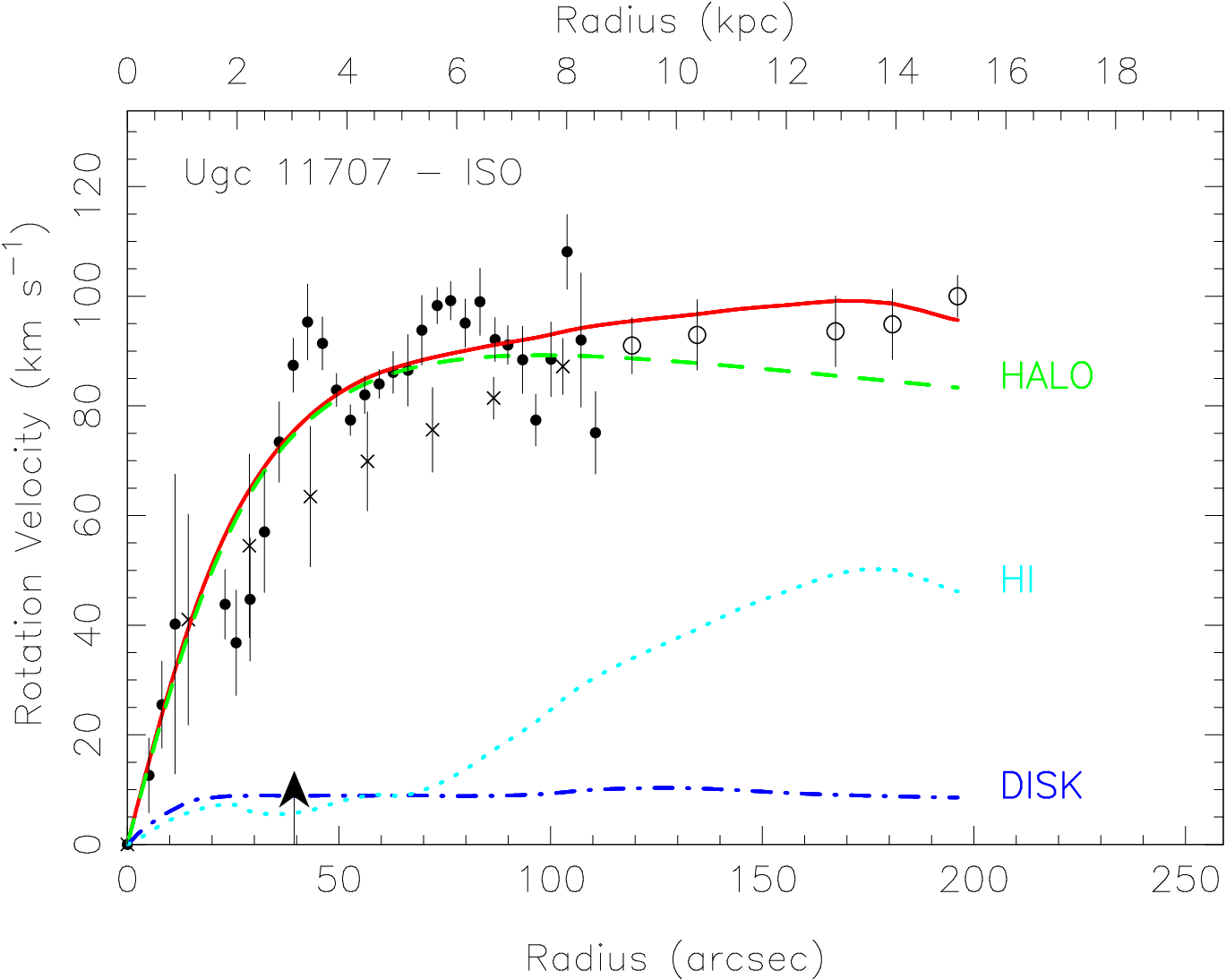}   \includegraphics[width=5.5cm,angle=-90]{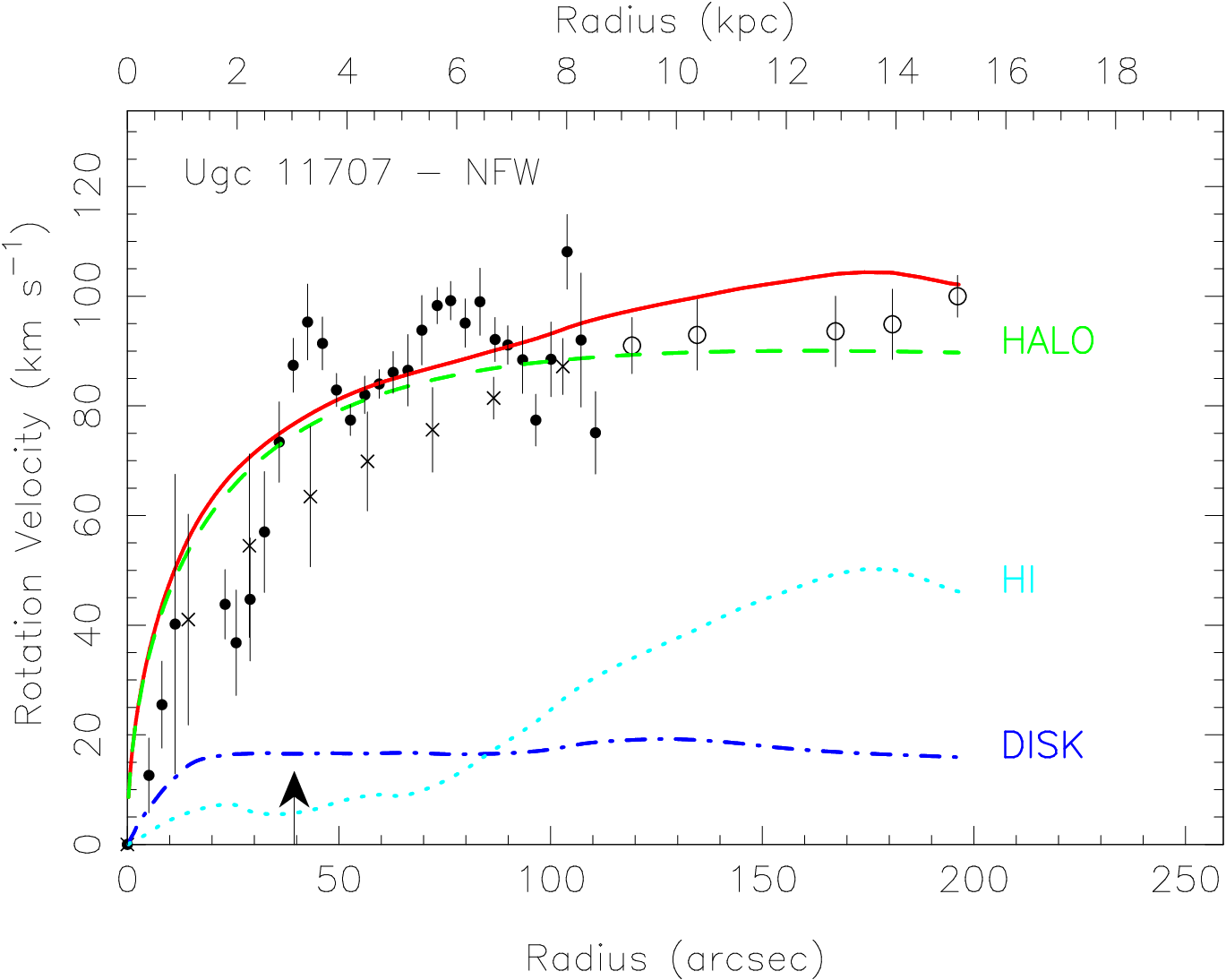} \\
\includegraphics[width=5.5cm,angle=-90]{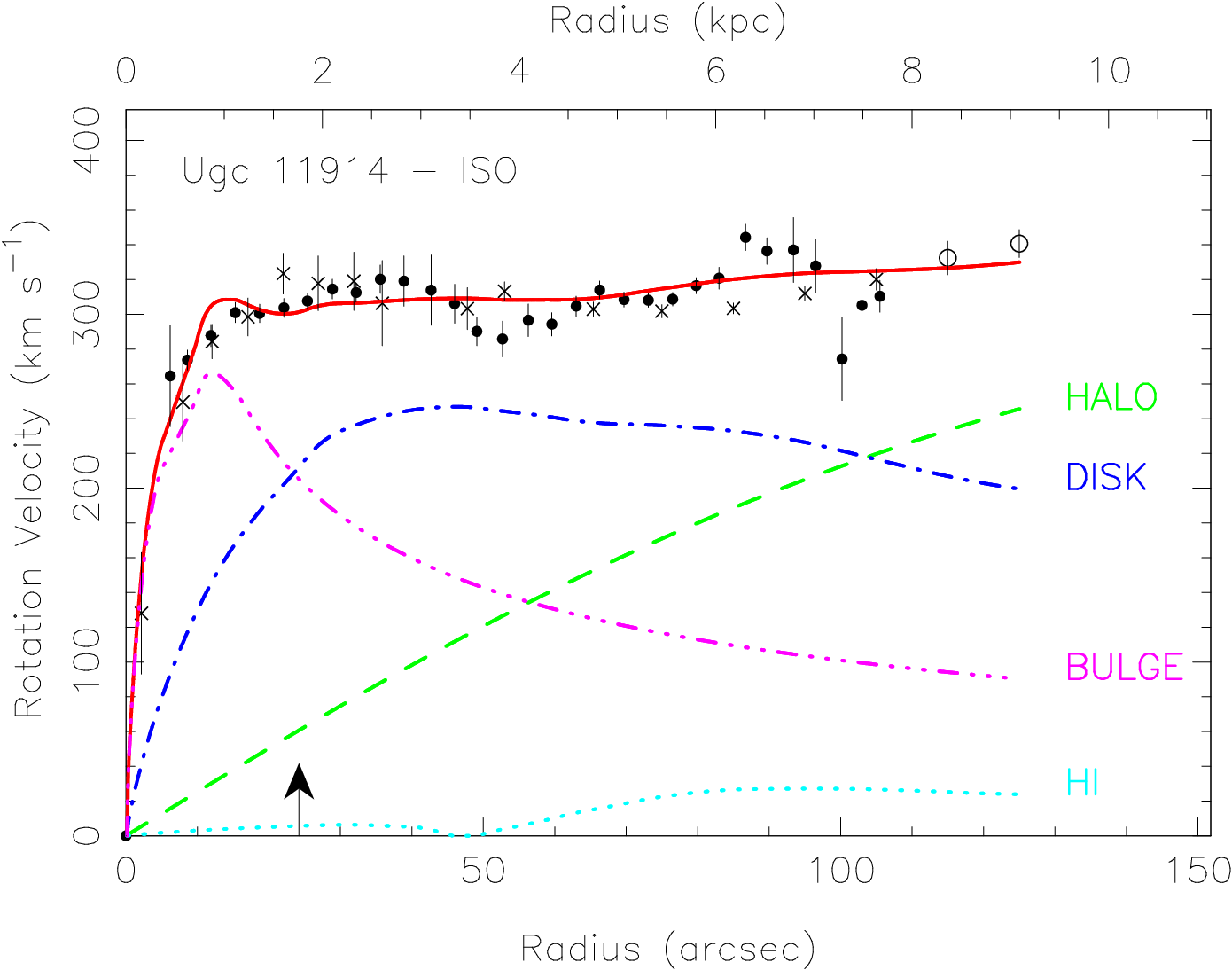}   \includegraphics[width=5.5cm,angle=-90]{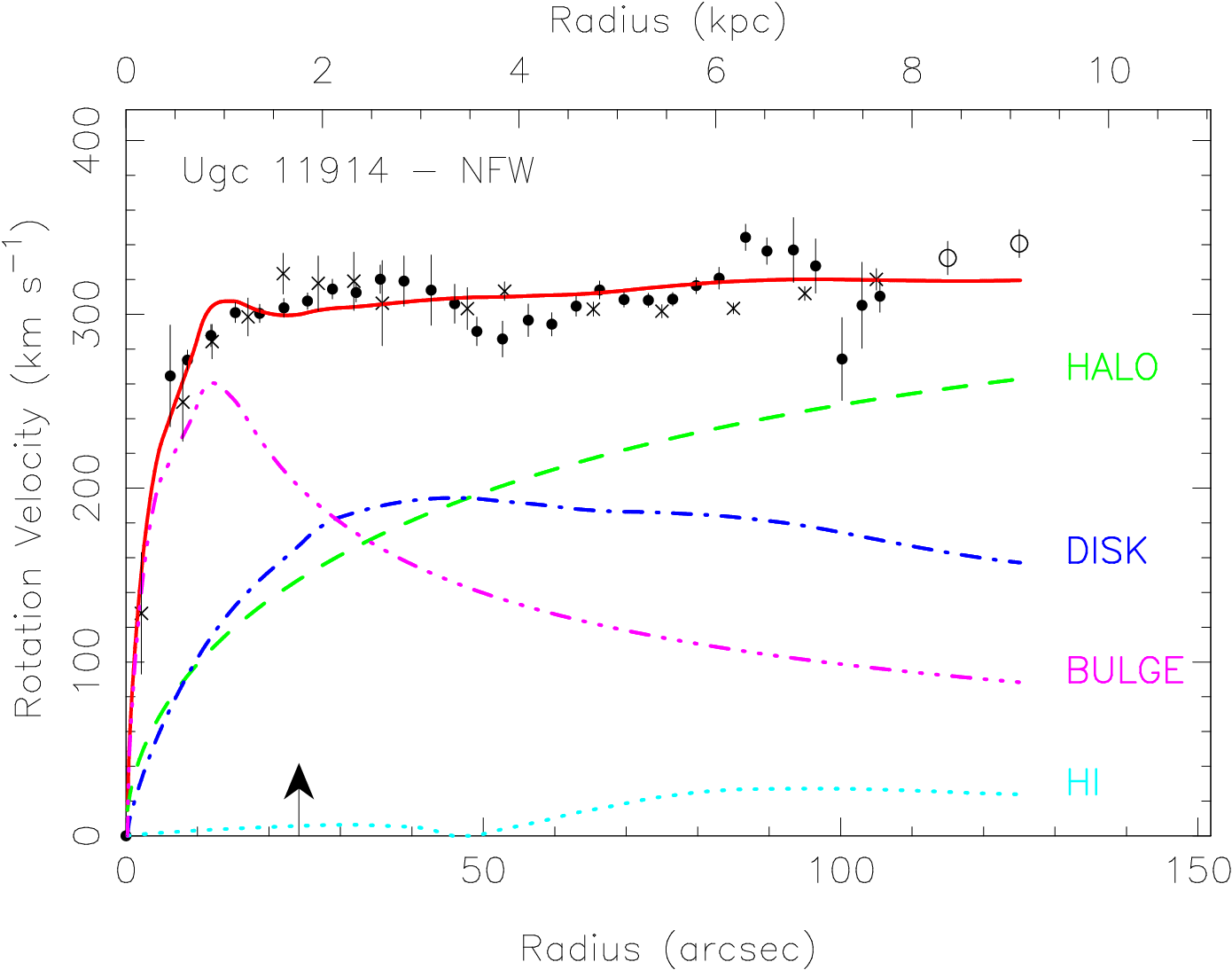}  \\
\includegraphics[width=5.5cm,angle=-90]{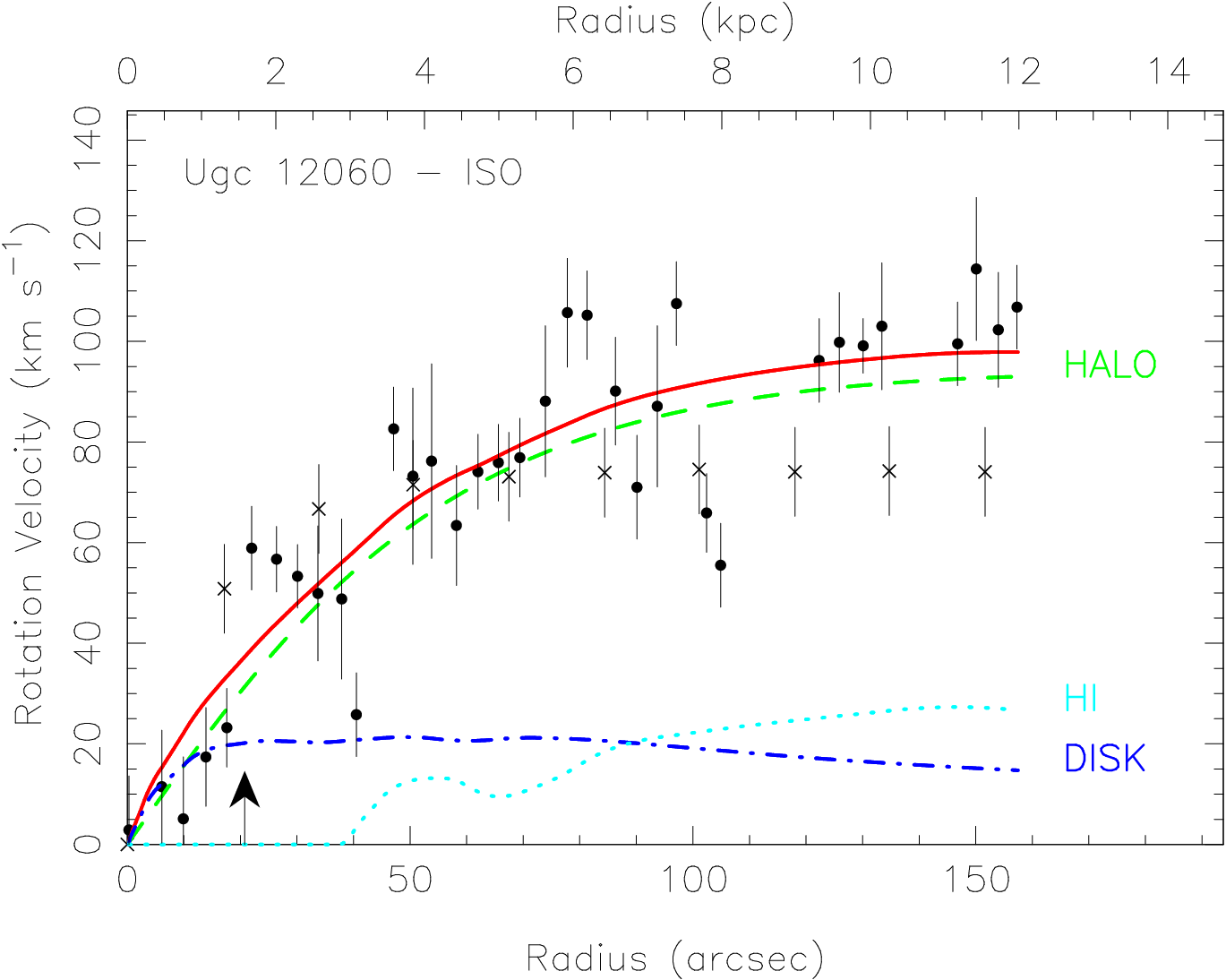}   \includegraphics[width=5.5cm,angle=-90]{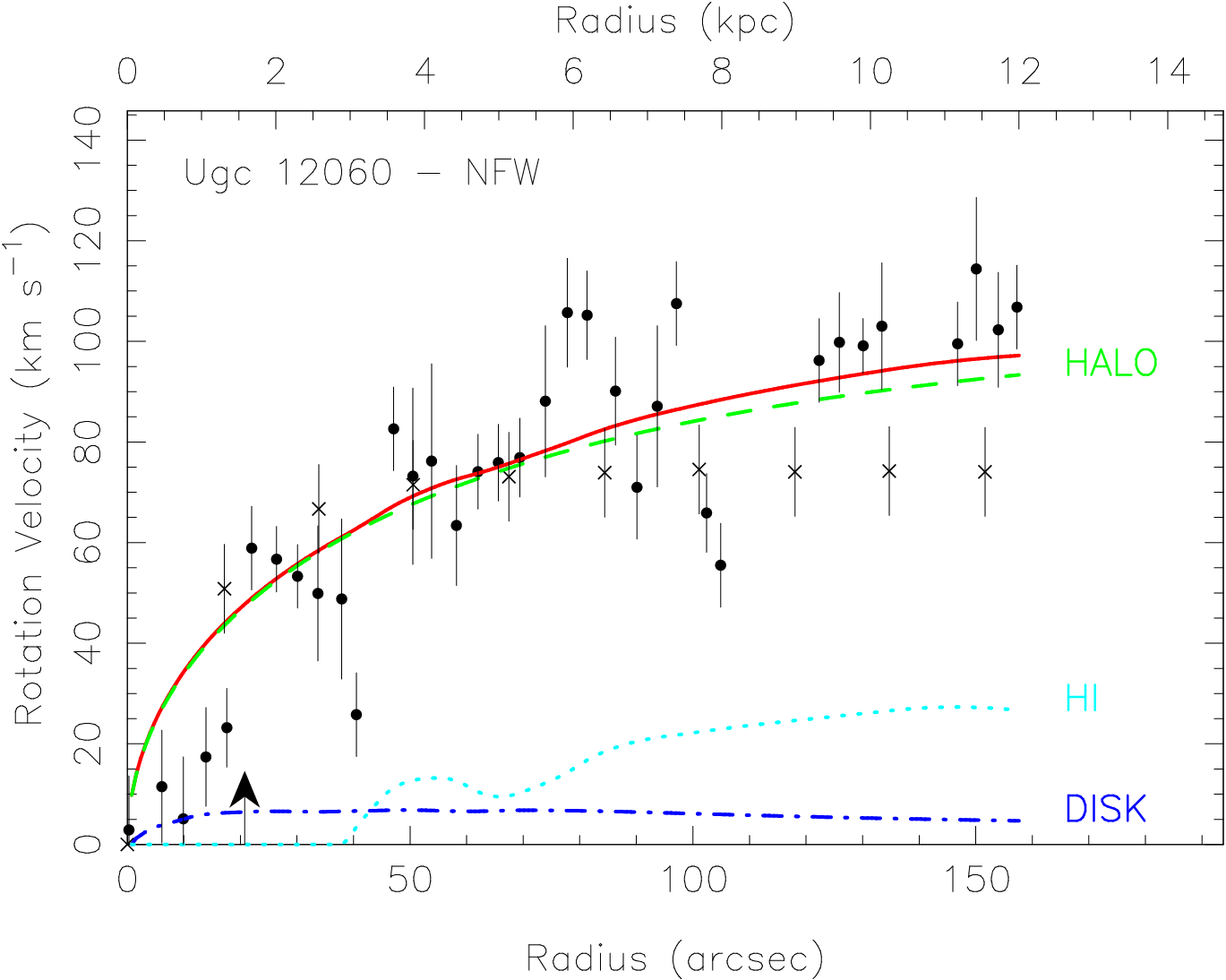}
\end{center}
\caption{Best fit model for the rotation curve of UGC11557,
UGC11707, UGC11914 and UGC12060  with isothermal sphere profile
(left) and NFW profile (right). Dots are for optical velocities,
open circles for HI velocities and crosses for HI in the central
part, not taken into account.The arrow on the X axis indicates the disk scale-length.}
\end{figure*}

\label{lastpage}

\end{document}